\documentclass[12pt,reqno]{amsart}
\usepackage{amssymb,amsmath,amsthm,amsfonts}
\usepackage{graphicx}
\usepackage{enumerate}
\usepackage{bm}
\usepackage{bbm} 
\usepackage{dsfont}
\usepackage[authoryear,longnamesfirst]{natbib}
\usepackage{fullpage}
\usepackage{multirow,multicol,tasks}
\usepackage{tikz}
\usepackage[shortlabels]{enumitem}

\numberwithin{equation}{section}
\newtheorem{theorem}{Theorem}
\newtheorem{algorithm}{Algorithm}

\newtheorem*{notation}{Notations}

\newtheorem{lemma}{Lemma}
\theoremstyle{definition}
\newtheorem{assumption}{Assumption}

\DeclareMathOperator{\Epsilon}{\mathcal{E}}

\newcommand{\norm}[1]{\left \|#1\right \|}
\newcommand{\abs}[1]{\left\vert #1 \right\vert}
\newcommand{\snorm}[1]{\lVert #1 \rVert}

\newcommand{\paren}[1]{\left(#1\right)}
\newcommand{\brac}[1]{\left[ #1 \right]}

\renewcommand{\tilde}{\widetilde}

\newcommand{\calP}[0]{\mathcal{P}}

\newcommand{\R}[0]{\mathbb{R}}

\newcommand{\xkj}{X_{I_k^c j}}

\newcommand{\dk}{D_{I_k^c}}
\newcommand{\zk}{Z_{I_k^c}}

\newcommand{\maxi}{\max_{1\le i \le N}}
\newcommand{\maxj}{\max_{1\le j\le p}}
\newcommand{\sumi}{\sum_{i=1}^N}

\DeclareMathOperator{\argmin}{argmin}
\DeclareMathOperator{\argmax}{argmax}

\renewcommand{\hat}{\widehat}
\renewcommand{\bar}{\overline}

\begin{document}

\title{Inference in High-Dimensional Regression Models \\without the Exact or $L^p$ sparsity}
\thanks{We thank two anonymous referees for very useful comments and suggestions. All the remaining errors are ours. H.D. Chiang is supported by the Office of the Vice Chancellor for Research and Graduate Education at the University of Wisconsin–Madison with funding from the Wisconsin Alumni Research Foundation.}
\author[J. Cha]{Jooyoung Cha}
\author[H. D. Chiang]{Harold D. Chiang}
\author[Y. Sasaki]{Yuya Sasaki}
\date{First arXiv version: This version: December 30, 2022}

\address[J. Cha]{
Department of Economics, Vanderbilt University\\
VU Station B \#351819, 2301 Vanderbilt Place,
Nashville, TN 37235-1819, USA.}
\email{jooyoung.cha@vanderbilt.edu}

\address[H. D. Chiang]{
Department of Economics, University of Wisconsin-Madison\\
 William H. Sewell Social Science Building, 1180 Observatory Drive,
Madison, WI 53706, USA.}
\email{hdchiang@wisc.edu}

\address[Y. Sasaki]{
Department of Economics, Vanderbilt University\\
VU Station B \#351819, 2301 Vanderbilt Place,
Nashville, TN 37235-1819, USA.}
\email{yuya.sasaki@vanderbilt.edu}

%%%%%%%%%%%%%%%%%%%%%%%%%%%%%%%%%%%%%%%%%%%%%%%%%%%%%%%%%%%%%%%%%%%%%
\begin{abstract}
We propose a new inference method in high-dimensional regression models and high-dimensional IV regression models.
The method is shown to be valid without requiring the exact sparsity or $L^p$ sparsity conditions.
Simulation studies demonstrate superior performance of this proposed method over those based on the LASSO or the random forest, especially under less sparse models.
We illustrate an application to production analysis with a panel of Chilean firms.
Our results are closer to the benchmark conventional estimates than the estimates by other machine learning methods.

\bigskip


\medskip
%\noindent {\bf JEL Codes:} 	C??
\end{abstract}

\maketitle
%%%%%%%%%%%%%%%%%%%%%%%%%%%%%%%%%%%%%%%%%%%%%%%%%%%%%%%%%%%%%%%%%%%%%

%%%%%%%%%%%%%%%%%%%%%%%%%%%%%%%%%%%%%%%%%%%%%%%%%%%%%%%%%%%%%%%%%%%%%
\section{Introduction}

The advent of modern machine learning\footnote{\label{foot:ml}We use the phrase ``machine learning'' following the recent related literature, but it is worthy to remark that it is a synonym of semiparametric or high-dimensional `estimation.'} techniques has significantly widened the class of analyzable regression models that include models with high-dimensional controls and/or models with flexible nonlinearity.
Perhaps the most popular and important machine learning approaches to estimation and inference in high-dimensional regression models today are those based on shrinkage and regularization, such as the least absolute shrinkage and selection operation (LASSO).
While they have practically appealing properties, these popular machine learning methods yet rely on a list of assumptions that may not be necessarily mild under certain applications.
In particular, the assumptions of the exact sparsity and the $L^p$ sparsity required for these methods are sometimes controversial and perceived to be strong for some applications.
As such, there still remains room in the literature for further widening the class of analyzable high-dimensional regression models if these assumption can be relaxed by a new machine learning method.

This paper proposes a method of inference in high-dimensional regression models without requiring the exact sparsity or the $L^p$ sparsity.
We set a low-dimensional parameter vector as the object of interest, and treat the remaining high-dimensional parameter vector as a nuisance component.
Under this common setting, our proposed method of inference works as follows.
First, use the orthogonal greedy algorithm \citep[OGA;][]{temlyakov2000weak} to order the high-dimensional regressors in a descending order of explanatory power.
Second, use the high-dimensional Akaike information criterion \citep[HDAIC;][]{ing2020} to select a model among the ordered list of models constructed in the first step.
Third, estimate the models selected in the second step.
Fourth, plug the estimated selected models in a Neyman orthogonal score and estimate the low-dimensional parameter vector of interest.

We take advantage of a number of recent methodological and theoretical developments, namely OGA, HDAIC, and DML, to derive asymptotic statistical properties of this new method of inference.
\citet{ing2020} investigates convergence rate properties of an estimator of high-dimensional regression models based on OGA and HDAIC (hereafter referred to as OGA+HDAIC).
Importantly, the setting imposes assumptions on the high-dimensional parameter vector that are weaker than the exact sparsity and the $L^p$ sparsity. 
Furthermore, this approach does not require to impose high-level conditions on the sample Gram matrix, such as a restricted eigenvalue type condition, that are often required in the literature.
Even under these weaker conditions, it is still possible to obtain similar rates of convergence to those based on existing machine learning techniques such as the LASSO.
Given the adequately fast convergence rates of the preliminary nuisance parameter estimators based on OGA+HDAIC, we can apply the post-double-selection approach \citep*{BCH2014RES} or the double/debiased machine learning (DML) framework \citep*{ccddhnr} to in turn obtain a root-$N$ convergence of the low-dimensional parameter vector of interest with a limit normal distribution.

Simulation studies demonstrate that this proposed method performs significantly better than those based on the LASSO or the random forest, especially when the data generating model becomes less sparse.
We apply the proposed method to an analysis of production functions using a panel of Chilean firms.
Our results are closer to the benchmark conventional estimates \citep{levinsohn2003estimating} than the estimates based on other machine learning methods, namely the LASSO and the random forest.

{\bf Relation to the literature:}
This paper is related to several branches of the econometrics and statistics literature.
First, it is closely related to the literature on inference in high-dimensional regression models and high-dimensional IV regression models, e.g., \cite{belloni2012sparse}, \cite{BCH2014RES}, \cite{javanmard2014confidence}, \cite{van2014asymptotically}, \cite{zhang2014confidence}, \cite{caner2018asymptotically}, \cite{caner2018high}, \cite{belloni2018high}, \cite{galbraith2020simple}, \cite{gold2020inference}, \cite{kueck2021estimation} to list a few. 
The majority of these papers focus on utilizing LASSO \citep{tibshirani1996regression} and its variants for estimation, and thus rely crucially on the exact sparsity, approximate sparsity\footnote{The approximate sparsity is closely related to the exact sparsity.} or $L^p$ sparsity.  
We contribute to this literature, as emphasized above, by relaxing the conventional assumptions of these notions of sparsity.
Second, also related is the literature on model selection methods for high-dimensional models. 
For extensive reviews of this vast literature, we refer readers to the monographs of \cite{buhlmann2011statistics}, \cite{giraud2015introduction}, and \cite{hastie2019statistical}.
In particular, we take advantage of the theoretical results of OGA+HDAIC by \cite{ing2020} as one of the main auxiliary steps to our goal as emphasized earlier. Methodologically, our paper benefits from the theoretical studies of various greedy algorithms in \cite{temlyakov2000weak},  \cite{tropp2004greed}, \cite{tropp2007signal}, \cite{ing2011stepwise}, and share ties with other iterated model selection methods such as the least absolute angle regression of \cite{efron2004least}, the $L_2$-boosting of \cite{buhlmann2003boosting}, the test-based forward model selection of \citet{kozbur2017testing,kozbur2020analysis}, and so forth.
Third, this paper is related to the literature on Neyman orthogonal scores or locally robust scores, e.g., \cite{belloni2015uniform}, \cite{chernozhukov2016locally}, \cite{BCCW2018}, and, in particular, the DML \citep*{ccddhnr}. 
We contribute to this literature by proposing to add OGA+HDAIC to the library of the list of preliminary estimators.
Fourth, the conditions that we impose in place of the exact sparsity and the $L^p$ sparsity concern the speed at which the absolute size of the regression parameters decays when descendingly ordered. These conditions are analogous to those that are used for model selection problems in autoregressive time series models \citep[e.g.,][]{shibata1980asymptotically,ing2007accumulated} as well as the ordinary- and super-smoothness on probability density functions that are used in the deconvolution literature \citep[e.g.,][]{fan1991optimal,fan1993nonparametric}.

%%%%%%%%%%%%%%%%%%%%%%%%%%%%%%%%%%%%%%%%%%%%%%%%%%%%%%%%%%%%%%%%%%%%%
\section{High-Dimensional Linear Regression Models}\label{sec:regression}
%%%%%%%%%%%%%%%%%%%%%%%%%%%%%%%%%%%%%%%%%%%%%%%%%%%%%%%%%%%%%%%%%%%%%
\subsection{The Model}
%%%%%%%%%%%%%%%%%%%%%%%%%%%%%%%%%%%%%%%%%%%%%%%%%%%%%%%%%%%%%%%%%%%%%
Consider the linear regression model
\begin{align}
	Y =& D\theta_0 + X'\Lambda_0 + U,&&E [U|X,D]=0, \label{eq:y}
\end{align}
where $Y$ denotes an outcome variable, $D$ denotes a treatment variable, $X$ denotes a $p$-dimensional vector of controls, and $U$ denotes unobserved factors. 
We allow for a high dimensionality in the sense that $p$ can be increasing in $N$ and may be even larger than $N$ -- more details will follow.
In this framework, we are interested in the partial effect $\theta_0$ of $D$ on $Y$.
Also write the linear projection of $D$ on $X$:
\begin{align}
	D =& X'\beta_0 + V,&&E [V|X]=0, \label{eq:d}
\end{align}
In Section \ref{sec:approx}, we consider an extended model in which we introduce approximation errors in \eqref{eq:y}--\eqref{eq:d}.

To construct a moment restriction under \eqref{eq:y}--\eqref{eq:d}, consider the orthogonal score function from \cite{Robinson}:
\begin{align}\label{eq:robinson}
    \psi(Y,D,X;\theta,\eta) := \left\{Y-X'\gamma-\theta(D-X'\beta) \right\}\left(D-X'\beta \right),
\end{align}
where $X'\gamma_0 = E[Y|X]$ and $\eta=(\gamma,\beta)$.
Note also that $X'\beta_0 = E[D|X]$ follows by construction from \eqref{eq:d}. 
%The true parameters $\theta_0$ and $\eta_0=(\gamma_0,\beta_0)$ satisfy the moment condition
%\begin{align*}
    %E[\psi(Y,D,X;\theta_0,\eta_0)]=0
%\end{align*}
%for this score function.
%Furthermore, it is locally robust (Neyman orthogonal) in the sense that
%\begin{align*}
    %\partial_\eta E[\psi(Y,D,X;\theta_0,\eta_0)][\eta-\eta_0] = 0. 
%\end{align*}
%(In fact, it is also doubly robust.)
%We can thus take advantage of the double/debiased machine learning \citep[DML;][]{ccddhnr} to estimate $\theta_0$ with a suitable machine learning method to estimate the nuisance parameters $\eta_0$, including the LASSO, random forest, and neural networks among others.

\begin{notation}
To proceed, we first fix basic notations.
We use subscripts $i$ and $j$ to denote indices of observations and coordinates, respectively. 
Define $X_{I j}=(X_{ij},\,i\in I)$ as a $\abs{I} \times 1$ vector, $X_{i J} = (X_{i j},\,j\in J)$ as a $\abs{J}\times 1$ vector, and $X_{I J} = (X_{i J},\,i\in I)'$ as a $\abs{I} \times \abs{J}$ matrix, 
where $I$ is a subset of observation indices $\{1,2,...,N\}$, $J$ is a subset of coordinate indices $\mathfrak{P} \equiv \{1,2,\dots,p\}$, and $\abs{.}$ denotes the set cardinality. 
For any vector, $\norm{.}$ refers to the Euclidean norm. 
The $L^q$ norm is defined by $\norm{\xi}_q=(\sum_{j=1}^{p} \xi_j^q)^{1/q}$ for $q < \infty$ and $\norm{\xi}_\infty = \max_{1\le j \le p}\abs{\xi_j}$.
\end{notation}

%%%%%%%%%%%%%%%%%%%%%%%%%%%%%%%%%%%%%%%%%%%%%%%%%%%%%%%%%%%%%%%%%%%%%
\subsection{The Method}
%%%%%%%%%%%%%%%%%%%%%%%%%%%%%%%%%%%%%%%%%%%%%%%%%%%%%%%%%%%%%%%%%%%%%
This section provides an overview of the method.
We propose the following procedure for a root-$N$ consistent estimation and inference about the partial effect $\theta_0$ without assuming sparsity on the high-dimensional parameters, $\beta_0$ or $\gamma_0$.

\begin{algorithm}[OGA+HDAIC with DML for high-dimensional linear models]\label{algorithm:dml_oga_hdaic} \hfill
\begin{enumerate}[label={Step \arabic*.}]
    \item Randomly split the sample indices $\{1,...,N\}$ into $K$ folds $(I_k)_{k=1}^K$. For simplicity, let the size of each fold be $n=N/K$ and the size of $I_k^c$ be $n^c$.
    \item For each fold $k \in \{1,...,K\}$, perform following procedure using $\{(X_i',D_i)'\}_{i\in I_k^c}$ to get $\hat\beta_k$.
    \begin{enumerate}
        \item Compute $\hat{\mu}_{0,j} = \xkj'\dk/\sqrt{n^c}\snorm{\xkj}$. Select the coordinate $\hat{j}_{1} = \argmax_{1\le j \le p} \abs{\hat{\mu}_{0,j}}$. Define $\hat{J}_1 = \{\hat{j}_1\}$.
        \item Compute $\hat{\mu}_{1,j} = \xkj'(I_{n^c} - H_1) \dk / \sqrt{n^c}\snorm{\xkj}$, where $H_1 = X_{I_k^c \hat{j}_1}(X_{I_k^c \hat{j}_1}'X_{I_k^c, \hat{j}_1})^{-1}X_{I_k^c, \hat{j}_1}'$. Select the coordinate $\hat{j}_2 = \argmax_{1\le j \le p,j\notin \hat{J}_1} \abs{\hat{\mu}_{1,j}}$. Update $\hat{J}_2 =\hat{J}_1\cup \{\hat{j}_2\}$.
        \item Given $m-1$ coordinates $\hat{J}_{m-1}$ that have been obtained, compute $\hat{\mu}_{m-1,j} = \xkj'(I_{n^c}- H_{m-1}) \dk / \sqrt{n^c}\snorm{\xkj}$, where $H_{m-1} = X_{I_k^c \hat{J}_{m-1}}(X_{I_k^c \hat{J}_{m-1}}'X_{I_k^c \hat{J}_{m-1}})^{-1}X_{I_k^c \hat{J}_{m-1}}'$. Select the coordinate $\hat{j}_m = \argmax_{1\le j\le p, j\notin \hat{J}_{m-1}}\abs{\hat{\mu}_{m,j}}$. Iteractively update $\hat{J}_m = \hat{J}_{m-1}\cup \{\hat{j}_m\}$.
        \item Compute $\textup{HDAIC}$ $(\hat{J}_m) = (1+C^* |\hat{J}_m|\log p/n^c)\hat{\sigma}_{m}^2$ for each $m$, where $C^*$ is from (\ref{Cstar}) in Appendix \ref{sec:regression:details} and $\hat{\sigma}_{m}^2 = 1/n^c \dk '(I-H_m)\dk$. Choose $\hat{m} = \argmin_{1\le m\le M_n^*}$ $\textup{HDAIC} (\hat{J}_m)$, where $M_n^*$ is defined in (\ref{Mnstar}) in Appendix \ref{sec:regression:details}.
		\item With coordinates $\hat{J}_{\hat{m}}$, run OLS of $D_i$ on $X_{i\hat{J}_{\hat{m}}}$ to get $\hat{\beta}_k.$
    \end{enumerate}
    \item Repeat Step 2 with $\{(X_i',Y_i)'\}_{i\in I_k^c}$ instead of $\{(X_i',D_i)'\}_{i\in I_k^c}$, to get $\hat{\gamma}_k$ for each fold $k \in \{1,...,K\}$.
    \item Obtain $\check{\theta}$ as a solution to $1/K \sum_{k=1}^K 1/n \sum_{i\in I_k}\psi(Y_i,D_i,X_i;\check{\theta},\hat{\eta}_k)=0$ where $\hat{\eta}_k=(\hat{\gamma}_k,\hat{\beta}_k)$ and $\psi$ is defined in \eqref{eq:robinson}.
    \item Compute $\hat{M} = -1/K\sum_{k=1}^K 1/n\sum_{i\in I_k}(D_i - X_i'\hat{\beta})^2$. Obtain a variance estimator of $\check{\theta}$ as $\hat{\Omega} = \hat{M}^{-1}\frac{1}{K} \sum_{k=1}^K \frac{1}{n}\sum_{i\in I_k} [\psi(Y,D,X;\check{\theta},\hat{\eta}_k)\psi(Y,D,X;\check{\theta},\hat{\eta}_k)'](\hat{M}^{-1})'$.
\end{enumerate}
\end{algorithm}

We highlight three notable elements of this algorithm.
First, the overall procedure (Steps 1--4) uses the cross fitting to remove an over-fitting bias.
Specifically, by using complementary sub-sample $I_k^c$ to estimate the nuisance parameters $\hat\eta_k=(\hat\gamma_k,\hat\beta_k)$ that are in turn evaluated in the $I_k$-mean of the score, we can circumvent a bias that arises from products of dependent factors in the score.
Our combined use of the orthogonal score \eqref{eq:robinson} and this cross-fitting method allows for the high-level theory of the double/debiased machine learning \citep[DML,][]{ccddhnr} to be applicable.
Section \ref{sec:without_cross_fitting} discusses an alternative algorithm that does not rely on the cross fitting at the cost of an additional assumption.
Second, the coordinates $\{\hat j_1,...,\hat j_{p}\}$ are ranked in Step 2 (a)--(c) in the order of decreasing importance after successive orthogonalization using OGA as in \cite{ing2020}.
Third, a subset $\hat J_{\hat m} = \{\hat j_1,...,\hat j_{\hat m}\}$ of the ordered set $\{\hat j_1,...,\hat j_{p}\}$ is selected in Step 2 (d) using HDAIC as in \cite{ing2020}.
Our combined use of these three elements (DML, OGA, and HDAIC) together allows for a novel root $N$ consistent estimation of $\theta_0$ without assuming traditional functional class restrictions (e.g., the sparsity) required by existing popular estimators (e.g., LASSO).
In Section \ref{sec:regression:theory}, we formally present theoretical arguments in support of this claim.

While Algorithm \ref{algorithm:dml_oga_hdaic} provides nearly full details of the proposed method, it omits a couple of details.
Specifically, Step 2 (d) on HDAIC uses two tuning parameters, $C^*$ and $M_n^*$.
We present details about these aspects of the algorithm in Appendix \ref{sec:regression:details}.

%%%%%%%%%%%%%%%%%%%%%%%%%%%%%%%%%%%%%%%%%%%%%%%%%%%%%%%%%%%%%%%%%%%%%
\subsection{The Theory}\label{sec:regression:theory}
%%%%%%%%%%%%%%%%%%%%%%%%%%%%%%%%%%%%%%%%%%%%%%%%%%%%%%%%%%%%%%%%%%%%%
This section proposes and discusses assumptions under which one can conduct an inference about $\check{\theta}$ based on root-$N$ asymptotic normality using the method described in Algorithm \ref{algorithm:dml_oga_hdaic}.
We use the notations $c,C,\bar{C},\bar{\tau}$ and $q$ for strictly positive constants such that their values can differ depending on the location. Let $q>4$ be a positive integer, $c_q$, $C_q$, $\lambda_1$ be some positive constants and $K_{N,q}$ be a positive sequence of constants such that $K_{N,q}\ge E[\max_{1\le j\le p}\abs{X_{ij}}^q]$.
Wherever there is no risk of confusion, 
% we also use the generic notation $\Epsilon$ to refer to both $\Epsilon_D = D - X'\beta$ and $\Epsilon_Y = Y - X'\gamma$ to avoid repetitive presentations. Similarly, 
we also use the generic notation $\xi$ to refer to both $\beta$ and $\gamma$ to avoid repetitions. All the random variables and parameter vectors are $N$-dependent unless otherwise specified. We abbreviate the $N$ index for brevity.

\begin{assumption}\label{a:1}
	For each $N\in \mathbb N$, it holds that
\begin{enumerate}[(a)]%[counter-format=(\alph*),item-indent = 0.55cm,label-offset = 5pt](2) 
    \item $(Y_i,D_i,X_i')_{i=1}^N$ are i.i.d. copies of $(Y,D,X')$. 
    \item \eqref{eq:y} and \eqref{eq:d} hold. 
    \item $E[\abs{Y}^q] + E[\abs{D}^q] \le C_q$. 
    \item $E[\abs{UV}^2]\ge c_q^2$ and $E[V^2]\ge c_q$.
    \item ${\max_{1\le j\le p}E[|X_{ij}|^q]}\le C_q$, $E[\abs{V}^q] \le C_q$, and $E[\abs{U}^q]\le C_q$. \label{qth_bdd}
\end{enumerate}
Furthermore, it holds asymptotically that
(f) $K_{N,q}^2 \log p/ N^{1-2/q}=o(1).$ \label{maxqth_bdd}
\end{assumption}

Assumption \ref{a:1} (a) requires a random sampling of data.
Assumption \ref{a:1} (b) requires that the correct model is given by \eqref{eq:y} and \eqref{eq:d}.
Assumption \ref{a:1} (c)--(e) requires bounded moments of various variables.
Assumption \ref{a:1} (f) requires constraints on the speed at which the dimensionality as well as the maximal of the covariate vector can grow. Vectors consist of independent subgaussian random variables with bounded variances, for example, are special cases satisfying this restriction, as the expectation of the maximum of $X_{ij}$ is bounded by a factor of $\sqrt{\log p}$. 
We emphasize that these assumptions are mild in comparison with the counterpart assumptions made in the high-dimensional regression literature.
%We also remark that Assumption \ref{a:1} (b)--(e) corresponds to assumption 4.1 (a)--(d) in \cite{ccddhnr}, and Assumption \ref{a:1} \ref{qth_bdd} implies (A1) and (A2) in \citet{ing2020}.

\begin{assumption}\label{a:2}
It holds over $N\in \mathbb N$ that
\begin{enumerate}[(a)]%[counter-format = (\alph*),item-indent=.55cm,label-offset=5pt]
    % \item $E[\max_{1\le j\le p}\abs{X_{ij}}^q]\le C,\, E[\abs{V}^q] \le C$, $E[\abs{Y-X'\gamma_0}^q]\le C$ and $\log p = o(N^{1/4}).$
    \item $\lambda_{\min} (\Gamma)\ge \lambda_1>0$ and $\lambda_{\max} (\Gamma) \le C_q$, where $\Gamma=E[X X']$. \label{additionalA}
    \item Define $\Gamma(J) = E[X_{iJ}X_{iJ}']$ and $d_{\ell}(J) = E[X_{i\ell}X_{iJ}]$ for a set of coordinate indices $J\subseteq \mathfrak{P}$. Then $$\max_{1\le \abs{J}\le \bar{C}(N/\log p)^{1/2},\,\ell \notin J} \abs{\Gamma^{-1}(J)d_\ell(J)}< C_q.$$ \label{IngA5}
\end{enumerate}
\end{assumption}\vspace{-.6cm}

Assumption \ref{a:2} \ref{additionalA} requires that the minimum eigenvalue $\lambda_{\min}(\Gamma)$ of the Gram matrix $\Gamma$ to be positive, and it is a very common restriction.
Assumption \ref{a:2} \ref{IngA5} is a restriction on the covariance structure of $X_{iJ}$. 
Observe that $\Gamma^{-1}(J)d_\ell(J)$ takes the form of regression coefficient of $X_{i\ell}$ on $X_{iJ}$, and so Assumption \ref{a:2} \ref{IngA5} means that $X_{i\ell}$ cannot be strongly correlated with $X_{iJ}$ for $\ell \notin J$. 
We remark that these conditions are imposed at the population level; unlike in the LASSO or Dantzig selector \citep{candes2007dantzig}, a restricted eigenvalue type condition for the sample Gram matrix \citep[see e.g.,][]{bickel2009simultaneous} is not required here -- see also the discussion in \citet[Sec. 3.2]{ing2020}. 
%Verifying these type of conditions is often not straightforward. Although some sufficient conditions for restricted eigenvalue type conditions have been derived for certain covariance structure under specific rate and tail assumptions in the literature -- such as in \cite{mendelson2008uniform} under a sub-gaussian condition and \cite{rudelson2008sparse} under a uniformly boundedness condition -- to our knowledge, a unified treatment of restricted eigenvalue condition for random designs under general distributional conditions remains unknown. In contrast, Assumption \ref{a:2} provides general yet tractable conditions that accommodate a large class of models. 
%We also remark that Assumption \ref{a:2} \ref{A5} corresponds to (A5) in \cite{ing2020}, and Assumption \ref{a:2}\ref{additionalA} corresponds to additional assumptions required in Theorem 3.1. in \citet{ing2020}.

The following assumption imposes restrictions on the function classes in terms the parameters $\beta_0$ and $\gamma_0$. 
We will use the generic notation $\xi_0$ to refer to $\beta_0$ and $\gamma_0$.
Note that $\xi_0 \in \R^p$ in both cases. 
Define $\xi(J) = (\xi_j)_{j\in J}$ to be a $\abs{J}\times 1$ vector, where recall that $\xi$ is a generic notation to refer to $\beta_0$ and $\gamma_0$.

\begin{assumption}\label{a:3}
	It holds over $N\in \mathbb N$ that
for each of $\xi_0 = \beta_0$ and $\gamma_0$, $\xi_0$ follows either (a) or (b) described below. 
\begin{enumerate}[(a)]%[counter-format = (\alph*),item-indent=.55cm,label-offset=5pt]
    \item Polynomial decay: $\log p = o(N^{1-2/q})$. Each $\xi_0$ is such that $\norm{\xi_0}_2^2 \le C_0$ for some $C_0>0$ and there exist $\alpha> 1$ such that for any $J \subseteq \mathfrak{P}$,$$\hspace{2cm}\norm{\xi_0(J)}_1 \le C \left( \norm{\xi_0(J)}_2^2\right)^{(\alpha-1)/(2\alpha-1)}.\label{poly} $$ 
    
    \item Exponential decay: $\log p= o(N^{1/4}).$ Each $\xi_0$ is such that $ \norm{\xi_0}_\infty \le C_0$ for some $C_0>0$ and there exists $C_1>1$ such that for any $J \subseteq \mathfrak{P}$, $$\norm{\xi_0(J)}_1 \le C_1 \norm{\xi_0(J)}_\infty.\label{expo}$$ 
\end{enumerate}
\end{assumption}

This is a key assumption in this paper, and defines admissible function classes for the high-dimensional linear models.
While the literature on LASSO requires the exact sparsity and the $L^p$ sparsity (including approximate sparsity) conditions, Assumption \ref{a:3} does not impose such conditions.
%{\color{red} [TO BE MODIFIED] Instead, it requires that the coefficients $\beta_0$ and $\beta_0$ after descending reordering follows either a polynomial decay (Assumption \ref{a:3} (a)) or an exponential decay (Assumption \ref{a:3} (b)). [HDC: WITH THESE GENERAL CONDITIONS, DOES (b) STILL IMPLY (a)???]}
 We remark that, if we rearrange the components of the parameter vector $\xi_0$ by their absolute values in a descending order (denote it again as $\xi_0$ with an abuse of notation), then Condition (a) contains special cases such as the conventional polynomial decay condition
\begin{align*}
Lj^{-\alpha} \le |\xi_{0j}|\le Uj^{-\alpha}, \quad 0<L\le U <\infty,
\end{align*}
as well as the polynomial summability condition
\begin{align*}
\sum_{j=1}^p |\xi_{0j}|^{1/\alpha} < M,\quad M\in(0,\infty),
\end{align*}
following the discussion in \citet[pp. 1962]{ing2020}. On the other hand, Condition (b) implies the conventional exponential decay condition that, for some $\alpha'>0$, 
\begin{align*}
L'\exp(-\alpha' j) \le |\xi_{0j}|\le U'\exp(-\alpha' j) ,\quad 0<L'\le U' <\infty,
\end{align*}
so long as the regressors have bounded second moments.
Hence  throughout the paper, Conditions (a) and (b) are referred to as the polynomial decay condition and the exponential decay condition, respectively, albeit their extra generality.  Clearly, the case of polynomial decay accommodates a larger function class, but we remark that there is a tradeoff in terms of how fast the dimension $p$ can diverge as the sample size $N$ increases.

Following \citet[][pp. 1960]{ing2020}, we now present a concrete example where our Assumption \ref{a:3} holds but the sparsity does not. 
Suppose that  
\begin{align*}
	L j^{-\alpha} \le \abs{\Lambda_{0(j)}\sigma_{(j)}} \le U j^{-\alpha},&&j=1,\dots,p,
\end{align*} 
for some $\alpha>1$, where $0< L \le U < \infty$, and $|\Lambda_{0(1)}\sigma_{(1)}|\ge |\Lambda_{0(2)}\sigma_{(2)}| \ge \dots \ge |\Lambda_{0(p)}\sigma_{(p)}|$ is a descending reordering of $\{\Lambda_{0,j}\sigma_j\}$ with $\sigma_j^2 = E[X_{ij}^2]$.
In this setting, our Assumption \ref{a:3} holds for the same $\alpha$, but $\sum_{j=1}^p \abs{\Lambda_{0,j}\sigma_j}^{1/\gamma}$ is now unbounded as $L(1+\log p)\le \sum_{j=1}^p \abs{\Lambda_{0,j}\sigma_j}^{1/\gamma} \le U(1+\log p).$

%\textcolor{red}{Add a discussion that this assumption is weaker than the exact or $L^p$ sparsity.}
%Assumption \ref{a:3} (a) and (b) corresponds to (A3) and (A4) in \citet{ing2020} respectively. 

\begin{theorem}\label{theorem:regression} Let $(\calP_N)_{N\in \mathbb N}$ be a sequence of sets of DGPs such that Assumptions \ref{a:1}--\ref{a:3} are satisfied on the model \eqref{eq:y}--\eqref{eq:d}. Then, the estimator $\check{\theta}$ satisfies
\begin{align*}
    \sqrt{N} \paren{\check{\theta}-\theta_0} \xrightarrow{d} N(0,\Omega),
\end{align*}
where $\Omega = (E[V^2])^{-1}E[V^2U^2](E[V^2])^{-1}$. 
Define $\hat{M} :=- 1/K\sum_{k=1}^K 1/n \sum_{i\in I_k} (D_i-X_i'\hat{\beta})^2$. Then, we can define the variance estimator \begin{align*}
    \hat{\Omega} = \hat{M}^{-1}\frac{1}{K} \sum_{k=1}^K \frac{1}{n}\sum_{i\in I_k} [\psi(Y,D,X;\check{\theta},\hat{\eta}_k)\psi(Y,D,X;\check{\theta},\hat{\eta}_k)'](\hat{M}^{-1})'
\end{align*}
and the confidence regions with significance level $a\in (0,1 )$ have uniform asymptotic validity:
\begin{align*}
    \sup_{P\in \calP_N} \abs{P\left(\theta_0 \in \left[\check{\theta} \pm \Phi^{-1}(1-a/2)\sqrt{\hat{\Omega}/N}\right]\right)-(1-a)} =o(1).
\end{align*}
\end{theorem}

A proof is provided in Appendix \ref{sec:theorem:regression}.
This theorem guarantees that the estimator $\check{\theta}$ of $\theta_0$ provided by Algorithm \ref{algorithm:dml_oga_hdaic} converges at the rate of $\sqrt{N}$ and asymptotically follows the normal distribution under Assumption \ref{a:1}--\ref{a:3}.
Furthermore, the sample-counterpart asymptotic variance estimator constructs an asymptotically valid confidence interval.
We emphasize that this result does not rely on the sparsity assumption which is used in the literature on high-dimensional linear models.

%%%%%%%%%%%%%%%%%%%%%%%%%%%%%%%%%%%%%%%%%%%%%%%%%%%%%%%%%%%%%%%%%%%%%
\section{Simulation Studies}\label{sec:simulations}
%%%%%%%%%%%%%%%%%%%%%%%%%%%%%%%%%%%%%%%%%%%%%%%%%%%%%%%%%%%%%%%%%%%%%
In this section, we investigate the finite sample properties of our proposed estimator $\check{\theta}$ and compare them with those of two existing estimators, namely the LASSO-based DML and random-forest-based DML.\footnote{For these two existing methods, we use the R package ``DoubleML : Double Machine Learning in R.''}

% Parameters: $C^*=2$
We follow \cite{BCH2014RES} in developing baseline data generating processes (DGPs).
The linear regression model is specified by
\begin{align*}
    Y =& D\theta_0 + X'\Lambda_0 + U,
\end{align*}    
where $\theta_0=0.5$ and $p = dim(X) = 500$.
Consistently with this specification, data are generated by the system
\begin{align*}
    Y =& \theta_0(D-X'\beta_0) + X'\gamma_0 + U, &&U\sim N(0,1),
		\\
    D =& X'\beta_0 + V, &&V \sim N(0,1),
\end{align*}
where the covariates are in turn generated by $X \sim N(0,\Sigma)$ with $\Sigma_{jk}=(0.5)^{\abs{k-j}}$.

For the high-dimensional nuisance parameters, $\eta_0 = (\gamma_0,\beta_0)$, we set $p=500$ throughout and consider a couple of alternative designs.
In the first design, each of $\beta_0$ and $\gamma_0$ has ten coordinates taking the value of 1 and $p-10$ coordinates taking the value of zero, i.e., sparse deign.
In the second design, both $\beta_0$ and $\gamma_0$ decay exponentially.
Specifically, the $j$-th coordinate of each of $\beta_0$ and $\gamma_0$ is set to $e^{-j}$.
The third design has both $\beta_0$ and $\gamma_0$ decaying at polynomial rates. 
Specifically, the $j$-th coordinate of each of $\beta_0$ and $\gamma_0$ is set to $j^{-2}$, $j^{-1.75}$, $j^{-1.5}$, $j^{-1.25}$ and $j^{-1}$ for five sets of simulations. 
For each of these sets of simulations, we experiment with the two sample sizes $N \in \{500,1000\}$.

%%%%%%%%%%%%%%%%%%%%%%%%%%%%%%%%%%%%%%%%%%%%%%%%%%%%%%%%%%%%%%%%%%%%%
\begin{table}
	\centering
	\renewcommand{\arraystretch}{0.54} 
	\scalebox{1}{
		\begin{tabular}{ccclcccc}
		\hline\hline
		  $\beta_{0,j},\gamma_{0,j}$ & $N$ & $p$ & Method of Preliminary Estimation & Bias & SD & RMSE & 95\%\\
		\hline
			Sparse & 500 & 500 & LASSO                       & 0.020 & 0.044 & 0.049 & 0.937\\
			           &     &     & Random Forest           & -0.481& 0.000 & 0.498 & 0.000\\
			           &     &     & OGA+HDAIC               & -0.003 & 0.045 & 0.046 & 0.943\\
		\cline{2-8}
			           &1000 & 500 & LASSO                       & 0.012 & 0.031 & 0.033 & 0.929\\
			           &     &     & Random Forest               & -0.347& 0.003 & 0.485 & 0.000\\
			           &     &     & OGA+HDAIC                & 0.000 & 0.032 & 0.032 & 0.947\\
		\hline
			$e^{-j}$   & 500 & 500 & LASSO                       & 0.006 & 0.044 & 0.045 & 0.934\\
			           &     &     & Random Forest               & 0.008 & 0.044 & 0.045 & 0.940\\
			           &     &     & OGA+HDAIC                & 0.000 & 0.045 & 0.045 & 0.941\\
		\cline{2-8}
			           &1000 & 500 & LASSO                       & 0.007 & 0.031 & 0.032 & 0.942\\
			           &     &     & Random Forest               & 0.006 & 0.031 & 0.031 & 0.950\\
			           &     &     & OGA+HDAIC                & 0.000 & 0.032 & 0.032 & 0.950\\
%		\hline
%			$j^{-4}$ & 500 & 500 & LASSO                       & 0.004 & 0.044 & 0.045 & 0.935\\
%			         &     &     & Random Forest               & 0.004 & 0.044 & 0.045 & 0.951\\
%			         &     &     & OGA+HDAIC                &-0.001 & 0.045 & 0.045 & 0.941\\
%		\cline{2-8}
%			         &1000 & 500 & LASSO                       & 0.006 & 0.031 & 0.032 & 0.944\\
%			         &     &     & Random Forest               & 0.002 & 0.031 & 0.031 & 0.945\\
%			         &     &     & OGA+HDAIC                & 0.001 & 0.032 & 0.032 & 0.950\\
		\hline
			$j^{-2}$ & 500 & 500 & LASSO                       & 0.010 & 0.044 & 0.046 & 0.934\\
			         &     &     & Random Forest               & 0.033 & 0.043 & 0.055 & 0.878\\
			         &     &     & OGA+HDAIC                &-0.002 & 0.045 & 0.046 & 0.938\\
		\cline{2-8}
			         &1000 & 500 & LASSO                       & 0.009 & 0.031 & 0.032 & 0.939\\
			         &     &     & Random Forest               & 0.025 & 0.031 & 0.039 & 0.882\\
			         &     &     & OGA+HDAIC                & 0.001 & 0.032 & 0.032 & 0.945\\
		\hline
		 $j^{-1.75}$& 500 & 500 & LASSO                       & 0.016 & 0.044 & 0.047 & 0.931\\
			         &     &     & Random Forest               & 0.046 & 0.043 & 0.063 & 0.818\\
			         &     &     & OGA+HDAIC                & -0.001 & 0.045 & 0.047 & 0.930\\
		\cline{2-8}
			         &1000 & 500 & LASSO                       & 0.011 & 0.031 & 0.033 & 0.932\\
			         &     &     & Random Forest               & 0.035 & 0.031 & 0.046 & 0.805\\
			         &     &     & OGA+HDAIC                & 0.001 & 0.031 & 0.032 & 0.951\\
		\hline
		 $j^{-1.5}$& 500 & 500 & LASSO                        & 0.020 & 0.044 & 0.049 & 0.931\\
			         &     &     & Random Forest               & 0.066 & 0.042 & 0.079 & 0.651\\
			         &     &     & OGA+HDAIC                & 0.001 & 0.045 & 0.046 & 0.936\\
		\cline{2-8}
			         &1000 & 500 & LASSO                       & 0.013 & 0.031 & 0.034 & 0.928\\
			         &     &     & Random Forest               & 0.053 & 0.030 & 0.060 & 0.576\\
			         &     &     & OGA+HDAIC                & 0.002 & 0.031 & 0.033 & 0.938\\
		\hline
		 $j^{-1.25}$& 500 & 500 & LASSO                       & 0.028 & 0.044 & 0.053 & 0.904\\
			         &     &     & Random Forest               & 0.108 & 0.041 & 0.115 & 0.245\\
			         &     &     & OGA+HDAIC                & 0.006 & 0.044 & 0.047 & 0.933\\
		\cline{2-8}
			         &1000 & 500 & LASSO                       & 0.018 & 0.031 & 0.036 & 0.909\\
			         &     &     & Random Forest               & 0.096 & 0.029 & 0.100 & 0.083\\
			         &     &     & OGA+HDAIC                & 0.004 & 0.031 & 0.034 & 0.923\\
		\hline
			$j^{-1}$ & 500 & 500 & LASSO                       & 0.038 & 0.043 & 0.061 & 0.827\\
			         &     &     & Random Forest               & 0.193 & 0.037 & 0.196 & 0.001\\
			         &     &     & OGA+HDAIC                & 0.022 & 0.043 & 0.053 & 0.893\\
		\cline{2-8}
			         &1000 & 500 & LASSO                       & 0.026 & 0.031 & 0.042 & 0.838\\
			         &     &     & Random Forest               & 0.179 & 0.026 & 0.181 & 0.000\\
			         &     &     & OGA+HDAIC                & 0.014 & 0.031 & 0.037 & 0.901\\
		\hline\hline
		\end{tabular}
	}
	\caption{\setlength{\baselineskip}{5.5mm}Monte Carlo simulation results. Displayed are Monte Carlo simulation statistics including the bias, standard deviation (SD), root mean square error (RMSE), and 95\% coverage frequency.}
	\label{tab:simulation}
\end{table}
%%%%%%%%%%%%%%%%%%%%%%%%%%%%%%%%%%%%%%%%%%%%%%%%%%%%%%%%%%%%%%%%%%%%%

Table \ref{tab:simulation} summarize simulation results.
Displayed are four Monte Carlo simulation statistics for each set of simulations, including the bias, standard deviation (SD), root mean square error (RMSE), and 95\% coverage frequency.
In the first row group of the table displaying the results the sparse design, both LASSO-based method and our proposed method based on the OGA and HDAIC work well, while that based on Random Forest significantly underperforms.
In the second row group of the table displaying the results under the exponential decay,
all the three machine learning methods yield desired results both in terms of all the displayed statistics.
There are no significant differences across the three methods under this sparse model.
In the subsequent row groups of the table displaying the results for the cases of the polynomial decays, however, observe that the performance varies across the three machine learning methods.
While our proposed method based on the OGA and HDAIC continues to perform well in terms of all the displayed statistics, the LASSO-based method slightly underperforms and the random-forest-based method significantly underperforms.
In particular, these differences in the finite-sample performance widen as the degree of polynomial decay becomes smaller, i.e., as the model becomes less sparse.
These results demonstrate the relative robustness of the method proposed in this paper under less sparse high-dimensional regression models.

We ran many other sets of simulations and present their results in Appendix \ref{sec:additional_simulations}.
In particular, Appendix \ref{sec:simulation:tuning} presents simulation results with various values of the tuning parameters, and demonstrate the robustness of the qualitative patterns observed above.
From these results, we recommend to use the method based on the OGA and HDAIC over the two alternative methods for its robust performance across various designs even including the sparse design.

%%%%%%%%%%%%%%%%%%%%%%%%%%%%%%%%%%%%%%%%%%%%%%%%%%%%%%%%%%%%%%%%%%%%%
\section{Extensions} \label{sec:extensions}
%%%%%%%%%%%%%%%%%%%%%%%%%%%%%%%%%%%%%%%%%%%%%%%%%%%%%%%%%%%%%%%%%%%%%

%%%%%%%%%%%%%%%%%%%%%%%%%%%%%%%%%%%%%%%%%%%%%%%%%%%%%%%%%%%%%%%%%%%%%
\subsection{Models with Approximation Errors}\label{sec:approx}
%%%%%%%%%%%%%%%%%%%%%%%%%%%%%%%%%%%%%%%%%%%%%%%%%%%%%%%%%%%%%%%%%%%%%
%\subsubsection{The Model}
%%%%%%%%%%%%%%%%%%%%%%%%%%%%%%%%%%%%%%%%%%%%%%%%%%%%%%%%%%%%%%%%%%%%%
% Consider the following linear regression models:
% \begin{align}
% 	Y =& D\theta_0 + X'\Lambda_0 + U,&&E [U|X,D]=0, \label{eq:y:original}\\
%     D =& X'\beta_0 + V,&&E [V|X]=0, \label{eq:d:original}
% \end{align}
% where $Y$ denotes an outcome variable, $D$ denotes a treatment variable, $X$ denotes a $p$-dimensional vector of controls, and $U$ and $V$ denotes unobserved factors. 
% We allow for a high dimensionality in the sense that $p$ can be increasing in $N$ and may be even larger than $N$ -- more details will follow.
% In this framework, we are interested in the partial effect $\theta_0$ of $D$ on $Y$.

% To incorporate the approximation errors, c

Extending the baseline model \eqref{eq:y}--\eqref{eq:d}, consider the following partially linear model motivated by \cite{BCH2014RES}:
\begin{align}
	Y =& D\theta_0 + f(X) + U,&&E [U|X]=0, \label{eq:y:approx:pl}\\
    D =& g(X) + V,&&E [V|X]=0, \label{eq:d:approx:pl}
\end{align}
where $Y$ denotes an outcome variable, $D$ denotes a treatment variable, $X$ denotes a $p$-dimensional vector of controls, and $U$ and $V$ denotes unobserved factors.
We do not directly impose any parametric restriction on $f$ or $g$ unlike the baseline model presented in Section \ref{sec:regression}.
This extension is useful in certain applications, such as the one we present in Section \ref{eq:application}.
In this semi-parametric framework, we are interested in the partial effect $\theta_0$ of $D$ on $Y$.
% {\color{blue}[TO DO: Motivate by BCH 2014 RES??]}

Now consider the followng reduced form regressions for \eqref{eq:y:approx:pl}--\eqref{eq:d:approx:pl}:
\begin{align}
	Y =& \underbrace{X'\gamma_0 + r_Y(X)}_{f(X)+\theta_0 g(X)} + \Epsilon,&&E [\Epsilon|X]=0, \label{eq:y:approx}\\
    D =& \underbrace{X'\beta_0 + r_D(X)}_{g(X)} + V,&&E [V|X]=0, \label{eq:d:approx}
\end{align}
where $X'\gamma_0$ and $X'\beta_0$ are approximations to $E[Y|X]$ and $E[D|X]$, and $r_{Y}(X)$ and $r_D(X)$ are approximation errors.
The functions $r_{Y}$ and $r_{D}$ are nonparametric as are $f$ and $g$. {\color{black}We will impose conditions on the magnitudes of $r_Y$ and $r_D$ below. Models under these conditions, along with certain sparsity conditions imposed on $\beta_0$ and $\gamma_{0}$, are said to be ``approximate sparse'' in \cite{belloni2012sparse,BCH2014RES}.} 
%As in the baseline model, $\gamma_0$ and $\beta_0$ are allowed to be non-sparse.
%Note that while it does not include the parameter of interest, \eqref{eq:y:approx} is the actual regression we estimate, since we adopt the orthogonal score to compute the parameter of interest in the end.

Recall the orthogonal score $\psi(Y,D,X;\theta,\eta)$ defined in \eqref{eq:robinson}.
%Following \cite{Robinson}, we consider the following orthogonal score function to construct a moment restriction under \eqref{eq:y:approx}--\eqref{eq:d:approx}:
%\begin{align}\label{eq:robinson:approx}
%    \psi(Y,D,X;\theta,\eta) := \left\{Y-X'\gamma-\theta(D-X'\beta) \right\}\left(D-X'\beta \right),
%\end{align}
%where $\eta=(\gamma,\beta)$.
%
%
%%%%%%%%%%%%%%%%%%%%%%%%%%%%%%%%%%%%%%%%%%%%%%%%%%%%%%%%%%%%%%%%%%%%%
%\subsubsection{The Method}
%%%%%%%%%%%%%%%%%%%%%%%%%%%%%%%%%%%%%%%%%%%%%%%%%%%%%%%%%%%%%%%%%%%%%
With this orthogonal score, we propose to obtain $\check\theta$ and $\hat\Omega$ via Algorithm \ref{algorithm:dml_oga_hdaic} presented in Section \ref{sec:regression} even under the current extended setting with approximation errors.

With the extended model \eqref{eq:y:approx}--\eqref{eq:d:approx}, a different set of assumptions are imposed from those in the baseline model.
First, we slightly modify Assumption \ref{a:1} as follows.

\begin{assumption}\label{a:1:approx}
	For each $N\in \mathbb N$, it holds that
\begin{enumerate}[(a)]%[counter-format=(\alph*),item-indent = 0.55cm,label-offset = 5pt](1) 
    \item $(Y_i,D_i,X_i')_{i=1}^N$ are i.i.d. copies of $(Y,D,X')$. 
    \item \eqref{eq:y:approx} and \eqref{eq:d:approx} hold. 
    \item $E[\abs{Y}^q] + E[\abs{D}^q] \le C_q$. 
    \item $E[\abs{UV}^2]\ge c_q^2$ and $E[V^2|(Y,D,X')]\ge c_q$. \label{Vsq_bdd_below}
    % \item $E[\max_{1\le i\le N} E [U^2|X]]\le C$ and \\$E[\max_{1\le i\le N}E [V^2|X]\le C$ 
    % \item ${E[X_{ij}^4]}\le C_4$, $E[V^4] \le C_4$, and $E[\Epsilon^4]\le C_4$. \label{4th_bdd}
    \item ${\max_{1\le j\le p}E[|X_{ij}|^q]}\le C_q$, $E[\abs{V}^q] \le C_q$, and $E[\abs{\Epsilon}^q]\le C_q$. \label{qth_bdd}
\end{enumerate}
Furthermore, it holds asymptotically that
(f) $K_{N,q}^2 C\log p/ N^{1-2/q}=o(1).$ \label{maxqth_bdd}
\end{assumption}

\noindent
In part \ref{Vsq_bdd_below}, we require the conditional variance of $V$ given $(Y,D,X')$ to be bounded away from zero whereas the counterpart in the baseline model assumed the unconditional variance to be bounded away from zero.

We continue to use Assumption \ref{a:2} from the baseline model.
However, it should be stressed that we now impose Assumption \ref{a:2} on \eqref{eq:y:approx}--\eqref{eq:d:approx} rather than \eqref{eq:y}--\eqref{eq:d}.
With the approximation errors introduced in the current extended model, we make the following assumption on the approximation error functions $r_Y$ and $r_D$.

\begin{assumption}\label{a:approx}
    For $r(X) = r_Y(X)$ and $r_D(X)$, it holds that
    \begin{enumerate}[(a)]%[counter-format=(\alph*),item-indent = 0.55cm,label-offset = 5pt](2) 
        \item $E[r^4(X)]\le C$. \label{r4_bdd}
        \item $E[r^2(X)]\le C\log p/N$. \label{rsq_bdd}
        \item $\max_{1\le j\le p}\left|E[r(X)X_{ij}]\right|\le C_{p,1}\sqrt{\log p}/N^{1/4}$.\label{rX_bdd}
    \end{enumerate}
\end{assumption}

\noindent
Assumption \ref{a:approx} \ref{r4_bdd} requires the fourth moment of the approximation error to be bounded, \ref{rsq_bdd} assumes the second moment to be of order $\log p/N$, and \ref{rX_bdd} bounds the maximum cross moment of the approximation and the covariates.
% \begin{assumption}\label{a:2:approx}
% It holds over $N\in \mathbb N$ that
% \begin{tasks}[counter-format = (\alph*),item-indent=.55cm,label-offset=5pt]
%     % \item $E[\max_{1\le j\le p}\abs{X_{ij}}^q]\le C,\, E[\abs{V}^q] \le C$, $E[\abs{Y-X'\gamma_0}^q]\le C$ and $\log p = o(N^{1/4}).$
%     \item $\lambda_{\min} (\Gamma)\ge \lambda_1>0$ and $\lambda_{\max} (\Gamma) \le C_q$, where $\Gamma=E[X X']$. \label{additionalA}
%     \item Define $\Gamma(J) = E[X_{iJ}X_{iJ}']$ and $d_{\ell}(J) = E[X_{i\ell}X_{iJ}]$ for a set of coordinate indices $J\subseteq \mathfrak{P}$. Then $$\max_{1\le \abs{J}\le \bar{C}(N/\log p)^{1/2},\,\ell \notin J} \abs{\Gamma^{-1}(J)d_\ell(J)}< C_q.$$ \label{IngA5}
% \end{tasks}
% \end{assumption}\vspace{-.6cm}

Finally, we focus on the more difficult case, namely the polynomial decay case, for brevity in this section.

\begin{assumption}\label{a:3:approx}
	It holds over $N\in \mathbb N$ that
for each of $\xi_0 = \beta_0$ and $\gamma_0$, $\xi_0$ follows
% either (a) or (b) described below. 
% \begin{tasks}[counter-format = (\alph*),item-indent=.55cm,label-offset=5pt]
%     \item 
    polynomial decay, i.e., $\log p = o(N^{1-2/q})$. Each $\xi_0$ is such that $\norm{\xi_0}_2^2 \le C_0$ for some $C_0>0$ and there exist $\alpha> 1$ and $C_{\alpha} > 0$ such that for any $J \subseteq \mathfrak{P}$,$$\hspace{2cm}\norm{\xi_0(J)}_1 \le C_\alpha \left( \norm{\xi_0(J)}_2^2\right)^{(\alpha-1)/(2\alpha-1)}.$$ 
    % \item Exponential decay: $\log p= o(N^{1/4}).$ Each $\xi_0$ is such that $ \norm{\xi_0}_\infty \le C_0$ for some $C_0>0$ and there exists $C_1>1$ such that for any $J \subseteq \mathfrak{P}$, $$\norm{\xi_0(J)}_1 \le C_1 \norm{\xi_0(J)}_\infty.\label{expo}$$ 
% \end{tasks}
\end{assumption}

The following theorem establishes the asymptotic normality of $\check\theta$ along with the asymptotic validity of inference under the extended model with approximation errors.

\begin{theorem}\label{theorem:approx} Let $(\calP_N)_{N\in \mathbb{N}}$ be a sequence of sets of DGPs such that Assumptions \ref{a:2} and \ref{a:1:approx}--\ref{a:3:approx} are satisfied on the model \eqref{eq:y:approx:pl}--\eqref{eq:d:approx:pl} entailing the reduced forms \eqref{eq:y:approx}--\eqref{eq:d:approx}. Then, the estimator $\check{\theta}$ defined in Algorithm \ref{algorithm:dml_oga_hdaic} satisfies
\begin{align*}
    \sqrt{N} \paren{\check{\theta}-\theta_0} \xrightarrow{d} N(0,\Omega),
\end{align*}
where $\Omega = (E[V^2])^{-1}E[V^2U^2](E[V^2])^{-1}$. 
% Define $\hat{M} :=- 1/K\sum_{k=1}^K 1/n \sum_{i\in I_k} (D_i-X_i'\hat{\beta})^2$. Then, we can define the variance estimator \begin{align*}
%     \hat{\Omega} = \hat{M}^{-1}\frac{1}{K} \sum_{k=1}^K \frac{1}{n}\sum_{i\in I_k} [\psi(Y,D,X;\check{\theta},\hat{\eta}_k)\psi(Y,D,X;\check{\theta},\hat{\eta}_k)'](\hat{M}^{-1})'
% \end{align*} and 
The confidence regions with significance level $a\in (0,1)$ have uniform asymptotic validity:
\begin{align*}
    \sup_{P\in \calP_N} \abs{P\left(\theta_0 \in \left[\check{\theta} \pm \Phi^{-1}(1-a/2)\sqrt{\hat{\Omega}/N}\right]\right)-(1-a)} =o(1),
\end{align*}
where $\hat{\Omega}$ is defined in Section \ref{sec:regression}.
\end{theorem}

\noindent
A proof is presented in Appendix \ref{sec:theorem:approx}.
The same remarks as those presented below the statement of Theorem \ref{theorem:regression} apply here. 

{\color{black} We want to stress that Theorem \ref{theorem:approx} is not an immediate consequence given Theorem \ref{theorem:regression}, because the original proofs for convergence rates of OGA+HDAIC in \cite{ing2020} do not permit approximately sparse models. In order to show Theorem \ref{theorem:approx}, we establish convergence rates for the OGA+HDAIC in approximately sparse regression models.}

%%%%%%%%%%%%%%%%%%%%%%%%%%%%%%%%%%%%%%%%%%%%%%%%%%%%%%%%%%%%%%%%%%%%%%%%%%%%%%%%%%%%%%%%%%%%%%%%%%%%%%%%%%%%%%%%%%%%%%%%%%%%%%%%%%%%%%%%%%
%%%%%%%%%%%%%%%%%%%%%%%%%%%%%%%%%%%%%%%%%%%%%%%%%%%%%%%%%%%%%%%%%%%%%%%%%%%%%%%%%%%%%%%%%%%%%%%%%%%%%%%%%%%%%%%%%%%%%%%%%%%%%%%%%%%%%%%%%%
%\clearpage
%%%%%%%%%%%%%%%%%%%%%%%%%%%%%%%%%%%%%%%%%%%%%%%%%%%%%%%%%%%%%%%%%%%%%
\subsection{Estimation and Inference without Cross Fitting}\label{sec:without_cross_fitting}
%%%%%%%%%%%%%%%%%%%%%%%%%%%%%%%%%%%%%%%%%%%%%%%%%%%%%%%%%%%%%%%%%%%%%
Thus far, our proposed procedures of estimation and inference are based on cross fitting.
A drawback of using the cross fitting is the randomness of estimates given the data.
To overcome this drawback, we provide an alternative procedure of estimation and inference without relying on cross fitting in this section.
However, we stress that this benefit comes with costs in some assumptions as discussed below. 

%%%%%%%%%%%%%%%%%%%%%%%%%%%%%%%%%%%%%%%%%%%%%%%%%%%%%%%%%%%%%%%%%%%%%
%\subsubsection{The Model}
%%%%%%%%%%%%%%%%%%%%%%%%%%%%%%%%%%%%%%%%%%%%%%%%%%%%%%%%%%%%%%%%%%%%%
We continue from Section \ref{sec:approx} to consider the partial linear model \eqref{eq:y:approx:pl}--\eqref{eq:d:approx:pl} entailing the reduced forms \eqref{eq:y:approx}--\eqref{eq:d:approx}.
Furthermore, we continue to use the same orthogonal score $\psi(Y,D,X;\theta,\eta)$ defined in \eqref{eq:robinson}.
However, we now replace Algorithm \ref{algorithm:dml_oga_hdaic} by the following algorithm which does not involve the cross-fitting procedure.
%In this section we provide an algorithm using the whole sample instead of sample splitting. 
Let $[N] = \{1,\dots,N\}$, so that $X_{[N] j} = \{X_{ij}, i\in [N]\}$ and $D_{[N]}=(D_1,\dots,D_N)'$. 
% {\color{blue} [TO DO] Change the $X_j$'s using bullets, and also in the original algorithm in section 2 and 3}
\begin{algorithm}[OGA+HDAIC with DML for high-dimensional linear models without cross fitting]\label{algorithm:dml_oga_hdaic:fullsample} \hfill
\begin{enumerate}[label={Step \arabic*.}] 
    % \item Randomly split the sample indices $\{1,...,N\}$ into $K$ folds $(I_k)_{k=1}^K$. For simplicity, let the size of each fold be $n=N/K$.
    % \item For each fold $k \in \{1,...,K\}$, 
    \item Perform following procedure using $\{(X_i',D_i)'\}_{i=1}^N$ to get $\hat{\beta}$.
    \begin{enumerate}
        \item Compute $\hat{\mu}_{0,j} = X_{[N] j}'D_{[N]}/\sqrt{N}\snorm{X_{[N] j}}$. Select the coordinate $\hat{j}_{1} = \argmax_{1\le j \le p} \abs{\hat{\mu}_{0,j}}$. Define $\hat{J}_1 = \{\hat{j}_1\}$.
        \item Compute $\hat{\mu}_{1,j} = X_{[N] j}'(I_N- H_1) D_{[N]} / \sqrt{N}\snorm{X_{[N] j}}$, where $H_1 = X_{[N] \hat{j}_1}(X_{[N] \hat{j}_1}'X_{[N] \hat{j}_1})^{-1}X_{[N] \hat{j}_1}'$. Select the coordinate $\hat{j}_2 = \argmax_{1\le j \le p,j\notin \hat{J}_1} \abs{\hat{\mu}_{1,j}}$. Update $\hat{J}_2 =\hat{J}_1\cup \{\hat{j}_2\}$.
        \item Given $m-1$ coordinates $\hat{J}_{m-1}$ that have been obtained, compute $\hat{\mu}_{m-1,j} = X_{[N] j}'(I_N - H_{m-1}) D_{[N]} / \sqrt{N}\snorm{X_{[N] j}}$, where $H_{m-1} = X_{[N] \hat{J}_{m-1}}(X_{[N] \hat{J}_{m-1}}'X_{[N] \hat{J}_{m-1}})^{-1}X_{[N] \hat{J}_{m-1}}'$. Select the coordinate $\hat{j}_m = \argmax_{1\le j\le p, j\notin \hat{J}_{m-1}}\abs{\hat{\mu}_{m,j}}$. Iteractively update $\hat{J}_m = \hat{J}_{m-1}\cup \{\hat{j}_m\}$.
        \item Compute $\textup{HDAIC}$ $(\hat{J}_m) = (1+C^* |\hat{J}_m|\log p/N)\hat{\sigma}_{m}^2$ for each $m$ and $\hat{\sigma}_{m}^2 = 1/N D '(I-H_m)D$. Choose $\hat{m} = \argmin_{1\le m\le M_n^*}$ $\textup{HDAIC} (\hat{J}_m)$, where $C^*$ and $M_n^*$ are defined in \eqref{Cstar} and \eqref{Mnstar} in Appendix \ref{sec:regression:details}.
		\item With coordinates $\hat{J}_{\hat{m}}$, run OLS of $D_i$ on $X_{i\hat{J}_{\hat{m}}}$ to get $\hat{\beta}.$
    \end{enumerate}
    \item Repeat Step 2 with $\{(X_i',Y_i)'\}_{i=1}^N$ instead of $\{(X_i',D_i)'\}_{i=1}^N$, to get $\hat{\gamma}$.
    \item Obtain $\widetilde{\theta}$ as a solution to $1/N \sumi \psi(Y_i,D_i,X_i;\widetilde{\theta},\hat{\eta})=0$ where $\hat{\eta} = (\hat{\gamma},\hat{\beta})$ and $\psi$ is defined in \eqref{eq:robinson}. 
    % \item 
    % \item Define $\widetilde{J}=\hat{J}_{\hat{m}}^D\cup \hat{J}_{\hat{m}}^Y$. Obtain $\widetilde{\theta}$ from $(\widetilde{\theta},\widetilde{\Lambda}) = \argmin_{\theta\in\R,\Lambda\in\R^p} \sum_{i=1}^N (Y_i - D_i\theta - X_i'\Lambda)^2$, where $\Lambda_j=0$ for $j\notin \widetilde{J}$.  
    % \item Compute $\hat{M} = -1/K\sum_{k=1}^K 1/n\sum_{i\in I_k}(D_i - X_i'\hat{\beta})^2$. Obtain a variance estimator of $\check{\theta}$ as $\hat{\Omega} = \hat{M}^{-1}\frac{1}{K} \sum_{k=1}^K \frac{1}{n}\sum_{i\in I_k} [\psi(Y,D,X;\check{\theta},\hat{\eta}_k)\psi(Y,D,X;\check{\theta},\hat{\eta}_k)'](\hat{M}^{-1})'$.
\end{enumerate}
\end{algorithm}

%%%%%%%%%%%%%%%%%%%%%%%%%%%%%%%%%%%%%%%%%%%%%%%%%%%%%%%%%%%%%%%%%%%%%
%\subsubsection{The Theory}\label{sec:fullsample:theory}
%%%%%%%%%%%%%%%%%%%%%%%%%%%%%%%%%%%%%%%%%%%%%%%%%%%%%%%%%%%%%%%%%%%%%

To establish asymptotic properties for this new estimator $\widetilde\theta$, we continue to impose Assumptions \ref{a:2}, \ref{a:1:approx}, and \ref{a:approx}.
As in Section \ref{sec:approx}, we focus on the more difficult case, namely the polynomial decay case, for brevity.

\begin{assumption}\label{a:3:fullsample}
	It holds over $N\in \mathbb N$ that
    $\abs{\theta_0} \le C$, and
for each of $\xi_0 = \beta_0$ and $\gamma_0$, $\xi_0$ follows
%  (a):
% either (a) or (b) described below. 
% \begin{tasks}[counter-format = (\alph*),item-indent=.55cm,label-offset=5pt]
%     \item 
    polynomial decay, i.e., 
    % $\log p = o(N^{1-2/q})$. 
    each $\xi_0$ is such that $\norm{\xi_0}_2^2 \le C_0$ for some $C_0>0$ and there exist $\alpha> 1$ such that $\log p = o(N^{(\alpha-1)/(3\alpha-1)})$ and for any $J \subseteq \mathfrak{P}$,$$\hspace{2cm}\norm{\xi_0(J)}_1 \le C \left( \norm{\xi_0(J)}_2^2\right)^{(\alpha-1)/(2\alpha-1)}.$$ 
    % \item Exponential decay: $\log p= o(N^{1/4}).$ Each $\xi_0$ is such that $ \norm{\xi_0}_\infty \le C_0$ for some $C_0>0$ and there exists $C_1>1$ such that for any $J \subseteq \mathfrak{P}$, $$\norm{\xi_0(J)}_1 \le C_1 \norm{\xi_0(J)}_\infty.$$ 
% \end{tasks}
\end{assumption}

\noindent
Unlike the previous sections, however, we now require $\log p = o(N^{(\alpha-1)/(3\alpha-1)})$ for $\alpha>1$.

The following theorem establishes the asymptotic normality of $\widetilde\theta$ defined without the cross-fitting procedure.

\begin{theorem}\label{theorem:fullsample} Let $(\calP_N)_{N\in \mathbb{N}}$ be a sequence of sets of DGPs such that Assumptions \ref{a:2}, \ref{a:1:approx}, \ref{a:approx}, and \ref{a:3:fullsample} are satisfied on the model \eqref{eq:y:approx:pl}--\eqref{eq:d:approx:pl} entailing the reduced forms \eqref{eq:y:approx}--\eqref{eq:d:approx}. Then, the estimator $\widetilde{\theta}$ satisfies
    \begin{align*}
        \sqrt{N} \paren{\widetilde{\theta}-\theta_0} \xrightarrow{d} N(0,\Omega),
    \end{align*}
    where $\Omega = (E[V^2])^{-1}E[V^2U^2](E[V^2])^{-1}$. 
    % Define $\hat{M} :=- 1/K\sum_{k=1}^K 1/n \sum_{i\in I_k} (D_i-X_i'\hat{\beta})^2$. Then, we can define the variance estimator \begin{align*}
    %     \hat{\Omega} = \hat{M}^{-1}\frac{1}{K} \sum_{k=1}^K \frac{1}{n}\sum_{i\in I_k} [\psi(Y,D,X;\check{\theta},\hat{\eta}_k)\psi(Y,D,X;\check{\theta},\hat{\eta}_k)'](\hat{M}^{-1})'
    % \end{align*}
    % and the confidence regions with significance level $a\in (0,1)$ have uniform asymptotic validity:
    % \begin{align*}
    %     \sup_{P\in \calP_N} \abs{P\left(\theta_0 \in \left[\check{\theta} \pm \Phi^{-1}(1-a/2)\sqrt{\hat{\Omega}/N}\right]\right)-(1-a)} =o(1).
    % \end{align*}
\end{theorem}  

\noindent
A proof is given in Appendix \ref{sec:theorem:fullsample}.

{\color{black}We stress that, although the proof builds on that of Theorem 1 in \cite{BCH2014RES}, it is far from being trivial as the lack of cross-fitting and $L^p$-sparsity creates extra challenges. Specifically, a key intermediate step is to control the $L^1$ distances between $\beta_0$ and $\tilde \beta(\tilde J)$, an oracle regression estimator defined in the proof of Theorem \ref{theorem:fullsample}. Due to the lack of exact or approximate sparsity of $\beta_0$, this is shown via different strategies from those employed in \cite{BCH2014RES}.
}

As emphasized at the beginning of the current subsection, the main advantage of the estimation procedure without cross fitting is that the estimate is now non-random given data.
Besides, this framework without cross fitting offers an additional advantage.
Recall that our main motivation to use the OGA+HDAIC is to weaken the sparsity assumptions required for conventional high-dimensional methods such as the LASSO.
In the current framework without cross fitting, there is another motivation to use the OGA+HDAIC.
Namely, it selects regressors based on their strength in explanatory power by the algorithm.
Hence, we have better interpretations of the model selected by the OGA+HDAIC than the conventional high-dimensional methods under the current framework without cross-fitting.
We highlight this additional advantage of our proposed method.

Appendix \ref{sec:simulation:no_cross_fitting} presents simulation results with this modified method without cross fitting.
The results are similar to those obtained for the baseline model presented in Section \ref{sec:simulations}.

%%%%%%%%%%%%%%%%%%%%%%%%%%%%%%%%%%%%%%%%%%%%%%%%%%%%%%%%%%%%%%%%%%%%%
\section{An Empirical Application}\label{eq:application}
%%%%%%%%%%%%%%%%%%%%%%%%%%%%%%%%%%%%%%%%%%%%%%%%%%%%%%%%%%%%%%%%%%%%%
In this section, we demonstrate an application of the proposed method to estimation of production functions.
The main challenge in the econometrics of production functions is the simultaneity in the choice of input firms \citep{marschak1944random}.
While early studies of production functions address this simultaneity problem by explicitly modeling rational choice structures of firms, \citet{OlleyPakes1996} more recently propose a novel idea to use the inverse of the reduced-form investment choice function as a control function.
\citet{levinsohn2003estimating} propose to use intermediate input, instead of investment, as a control variable for a number of advantages.

The use of the control function \textit{a la} \citet{levinsohn2003estimating} entails the partial linear estimating equation for the labor elasticity of the form
\begin{equation}\label{eq:production_first_stage}
y_{it} = \ell_{it}\theta + f(k_{it},m_{it}) + u_{it},
\end{equation}
where 
$y_{it}$ denotes the logarithm of output,
$\ell_{it}$ denotes the logarithm of labor input,
$k_{it}$ denotes the logarithm of capital input,
$m_{it}$ denotes the logarithm of intermediate input,
$g$ is a nonparametric function that subsumes a part of the production function and the control function, and
$u_{it}$ denotes a mean-orthogonal reduced-form composite error.
See \citet{OlleyPakes1996} and \citet{levinsohn2003estimating} for details.

In light of the partial linear form \eqref{eq:production_first_stage}, \citet{OlleyPakes1996} and \citet{levinsohn2003estimating} propose to use the estimator of \citet{Robinson} which is semiparametric root-$n$ consistent for $\theta$.
Following these seminal papers, numerous researchers have estimated production functions.
That said, many of these subsequent studies follow the Stata command \citep{petrin2004production} which implements estimation of \eqref{eq:production_first_stage} via the parametric third-degree polynomial approximation
\begin{equation}\label{eq:production_first_stage_degree_three}
y_{it} = \ell_{it}\theta + \sum_{\rho_1=0}^3\sum_{\rho_2=0}^{3-\rho_1} \delta_{\rho_1 \rho_2} k_{it}^{\rho_1} m_{it}^{\rho_2} + u_{it}.
\end{equation}
See \citet[][pages 116--118]{petrin2004production}.

To mitigate the approximation bias asymptotically, we consider a higher-dimensional approximation
\begin{equation}\label{eq:production_first_stage_high_dimensional}
y_{it} = \ell_{it}\theta + \underbrace{\sum_{j=1}^p \delta_{j} \phi_j(k_{it},m_{it}) + r_p(k_{it},m_{it})}_{f(k_{it},m_{it})} + u_{it}
\end{equation}
with an error $r_p(k_{it},m_{it})$ in approximation,
where $\phi = (\phi_1,\phi_2,\phi_3,\cdots)$ is a basis and $p$ can be large and increasing with the sample size.
The basis $\phi$ could be defined as the Cartesian product of polynomials, i.e.,
$(\phi_1(k,m),\phi_2(k,m),$ $\phi_3(k,m),\cdots)=(1,k,m,k^2,m^2,km,\cdots)$, as a generalization of the popular estimating equation \eqref{eq:production_first_stage_degree_three} in the Stata command.
More generally, we can define the basis $\phi$ as the tensor product of orthonomal bases.
We employ the tensor product of Hermite bases \citep{gallant1987semi,chen2007large} for our basis $\phi$, and apply our proposed method to \eqref{eq:production_first_stage_high_dimensional} to get an estimate of $\theta$ and its standard error.

Following \citet{levinsohn2003estimating}, we use a plant-level panel of Chilean firms from 1979 to 1986.
See \citet{liu1991entry} for details about the construction of the data.
Among others, we focus on the 3-digit level industry of food products (311) because of its large sample size compared to other industries.
We are interested in the elasticity with respect to unskilled labor input $\ell^u_{it}$ and skilled labor input $\ell^s_{it}$.
The intermediate input variables include electricity $m^e_{it}$, fuels $m^f_{it}$, and materials $m^m_{it}$.
To estimate the elasticity with respect to unskilled labor input $\ell^u_{it}$ using $m^m_{it}$ as a proxy following \citet{levinsohn2003estimating}, we consider the estimating equation of the form 
\begin{equation}\label{eq:unskilled}
\underbrace{y_{it}}_{Y} = \underbrace{\ell^u_{it}\theta^u}_{D\theta_0} + \underbrace{\ell^s_{it}\theta^s + m^e_{it}\theta^e + m^f_{it}\theta^f + \sum_{j=1}^p \delta_{j} \phi_j(k_{it},m^m_{it}) + \tau_{t} + r_p(k_{it},m^m_{it})}_{f(X) } + \underbrace{u_{it}}_{U}
\end{equation}
as in \eqref{eq:y:approx:pl}.
To estimate the elasticity with respect to skilled labor input $\ell^s_{it}$, we swap $\ell^u_{it}\theta^u$ and $\ell^s_{it}\theta^s$ in the above estimating equation:
\begin{equation}\label{eq:skilled}
\underbrace{y_{it}}_{Y} = \underbrace{\ell^s_{it}\theta^s}_{D\theta_0} + \underbrace{\ell^u_{it}\theta^u + m^e_{it}\theta^e + m^f_{it}\theta^f + \sum_{j=1}^p \delta_{j} \phi_j(k_{it},m^m_{it}) + \tau_{t} + r_p(k_{it},m^m_{it})}_{f(X) } + \underbrace{u_{it}}_{U}
\end{equation}
as in \eqref{eq:y:approx:pl}.

The term $\tau_{t}$ represents time effects.
Following \citet{levinsohn2003estimating}, we include the indicator for year groups 1979--1981, 1982--1983, and 1984--1986.

For estimation of \eqref{eq:unskilled} using a polynomial basis, we let $X$ consist of 
(i) $\ell_{it}^s$ 
(ii) $m^e_{it}$,
(iii) $m^f_{it}$,
(iv) $k_{it}$, $\ldots$, $k_{it}^{10}$,
(v) $m^m_{it}$, $\ldots$, $(m^m_{it})^{10}$,
(vi) dummy for 1979--1981,
(vii) dummy for 1982--1983, and
(viii) interactions of the terms in (iv) and (v).
We also consider an estimation of \eqref{eq:unskilled} using a Hermite basis $(\psi_0,\ldots,\psi_9)$, we let $X$ consist of 
(i) $\ell_{it}^u$ 
(ii) $m^e_{it}$,
(iii) $m^f_{it}$,
(iv) $\psi_0(k_{it})$, $\ldots$, $\psi_9(k_{it})$,
(v) $\psi_0(m^m_{it})$, $\ldots$, $\psi_9(m^m_{it})$, and
(vi) dummy for 1979--1981,
(vii) dummy for 1982--1983, and
(viii) interactions of the terms in (iv) and (v).
We use a finite-sample adjusted version of the DML estimates following \citet[][Sec. 3.4]{ccddhnr} -- see Appendix \ref{eq:finite} for details.
See Appendix \ref{eq:hermite} for details about the Hermite basis.
We repeat analogous estimation procedures for \eqref{eq:unskilled}. 

%\textcolor{blue}{model specifications are number of folds $=10$, num.trees $=100$, min.node.size $= 2$,max.depth $= 5$, number of repetitions $= 20$. I followed the default settings in the example code, other than the number of folds and the number of repetition.}

Table \ref{tab:empirical} summarizes estimation results.
Row (I) copies estimates from \citet{levinsohn2003estimating}.
%Row (II) reports estimates based on  the degree-three polynomial using the Stata command \texttt{levpet}.
Rows (II), (III), and (IV) report results based on the DML with LASSO, DML with random forest, and DML with the OGA and HDAIC (the estimator proposed in this paper), respectively.\footnote{For (II) and (III), we use the R package ``DoubleML : Double Machine Learning in R.'' We set the parameters as folds $=10$, num.trees $=100$, min.node.size $= 2$, max.depth $= 5$, and the number of repetitions $= 20$}
The first two columns show results based on the tensor product of polynomial bases, while the last two columns show results based on the tensor product of Hermite bases.

\begin{table}
	\centering
	\renewcommand{\arraystretch}{0.85} 
		\begin{tabular}{clccccc}
		\hline\hline
		&& \multicolumn{2}{c}{Polynomial Basis} && \multicolumn{2}{c}{Hermite Basis} \\
		\cline{3-4}\cline{6-7}
		&& Unskilled & Skilled && Unskilled & Skilled \\
		&& Labor & Labor && Labor & Labor \\
		\hline
(I) 	&\citet{levinsohn2003estimating}& 0.139 & 0.051 && --- & ---\\
		&	                             &(0.010)&(0.009)\\
		\hline
%(II)	&	Degree-Three Polynomial      & ?.??? & ?.??? && --- & ---\\
%		&	(Stata Command \texttt{levpet} for LP)       &(?.???)&(?.???)\\
%		\hline
(II)	&	Double Machine Learning with & 0.170 & 0.063 && 0.196 & 0.085\\
		&	LASSO Preliminary Estimation &(0.011)&(0.008)&&(0.011)&(0.010)\\
		\hline
(III)	&	Double Machine Learning with & 0.185 & 0.061 && 0.189 & 0.062\\
		&	Random Forest Preliminary Estimation 
			                             &(0.013)&(0.010)&&(0.013)&(0.011)\\
		\hline
(IV)	&	Double Machine Learning with & 0.279 & 0.161 && 0.165 & 0.038\\
		&	OGA+HDAIC Preliminary Estimation &(0.017)&(0.012)&&(0.011)&(0.010)\\
		\hline\hline
		\end{tabular}
	\caption{Estimates of labor elasticities in the 3-digit level industry of food products (311) in Chile.}
	\label{tab:empirical}
\end{table}

First, observe that all the three machine learning estimates, (II), (III), and (IV), based on the polynomial basis yield larger point estimates than the low-dimensional estimates (I).
This may indicate a potential bias of the conventional estimator based on a low-dimensional polynomial approximation.
However, it is also worthy of remarking that polynomial bases (including the Legendre bases) are not suitable to approximating functions of variables that have unbounded supports.
Hermite bases, on the other hand, are capable of approximating functions of variables with unbounded supports.\footnote{\label{foot:approximation}Namely, the set of functions spanned by the standard polynomials is dense in the set $C^0(K)$ of continuous functions (or $L^2(K)$ of square integrable functions) defined only on a \textit{compact} support $K \subset \mathbb{R}$, whereas the set of functions spanned by the Hermite polynomials is dense in the set $L^1(\mathbb{R})$ of integrable functions defined on the \textit{entire} real line $\mathbb{R}$. As such, when the regressor(s) are infinitely supported, as is likely the case in the current application, the approximation by the standard polynomial basis is not credible but that by the Hermite polynomial basis is credible.}
We therefore focus on the results based on the Hermite basis.\footnote{\label{foot:not_invariant}In addition to this difference in the approximation theoretic properties between the polynomial and Hermite bases, we also remark that the difference may be due to the fact that the proposed method is not invariant to an invertible linear transformation of the regressors like the LASSO.}
For the unskilled labor coefficient, all the three machine learning methods, (II), (III), and (IV), still yield larger point estimates than that of the low-dimensional method (I), but the estimate based on our proposed method (IV) is relatively smaller and closer to that of (I).
For the skilled labor coefficient, the two machine learning methods, (II) and (III), yield slightly larger estimates than that of (I), while our proposed machine learning method (IV) yields a slightly smaller estimate than that of (I).
In summary, the results of the DML based on the LASSO or the random forest significantly differ from the result of the conventional low-dimensional method, but our proposed method also yields slightly different results from those of the DML based on the LASSO or the random forest, as is also the case in our simulation studies presented in Section \ref{sec:simulations}.

We ran several other estimates for robustness checks.
Appendix \ref{sec:additional_empirical} presents estimation results based on alternative values of the tuning parameters.
Appendix \ref{sec:additional_empirical} also presents results based on the method without cross fitting introduced in Section \ref{sec:without_cross_fitting}.
It turns out that the qualitative pattern of the results summarized above remain robust.

Finally, we conclude this section with a few remarks about the validity of the estimation approach employed in this empirical application following \citet{OlleyPakes1996} and \citet{levinsohn2003estimating}.
It is well known today that the estimating equation \eqref{eq:production_first_stage} fails to identify the parameter $\theta$ in general, as first pointed out by \citet{ackerberg2015identification}.
That said, they also suggest that $\theta$ can be correctly identified by \eqref{eq:production_first_stage} under certain DGPs.
They include DGPs with: (1) i.i.d. optimization error in $\ell_{it}$ and not in $m_{it}$; or (2) i.i.d. shocks to the price of labor or output after $m_{it}$; for instance \citep[][Sec. 3.1]{ackerberg2015identification}.
As such, we stress that the validity of the estimation method present above is contingent on these assumptions about the underlying DGPs.

%%%%%%%%%%%%%%%%%%%%%%%%%%%%%%%%%%%%%%%%%%%%%%%%%%%%%%%%%%%%%%%%%%%%%
\section{Summary and Discussions}\label{sec:summary_discussions}
%%%%%%%%%%%%%%%%%%%%%%%%%%%%%%%%%%%%%%%%%%%%%%%%%%%%%%%%%%%%%%%%%%%%%

In this paper, we propose a new method of inference in high-dimensional regression models.
%  and high-dimensional IV regression models
The estimation procedure is based on a combined use of the OGA, HDAIC, and DML.
The method of inference about any low-dimensional subvector of high-dimensional parameters is based on a root-$N$ asymptotic normality, which does not require the exact sparsity condition or the $L^p$ sparsity condition.
In stead imposed are conditions on the rate at which the absolute size of parameters decays when descendingly ordered.
We demonstrate through simulation studies superior finite sample performance of this proposed method over those based on two popular alternatives, namely the LASSO and the random forest.
The extent of this outperformance is more prominent under less sparse models characterized by slower polynomial decays.
Finally, we illustrate an application of the method to production analysis using a panel of Chilean firms.
Using the tensor product of Hermite basis as high-dimensional controls, we find that estimates based on our proposed method differ from those based on the LASSO and random forest, similarly to what we observe in the simulation studies.

We close this paper with discussions of limitations, omitted extensions and potential directions for future research.
First, unlike regressions and like the LASSO, the method is not invariant to invertible linear transformation of the regressors $X$.
Practitioners should be aware of this drawback in our proposed method.
Second, as is the case with other DML methods, our proposed method based on DML is subject to random estimates.
To overcome this problem, we present an alternative procedure without cross fitting in Section \ref{sec:without_cross_fitting}, but this comes at the expense of an alternative set of assumptions.
Again, practitioners should be aware of these tradeoffs in choosing an appropriate method.
Third, we focus on high-dimensional linear regression models with exogenous regressors throughout the main text.
We provide an extension to high-dimensional linear IV models in Section \ref{sec:iv}.
Extensions to other important models are left for future research.

\vspace{1cm}
\appendix
%%%%%%%%%%%%%%%%%%%%%%%%%%%%%%%%%%%%%%%%%%%%%%%%%%%%%%%%%%%%%%%%%%%%%
\section{Proofs of the main Result and Auxiliary Lemmas}
%%%%%%%%%%%%%%%%%%%%%%%%%%%%%%%%%%%%%%%%%%%%%%%%%%%%%%%%%%%%%%%%%%%%%

This appendix section collects a proof of the main theorem and an auxiliary lemma.

%%%%%%%%%%%%%%%%%%%%%%%%%%%%%%%%%%%%%%%%%%%%%%%%%%%%%%%%%%%%%%%%%%%%%
\subsection{Proof of Theorem \ref{theorem:regression}}\label{sec:theorem:regression}
%%%%%%%%%%%%%%%%%%%%%%%%%%%%%%%%%%%%%%%%%%%%%%%%%%%%%%%%%%%%%%%%%%%%%
\begin{proof}
In this proof, we first show that the convergence rates from Theorem 3.1. of \cite{ing2020} can be guaranteed under our Assumptions \ref{a:1}-\ref{a:3}. It then suffices to demonstrate that Theorem 4.1. in \cite{ccddhnr} can be applied under our assumptions and with such convergence rates. Following Corollary 1 in \cite{BCH2014RES}, we fix a sequence of DGPs $(P_N)_{N\in \mathbb N}$, $P_N\in \calP_N$ and establish the asymptotic statements in order to show uniformity over the sequence of sets of DGPs.

\textit{Step 1.} We shall verify all the conditions for Theorem 3.1. in \cite{ing2020}.
Let us first examine each of the regularity conditions (A1)-(A5) for Theorem 3 in \cite{ing2020}.
Note that 
Assumption \ref{a:3} (a) and (b), correspond to the conditions (A3) and (A4) in \cite{ing2020}, respectively;
Assumption \ref{a:2} (b) corresponds to the condition (A5) in \cite{ing2020}.
Assumption \ref{a:2} (a) is the additional conditions listed in Theorem 2.1, equations (2.19)-(2.21) in \cite{ing2020}.

Now, we show that Assumption \ref{a:1}\ref{qth_bdd} and Assumption \ref{a:3} imply (A1) in \cite{ing2020}.
We shall verify that there exists a strictly positive constant $c_1$ such that
\begin{align}
    P\left(\max_{1\le j \le p}\abs{\frac{1}{N}\sum_{i=1}^N X_{ij}V_i} \ge c_1 \sqrt{\frac{\log p}{N}} \right)=o(1). \label{A1}
\end{align}
uniformly over $P\in\calP_N$.
Set to $Z_{ij} = X_{ij} V_i$, $F=\max_{1\le i\le N}\max_{1\le j\le p} X_{ij}V_i$, and 
$$
\sigma^2 = \max_{1\le j\le p} E[X_{ij}^2 V_i^2] \le \max_{1\le j\le p} \sqrt{E[X_{ij}^4]E[V_i^4]} \le C_q
$$ 
by Assumption \ref{a:1}\ref{qth_bdd}.  Then Jensen's inequality and some calculations yield 
\begingroup
\allowdisplaybreaks
\begin{align*}
    \sqrt{E[F^2]} =& \sqrt{E\brac{\maxi \maxj X_{ij}^2 V_i^2}}
    \le \paren{E\brac{\maxi \maxj \abs{X_{ij}^{q/2} V_i^{q/2}}}}^{2/q}\\
    \le& \paren{\sumi E\brac{\maxj \abs{X_{ij}^{q/2} V_i^{q/2}}}}^{2/q}
    \le \paren{\sumi \paren{E\brac{\maxj \abs{X_{ij}}^q E\brac{\abs{V_i}^q}}}^{1/2}}^{2/q}\\
    \le& N^{2/q} \paren{E\brac{\maxj \abs{X_{ij}}^q}E\brac{\abs{V_i}^q}}^{1/q}.
\end{align*}
\endgroup
By applying Lemma \ref{lemma:lmi}, we have 
\begin{align*}
    E\left[\max_{1\le j \le p} \abs{\frac{1}{N} \sum_{i=1}^N X_{ij}V_i}\right] \le C\left\{ \sqrt{\frac{\log p}{N}} + K_{N,q}^{1/q} \frac{\log p}{N^{1-2/q}}\right\}.
\end{align*}
%where the first term dominates the second term since $q>4$ and the RHS is $o(1)$ 
Note that the constant $C$ is universal and the bound depends on the DGP $P\in\calP_N$ only via $q$, $N$, $p$, $C_q$, and $K_{N,1/q}$. Thus the right hand side of the bound is $o(1)$ uniformly as $N\to \infty$ by Assumption \ref{a:1}\ref{maxqth_bdd}.
A similar argument applies to the case of estimating $\gamma_0$, where we replace $V$ with $\Epsilon$. 
Therefore, Assumption \ref{a:1}\ref{qth_bdd},  Assumption \ref{a:3}, and Markov's inequality imply (\ref{A1}).

Next, we show that Assumption \ref{a:1}\ref{qth_bdd} and Assumption \ref{a:3} imply (A2) in \cite{ing2020}.
We shall illustrate that there exists a strictly positive constant $c_2$ such that
\begin{align}
    P\left(\max_{1\le j,\ell \le p}\abs{\frac{1}{N}\sum_{i=1}^N X_{ij}X_{i\ell}-E[X_{1j}X_{1\ell}]}\ge c_2 \sqrt{\frac{\log p}{N}} \right)=o(1) . \label{A2}
\end{align}
We apply Lemma \ref{lemma:lmi} with $Z_{ij\ell} = X_{ij}X_{i\ell}$, $F= \max_{1\le i\le N} \max_{1\le j,\ell \le p} \abs{Z_{ij\ell}}$,  and $\sigma^2 = \max_{1\le j,\ell \le p}$  $E[X_{ij}^2 X_{i\ell}^2]\le \max_{1\le j\le p} E[X_{ij}^4] \le C_q$ by Assumption \ref{a:1}\ref{qth_bdd}. Also note that
\begingroup
\allowdisplaybreaks
\begin{align*}
    \sqrt{E[F^2]} =& \sqrt{E\left[\max_{1\le i \le N} \max_{1\le j,\ell \le p} X_{ij}^2 X_{i\ell}^2\right]}		
    \le
    \left(E\left[\max_{1\le i \le N}\max_{1\le j,\ell \le p} \abs{X_{ij}^{q/2} X_{i\ell}^{q/2}}\right]\right)^{2/q}\\
    \le& \left(\sum_{i=1}^N E\left[\max_{1\le j,\ell \le p} \abs{X_{ij}^{q/2} X_{i\ell}^{q/2}}\right]\right)^{2/q}
    \le \left(\sum_{i=1}^N \left(E\left[\max_{1\le j\le p} \abs{X_{ij}}^{q}\right] E\left[\max_{1\le \ell \le p} \abs{X_{i\ell}}^{q}\right]\right)^{1/2}\right)^{2/q}\\
    \le& N^{2/q} \left(E\left[\max_{1\le j \le p}  \abs{X_{ij}}^q\right]\right)^{2/q}.  %\lesssim  N^{2/q}
\end{align*}
\endgroup
An application of Lemma  \ref{lemma:lmi} yields that
\begin{align*}
    E\left[\max_{1\le j,\ell \le p} \abs{\frac{1}{N} \sum_{i=1}^N X_{ij}X_{i\ell} - E[X_{1j}X_{1\ell}]}\right] \le C \left\{ \sqrt{\frac{\log p}{N}} + K_{N,q}^{2/q} \frac{\log p}{N^{1-2/q}}\right\},
\end{align*}
%where the first term dominates the second term and the RHS is $o(1)$ 
where the last equality is $o(1)$ uniformly as $N\to \infty$ following Assumptions \ref{a:1}\ref{maxqth_bdd} and \ref{a:3}. 
By Markov's inequality, we conclude that Assumption \ref{a:1}\ref{qth_bdd} and Assumption \ref{a:3} imply (\ref{A2}).

Now, by invoking Theorem 3.1. in \cite{ing2020} and Equation (3.16), we obtain the convergence rates 
%$$\frac{1}{\abs{I_k^c}}\sum_{i\in I_k^c}(D_i-X_i'\hat{\beta})^2 = O_p\left(\frac{\log p}{N}\right)^{1-1/2\alpha}$$
%and,
\begin{align}
    \snorm{\hat{\beta}-\beta_0 }^2 = O_p\left(\left(\frac{\log p}{N}\right)^{1-1/2\alpha}\right) \label{poly_convgrate}
\end{align}
for the polynomial decay case, and 
%$$\frac{1}{\abs{I_k^c}}\sum_{i\in I_k^c}(D_i -X_i'\hat{\beta})^2 = O_p{\frac{\log p\log N}{N}}$$ and 
\begin{align}
    \snorm{\hat{\beta}-\beta_0 }^2 = O_p\left(\frac{\log p\log N}{N}\right) \label{exp_convgrate}
\end{align}
for the exponential decay case., Note that we use $N$ instead of $\abs{I_k^c}$ since the number of folds $K$ is fixed and the fold size is proportional to $N$. 

\textit{Step 2}. Given the convergence rates obtained in Step 1, we shall next show that Theorem 4.1. in \cite{ccddhnr} can be applied.
To this end, it suffices to verify Assumption 4.1. in \cite{ccddhnr}. 
First, note that Assumption \ref{a:1} (b)-(e) corresponds to Assumption 4.1 (a)-(d) in \cite{ccddhnr}. 

%{\color{blue}
%\begin{rem}
%Note that I was using $\norm{\hat{\eta} - \eta_0}_2^2$ as $\norm{\hat{\eta}-\eta_0}_{P,2}$, while it is not necessarily true. I should use $\norm{X'(\hat{\beta}-\beta_0)}_{P,2}$ instead of $\norm{\hat{\beta}-\beta_0}_{P,2}$ and note
%\begin{align*}
%    E\Big[\norm{X'(\hat{\beta} -\beta_0)}^2\Big] =& E[E[(X'(\hat{\beta}-\beta_0))^2|\hat{\beta}]]\\
%    =& E[(\hat{\beta}-\beta_0)'E[XX'|\hat{\beta}](\hat{\beta}-\beta_0)]\\
%    =& E[(\hat{\beta}-\beta_0)'E[XX'](\hat{\beta}-\beta_0)]\\
%    =& \norm{E(XX')}\norm{\hat{\beta}-\beta_0}^2 = \lambda_{\max}(E[XX']) \norm{\hat{\beta} - \beta_0}^2 < \infty,
%\end{align*}
%where the second equality holds because of sample splitting.
%Therefore, $\norm{.}_{P,2}\lesssim \norm{.}$. 
%\end{rem}
%}

Let us now vindicate that Assumption 4.1. (e) in \cite{ccddhnr} is implied by our (\ref{poly_convgrate}) and (\ref{exp_convgrate}), obtained in Step 1.
Observe that (\ref{poly_convgrate}) implies that there exists a sequence of events $(A_N)_{N}$ with probability $P(A_N)=1-o(1)$ such that conditionally on $A_N$, we have
\begin{align*}
\snorm{\hat{\beta}-\beta_0 }^2 \lesssim \left(\frac{\log p}{N}\right)^{1-1/2\alpha}.
\end{align*}
Therefore, for $X$, an independent copy of the regressor vector $X_i$, we have
\begingroup
\allowdisplaybreaks
\begin{align}
E\left[\norm{X'(\hat{\beta} -\beta_0)}^2\mid A_N\right] =& E[E[(X'(\hat{\beta}-\beta_0))^2|\hat{\beta},A_N]\mid A_N] \label{A5}\\
=& E[(\hat{\beta}-\beta_0)'E[XX'\mid \hat{\beta},A_N](\hat{\beta}-\beta_0)\mid A_N]\\
=& E[(\hat{\beta}-\beta_0)'E[XX'](\hat{\beta}-\beta_0)\mid A_N]\\
=&  \lambda_{\max}(E[XX']) \|\hat{\beta}-\beta_0\|^2, \label{A8}
\end{align}
\endgroup
where the third equality follows from the fact that the sigma-algebra generated by $X$ is independent of $A_N$. Observe that (\ref{A8}) is bounded by Assumption \ref{a:2}\ref{additionalA}, (\ref{poly_convgrate}), and (\ref{exp_convgrate}).
Thus the above bound holds with probability at least $1-\Delta_N$ for some $\Delta_N =o(1)$.

% We want to show that with probability at least $1-\Delta_N$, $E[(\hat{\eta}-\eta_0)^2]\le \delta_N$ and $E[(\hat{\beta}-\beta_0)^2]\times(E[(\hat{\beta}-\beta_0)^2]+E[(\hat{\gamma}-\gamma_0)^2])\le \delta_N N^{-1/4}$ is satisfied for some $\delta_N$ such that $\delta_N$ is a sequence of positive constants converging to zero, where $\Delta_N = o(1)$.
Define $\norm{\cdot}_{P,q}$ as the $L^q(P)$ norm, where $\norm{f}_{P,q} = (\int \abs{f(w)}^{q} dP(w))^{1/q}$ with $P$ being the law with respect to $(Y,D,X)$.
We shall now establish that with $P$-probability no less than $1-\Delta_N$, $\|\hat \eta - \eta_0\|_{P,q}\le C$, $\norm{\hat{\eta}-\eta_0}_{P,2}\le \delta_N$, and $\snorm{\hat{\beta}-\beta_0}_{P,2}\times (\snorm{\hat{\beta}-\beta_0}_{P,2}+\snorm{\hat{\gamma}-\gamma_{0}}_{P,2} ) \le \delta_N N^{-1/2}$ is satisfied for some $\Delta_N,\,\delta_N$ such that both are sequences of strictly positive constants converging to zero.

% Define $q' := \lfloor q/2\rfloor$, where $\lfloor \cdot \rfloor$ denotes the floor function. 
From a similar argument as in (\ref{A5})--(\ref{A8}), 
\begin{align*}
    \norm{\hat{\eta} - \eta_0}_{P,q} =& \snorm{\hat{\beta} - \beta_0}_{P,q} \vee \snorm{\hat{\gamma} - \gamma_0}_{P,q}
    =\paren{E\left[\norm{X'(\hat{\beta} -\beta_0)}^q\mid A_N\right]}^{1/q}\\
    \le& \paren{E[\paren{(\hat{\beta}-\beta_0)'E[XX'](\hat{\beta}-\beta_0)}^{q/2}\mid A_N]}^{1/q}
    = \paren{\paren{\lambda_{\max}(E[XX']) \|\hat{\beta}-\beta_0\|^2}^{q/2}}^{1/q}\\
    =& \paren{\lambda_{\max}(E[XX']) \|\hat{\beta}-\beta_0\|^2}^{1/2},
\end{align*}
where it is bounded by Assumption \ref{a:2}\ref{additionalA}, (\ref{poly_convgrate}), and (\ref{exp_convgrate}).
Thus, $\norm{\hat{\eta} - \eta_0}_{P,q}\le C$ holds with probability at least $1-\Delta_N$ for some $\Delta_N = o(1)$.

Since we use the identical procedure, namely the OGA and HDAIC, in estimating both of $\beta_0$ and $\gamma_0$, the convergence rates apply to both. 
We consider two cases where both parameters follow the polynomial decay case or the exponential decay case.
\begin{enumerate}[label={Case \arabic*.}]
    \item For the polynomial decay case, let $\delta_N = (\log p)^{1-1/2\alpha}N^{1/2\alpha - 1/2}$. Then $\delta_N = o(1)$ since $\alpha$ is assumed to be strictly larger than $1$ in Assumption \ref{a:3} (a). We have
\begin{align*}
    \snorm{\hat{\eta}-\eta_0}_{P,2} =& \snorm{\hat{\beta}-\beta_0}_{P,2}\vee \snorm{\hat{\gamma}-\gamma_0}_{P,2}   \lesssim \left(\frac{\log p}{N}\right)^{1/2-1/4\alpha} \le \delta_N,
\end{align*}
where the last inequality holds since $(\log p/N)^{1/4\alpha}\le \log p$ holds with $\alpha>1$.
Also,
\begin{align*}
    \snorm{\hat{\beta}-\beta_0}_{P,2}\times \paren{\snorm{\hat{\beta}-\beta_0}_{P,2}+\snorm{\hat{\gamma}-\gamma_0}_{P,2}} \lesssim \paren{\frac{\log p}{N}}^{1-1/2\alpha} = \delta_N N^{-1/2}
\end{align*}
holds. 
    \item For the exponential decay case, let $\delta_N = N^{-1/2}\log p\log N$. By the assumption $\log p = o(N^{1/4})$, $\delta_N = o(1)$. 
We have
\begin{align*}
    \snorm{\hat{\eta}-\eta_0}_{P,2} \lesssim \sqrt{\frac{\log p \log N}{N}} < \delta_N
\end{align*}
and
\begin{align*}
    \snorm{\hat{\beta}-\beta_0}_{P,2}\times \paren{\snorm{\hat{\beta}-\beta_0}_{P,2}+\snorm{\hat{\gamma}-\gamma_0}_{P,2}} \lesssim \frac{\log p \log N}{N} = \delta_N N^{-1/2}.
\end{align*}
\end{enumerate}
A similar argument applies to the cross cases.

We have shown that Assumption 4.1 in \cite{ccddhnr} holds for both sparsity assumptions. 
Therefore, applying Theorem 4.1. in \cite{ccddhnr}, we get the desired results.
\end{proof}

%%%%%%%%%%%%%%%%%%%%%%%%%%%%%%%%%%%%%%%%%%%%%%%%%%%%%%%%%%%%%%%%%%%%%
\subsection{Local Maximum Inequality}
%%%%%%%%%%%%%%%%%%%%%%%%%%%%%%%%%%%%%%%%%%%%%%%%%%%%%%%%%%%%%%%%%%%%%
\begin{lemma}[\cite{CCK2015PTRF}, Lemma 8]\label{lemma:lmi}
Let $(Z_i)_{i=1}^N$ be i.i.d. random vectors, where $Z_i\in \R^p$ with $p\ge 2$. Define $F = \max_{1\le i\le N}\max_{1\le j \le p} \abs{Z_{ij}}$ and $\sigma^2 = \max_{1\le j\le p} E[Z_{ij}^2].$ Then there exists a universal constant $C>0$ such that
\begin{align*}
    E\left[\max_{1\le j \le p} \Big| \frac{1}{N}\sum_{i=1}^N Z_{ij} - E[Z_{ij}] \Big|\right] \le C\left\{ \sigma \sqrt{\frac{\log p}{N}} + \frac{\sqrt{E[F^2]}\log p}{N}\right\}.
\end{align*}
% Let $\calF$ be a class of measurable functions $f$ such that $E\abs{f}<\infty$ for all $f\in \calF$. Let $G_N f = 1/\sqrt{N}\sum_i\{f(X_i)-Ef\}$. Let $F$ be an envelope function. Suppose $0<\norm{F}_{P,2}<\infty$ and let $\sigma^2$ be a fixed constant such that $\sup_{f\in\calF}Ef^2\le \sigma^2\le \norm{F}^2_{P,2}$. Let $B=\sqrt{E[\max_i F^2(X_i)]}$. 
% If $\calF$ is finite, then
% \begin{align*}
%     E[\norm{G_N f}_\calF] \lesssim \sigma \sqrt{1+\log \abs{\calF}} + \frac{B(1+\log \abs{\calF})}{\sqrt{N}}.
% \end{align*}
\end{lemma}

\section{High-Dimensional Linear IV Regression Models}\label{sec:iv}
%%%%%%%%%%%%%%%%%%%%%%%%%%%%%%%%%%%%%%%%%%%%%%%%%%%%%%%%%%%%%%%%%%%%%
Section \ref{sec:regression} in the main text presented the method for high-dimensional linear regression models.
In this section, we extend the method by accommodating high-dimensional linear IV regression models.

%%%%%%%%%%%%%%%%%%%%%%%%%%%%%%%%%%%%%%%%%%%%%%%%%%%%%%%%%%%%%%%%%%%%%
\subsection{The Model}
%%%%%%%%%%%%%%%%%%%%%%%%%%%%%%%%%%%%%%%%%%%%%%%%%%%%%%%%%%%%%%%%%%%%%
Consider the high-dimensional linear IV model
\begin{align}
    Y =& D\theta_0 + X'\Lambda_0 +U, &&E [U|X,Z]=0, \label{eq:iv:y}\\
    Z =& X'\beta_0 + V, &&E [V|X]=0, \label{eq:iv:z}
\end{align}
where $Z$ denotes an instrumental variable and the parameter of interest is the partial effect $\theta_0$ of the endogenous treatment variable $D$ on the outcome variable $Y$. 
To construct a moment restriction under \eqref{eq:iv:y}--\eqref{eq:iv:z}, consider the orthogonal score function
\begin{align}
    \psi(Y,D,X,Z;\theta,\eta) := \{Y-X'\gamma - \theta (D - X'\zeta)\} (Z-X'\beta )\label{eq:robinson_iv},
\end{align}
where $X'\gamma_0 = E[Y|X]$, $X'\zeta_0 = E[D|X]$ and $\eta = (\gamma,\zeta,\beta)$.

%%%%%%%%%%%%%%%%%%%%%%%%%%%%%%%%%%%%%%%%%%%%%%%%%%%%%%%%%%%%%%%%%%%%%
\subsection{The Method}
%%%%%%%%%%%%%%%%%%%%%%%%%%%%%%%%%%%%%%%%%%%%%%%%%%%%%%%%%%%%%%%%%%%%%
This section describes the algorithm for estimation and inference about $\theta_0$ in the high-dimensional linear IV regression model \eqref{eq:iv:y}--\eqref{eq:iv:z}.

\begin{algorithm}[OGA+HDAIC with DML for high-dimensional linear IV models]\label{algorithm:dml_oga_hdaic_iv} \hfill
\begin{enumerate}[label={Step \arabic*.}]
    \item Randomly split the sample indices $\{1,...,N\}$ into $K$ folds $(I_k)_{k=1}^K$. For simplicity, let the size of each fold be $n=N/K$ and the size of $I_k^c$ be $n^c$.
    \item For each fold $k \in \{1,...,K\}$, perform following procedure using $\{(X_i',Z_i)'\}_{i\in I_k^c}$ to get $\hat\beta_k$.
    \begin{enumerate}
        \item Compute $\hat{\mu}_{0,j} = \xkj'\zk/\sqrt{n^c}\snorm{\xkj}$. Select the coordinate $\hat{j}_{1} = \argmax_{1\le j \le p} \abs{\hat{\mu}_{0,j}}$. Define $\hat{J}_1 = \{\hat{j}_1\}$.
        \item Compute $\hat{\mu}_{1,j} = \xkj'(I_{n^c}- H_1) \zk / \sqrt{n^c}\snorm{\xkj}$, where $H_1 = X_{I_k^c \hat{j}_1}(X_{I_k^c \hat{j}_1}'X_{I_k^c, \hat{j}_1})^{-1}X_{I_k^c, \hat{j}_1}'$. Select the coordinate $\hat{j}_2 = \argmax_{1\le j \le p,j\notin \hat{J}_1} \abs{\hat{\mu}_{1,j}}$. Update $\hat{J}_2 =\hat{J}_1\cup \{\hat{j}_2\}$.
        \item Given $m-1$ coordinates $\hat{J}_{m-1}$ that have been obtained, compute $\hat{\mu}_{m-1,j} = \xkj'(I_{n^c}- H_{m-1}) \zk / \sqrt{n^c}\snorm{\xkj}$, where $H_{m-1} = X_{I_k^c \hat{J}_{m-1}}(X_{I_k^c \hat{J}_{m-1}}'X_{I_k^c \hat{J}_{m-1}})^{-1}X_{I_k^c \hat{J}_{m-1}}'$. Select the coordinate $\hat{j}_m = \argmax_{1\le j\le p, j\notin \hat{J}_{m-1}}\abs{\hat{\mu}_{m,j}}$. Iteractively update $\hat{J}_m = \hat{J}_{m-1}\cup \{\hat{j}_m\}$.
        \item Compute $\textup{HDAIC}(\hat{J}_m) = (1+C^* |\hat{J}_m|\log p/n)\hat{\sigma}_{m}^2$ for each $m$, where $C^*$ is from (\ref{Cstar}) in Appendix \ref{sec:regression:details} and $\hat{\sigma}_{m}^2 = 1/n \zk '(I-H_m)\zk$. Choose $\hat{m} = \argmin_{1\le m\le M_n^*}$ $\textup{HDAIC}(\hat{J}_m)$, where $M_n^*$ is defined in (\ref{Mnstar}) in Appendix \ref{sec:regression:details}.
		\item With coordinates $\hat{J}_{\hat{m}}$, run OLS of $Z_i$ on $X_{i\hat{J}_{\hat{m}}}$ to get $\hat{\beta}_k.$
    \end{enumerate}
    \item Repeat Step 2 with $\{(X_i',D_i)'\}_{i\in I_k^c}$ in place of $\{(X_i',Z_i)'\}_{i\in I_k^c}$, to get $\hat{\zeta}_k$ for each fold $k \in \{1,...,K\}$.
    \item Repeat Step 2 using $\{(X_i',Y_i)'\}$ to get $\hat{\gamma}_k$ for each fold $k \in \{1,...,K\}$.
    \item Obtain $\check{\theta}$ as a solution to $1/K \sum_{k=1}^K 1/n \sum_{i\in I_k}\psi(Y_i,D_i,X_i,Z_i;\check{\theta},\hat{\eta}_k)=0$ where $\hat{\eta}_k=(\hat{\gamma}_k,\hat{\beta}_k,\hat{\zeta}_k)$ and $\psi$ is defined in \eqref{eq:robinson_iv}.
    \item Compute $\hat{M} = -1/K\sum_{k=1}^K 1/n\sum_{i\in I_k}(D_i - X_i'\hat{\zeta})(Z_i - X_i'\hat{\beta})$. Obtain a variance estimator of $\check{\theta}$ as $\hat{\Omega} = \hat{M}^{-1} \frac{1}{K} \sum_{k=1}^K \frac{1}{n} \sum_{i\in I_k} [\psi(Y,D,X,Z;\check{\theta},\hat{\eta}_k)\psi(Y,D,X,Z;\check{\theta},\hat{\eta}_k)'](\hat{M}^{-1})'$.
\end{enumerate}
\end{algorithm}

Observe that this algorithm parallels Algorithm \ref{algorithm:dml_oga_hdaic}, and hence similar remarks are in order.
First, the procedure (Steps 1--4) uses the cross fitting to remove an over-fitting bias.
Second, the coordinates $\{\hat j_1,...,\hat j_{p}\}$ are ranked in Step 2 (a)--(c) in the order of decreasing importance after successive orthogonalization using OGA as in \cite{ing2020}.
Third, a subset $\hat J_{\hat m} = \{\hat j_1,...,\hat j_{\hat m}\}$ of the ordered set $\{\hat j_1,...,\hat j_{p}\}$ is selected in Step 2 (d) using HDAIC as in \cite{ing2020}.
The combined use of these three elements (DML, OGA, and HDAIC) together allows for a novel root $N$ consistent estimation of $\theta_0$ without assuming traditional functional class restrictions (e.g., the sparsity) required by existing popular estimators (e.g., LASSO).
Section \ref{sec:ivregression:theory} formally presents theoretical arguments in support of this claim.

%%%%%%%%%%%%%%%%%%%%%%%%%%%%%%%%%%%%%%%%%%%%%%%%%%%%%%%%%%%%%%%%%%%%%
\subsection{The Theory}\label{sec:ivregression:theory}
%%%%%%%%%%%%%%%%%%%%%%%%%%%%%%%%%%%%%%%%%%%%%%%%%%%%%%%%%%%%%%%%%%%%%
This section follows as a corollary to the main theory in section 2. {Again, we use a generic notation $\Epsilon$ to refer to $\Epsilon_D = D - X'\zeta_0$ and $\Epsilon_Y = Y - X'\gamma_0$.} 

\begin{assumption}\label{a:1:iv}
	For each $N\in \mathbb N$, it holds that
\begin{enumerate}[(a)]%[counter-format=(\alph*),item-indent = 0.55cm,label-offset = 5pt](1)
    \item $(Y_i,D_i,X_i',Z_i)_{i=1}^N$ are i.i.d. copies of $(Y,D,X',Z)$.
    \item \eqref{eq:iv:y} and \eqref{eq:iv:z} hold. 
    \item $E[\abs{Y}^q] + E[\abs{D}^q] + E[\abs{Z}^q] \le C_q$. 
    \item $E[\abs{UV}^2]\ge c_q^2$ and $E[DV]\ge c_q$.
    % \item $E[\max_{1\le i\le N} E [U^2|X]]\le C$ and \\$E[\max_{1\le i\le N}E [V^2|X]\le C$ 
    % \item ${E[X_{ij}^4]}\le C_4$, $E[V^4] \le C_4$, and $E[\Epsilon^4]\le C_4$. \label{4th_bdd:iv}
    \item ${\max_{1\le j\le p}E[|X_{ij}|^q]}\le C_q$, $E[\abs{V}^q] \le C_q$, and $E[\abs{\Epsilon}^q]\le C_q$. \label{qth_bdd:iv}
\end{enumerate}
Furthermore, it holds asymptotically that (f) $K_{N,q}^2 \log p/ N^{1-2/q}=o(1).$ \label{maxqth_bdd:iv} 
\end{assumption}

\begin{assumption}\label{a:2:iv}
It holds over $N\in \mathbb N$ that
\begin{enumerate}[(a)]%[counter-format = (\alph*),item-indent=.55cm,label-offset=5pt]
    % \item $E[\max_{1\le j\le p}\abs{X_{ij}}^q]\le C,\, E[\abs{V}^q] \le C$, $E[\abs{Y-X'\gamma_0}^q]\le C$ and $\log p = o(N^{1/4}).$
    \item $\lambda_{\min} (\Gamma)\ge \lambda_1>0$ and $\lambda_{\max} (\Gamma) \le C_q$, where $\Gamma=E[X X']$. \label{additionalA:iv}
    \item Define $\Gamma(J) = E[X_{iJ}X_{iJ}']$ and $d_{\ell}(J) = E[X_{i\ell}X_{iJ}]$ for the set of coordinate indices $J$.$$\max_{1\le \abs{J}\le \bar{C}(N/\log p)^{1/2},\,\ell \notin J} \abs{\Gamma^{-1}(J)d_\ell(J)}< C_q.$$ \label{A5:iv}
\end{enumerate}
\end{assumption}\vspace{-.6cm}

\begin{assumption}\label{a:3:iv}
For each of $\xi_0 = \beta_0$, $\zeta_0$ and $\gamma_0$, $\xi_0$ follows either (a) or (b) described below. 
\begin{enumerate}[(a)]%[counter-format = (\alph*),item-indent=.55cm,label-offset=5pt]
    \item Polynomial decay: $\log p = o(N^{1-2/q})$. Each $\xi_0$ is such that $ \norm{\xi_0}_2^2 \le C_0$ for some $C_0>0$, there exist $\alpha> 1$ such that for any $J \subseteq \mathfrak{P}$,$$\hspace{2cm}\norm{\xi_0(J)}_1 \le C \left( \norm{\xi_0(J)}_2^2\right)^{(\alpha-1)/(2\alpha-1)}.\label{poly} $$ 
    
    \item Exponential decay: $\log p= o(N^{1/4}).$ Each $\xi_0$ is such that $ \norm{\xi_0}_\infty \le C_0$ for some $C_0>0$ and there exists $C_1>1$ such that for any $J \subseteq \mathfrak{P}$, $$\norm{\xi_0(J)}_1 \le C_1 \norm{\xi_0(J)}_\infty.\label{expo}$$ 
\end{enumerate}
\end{assumption}

Assumptions \ref{a:1:iv}--\ref{a:3:iv} closely parallel Assumptions \ref{a:1}--\ref{a:3}, and thus similar remarks apply here.
The following theorem supports the estimation and inference procedure presented in Algorithm \ref{algorithm:dml_oga_hdaic_iv}.

\begin{theorem}\label{theorem:iv:regression}  Let $(\calP_N)_{N\in \mathbb N}$ be a sequence of sets of DGPs such that Assumptions
	\ref{a:1:iv}--\ref{a:3:iv}
 %\ref{a:4}--\ref{a:6} 
	are satisfied on the model \eqref{eq:iv:y}--\eqref{eq:iv:z}. Then, the estimator $\check{\theta}$ follows
\begin{align*}
    \sqrt{N} \paren{\check{\theta}-\theta_0} \xrightarrow{d} N(0,\Omega),
\end{align*}
where $\Omega = (E[DV])^{-1}E[V^2U^2](E[DV])^{-1}$. 
Define $\hat{M} :=- 1/K\sum_{k=1}^K 1/n \sum_{i\in I_k} (D_i-X_i'\hat{\zeta})(Z_i - X_i'\hat{\beta})$. Then, we can define the variance estimator \begin{align*}
    \hat{\Omega} = \hat{M}^{-1}\frac{1}{K} \sum_{k=1}^K \frac{1}{n}\sum_{i\in I_k} [\psi(Y,D,X,Z;\check{\theta},\hat{\eta}_k)\psi(Y,D,X,Z;\check{\theta},\hat{\eta}_k)'](\hat{M}^{-1})'
\end{align*}
and the confidence regions with significance level $a\in (0,1 )$ have uniform asymptotic validity:
\begin{align*}
    \sup_{P\in \calP_N} \abs{P\left(\theta_0 \in \left[\check{\theta} \pm \Phi^{-1}(1-a/2)\sqrt{\hat{\Omega}/N}\right]\right)-(1-a)} =o(1).
\end{align*}
\end{theorem}

\noindent
A proof is provided in Appendix \ref{sec:theorem:iv:regression}.
As in the case of the baseline regression model, we once again emphasize that this result does not rely on the sparsity assumption which is used in the literature on high-dimensional linear models.

Appendix \ref{sec:simulation:iv} presents simulation designs and results for high-dimensional IV regression models studied in the current appendix section.
The results are similar to those obtained for the baseline model presented in Section \ref{sec:simulations}.
Namely, while our proposed method based on the OGA and HDAIC perform well in terms of all the simulation statistics, the LASSO-based method slightly underperforms and the random-forest-based method significantly underperforms.
These differences in the finite-sample performance widen as the degree of polynomial decay becomes smaller.

%%%%%%%%%%%%%%%%%%%%%%%%%%%%%%%%%%%%%%%%%%%%%%%%%%%%%%%%%%%%%%%%%%%%%
\section{Proofs for the Extensions}
%%%%%%%%%%%%%%%%%%%%%%%%%%%%%%%%%%%%%%%%%%%%%%%%%%%%%%%%%%%%%%%%%%%%%

%%%%%%%%%%%%%%%%%%%%%%%%%%%%%%%%%%%%%%%%%%%%%%%%%%%%%%%%%%%%%%%%%%%%%
\subsection{Proof of Theorem \ref{theorem:approx}}\label{sec:theorem:approx}
%%%%%%%%%%%%%%%%%%%%%%%%%%%%%%%%%%%%%%%%%%%%%%%%%%%%%%%%%%%%%%%%%%%%%

\begin{proof}
    Throughout this proof, write $\|v\|_{A}^{2}=v^{\top} A v$ for a vector $v$ and nonnegative definite matrix $A$.
    In this proof, we show the prediction norm rates of the nuisance parameters, that are subsequently used in the proof of Theorem \ref{theorem:regression}, to be attainable with the reduced form models \eqref{eq:y:approx} and \eqref{eq:d:approx}.
    The proof consists of two parts: first we establish the convergence rates of OGA under the setting with approximation errors and, in the second part, the convergence rates of OGA coupled with HDAIC. These two parts correspond to Theorems 2.1 and 3.1. in \cite{ing2020}, respectively. Hence the proof strategies follow closely the proofs of these two results with appropriate modifications. 
    %{\color{blue} We provide proof under Assumption \ref{a:3} \ref{poly} only. The proof under Assumption \ref{a:3} \ref{expo} can be derived analogously to the proof here.} 

    We consider the estimation of $\beta_0$ from \eqref{eq:d:approx} within a partition with sample size $n$. The same logic applies to $\gamma_0$ in \eqref{eq:y:approx}. 

    \textit{Step 1.} (OGA part 1: Definitions) In this step, we show how to modify the definitions of some objects and sets of events in \cite{ing2020} to accommodate the presence of the extra approximation errors.
    Define $X_{\cdot j}$ as the $j-$th coordinate of $X$. Consider the model \eqref{eq:d:approx} and define
    \begin{align}
        D(X) =& \sum_{j=1}^p \beta_j X_{\cdot j}, \quad D_J(X) = \sum_{j\in J} \beta_j X_{\cdot j},\nonumber\\
        \hat{D}_m(X) \equiv& \hat{D}_{\hat{J}_m}(X)= \sum_{j\in\hat{J}_m}\hat{\beta}_jX_{\cdot j}, \quad \hat{D}_{i;\hat{J}_m} = \sum_{j\in\hat{J}_m}\hat{\beta}_j X_{ij}, \nonumber\\
        \mu_{J,k} =& E[(D(X)-D_J(X))X_{\cdot k}]/\sigma_k,\quad \sigma_k =\sqrt{E[X_{ik}^2]}, \nonumber\\
        \hat{\mu}_{J,k} =& \frac{1/n\sum_{i=1}^n(D_i - \hat{D}_{i;J})X_{ik}}{(1/n\sum_{i=1}^n X_{ik}^2)^{1/2}}, \nonumber
    \end{align}
    and the collections of events 
    \begin{align}
        A_n(m) =& \{ \max_{(J,k):\abs{J}\le m-1, k\notin J} \abs{\hat{\mu}_{J,k} - \mu_{J,k}} \le C(\log p/n)^{1/2}\},\text{ and}\label{eq:Anm}\\
        B_n(m) =& \{ \min_{0\le j\le m-1} \max_{1\le k\le p} \abs{\mu_{\hat{J}_j,k}}> \bar{\xi} C(\log p/n)^{1/2}\}\label{eq:Bnm},
    \end{align}
    where $\bar{\xi}, C>0$ are some large constants. 

    Now, define the corresponding variables with the approximation errors:
    \begin{align}
        D_i^r =& \sum_{j=1}^p \beta_j X_{ij} + r_D(X_i) + V_i, \quad D^r(X) = \sum_{j=1}^p \beta_j X_{\cdot j} + r_D(X),\nonumber\\
        \mu^r_{J,k} =& E[(D^r(X)-D_J(X))X_{\cdot k}]/\sigma_k = \mu_{J,k} + E[r(X)X_{\cdot k}]/\sigma_k, \nonumber\\
        \hat{\mu}^r_{J,k} =& \frac{1/n\sum_{i=1}^n(D_i^r - \hat{D}_{i;J})X_{ik}}{(1/n\sum_{i=1}^n X_{ik}^2)^{1/2}} =\hat{\mu}_{J,k} + \frac{1/n\sum_{i=1}^n r(X_i)X_{ik}}{(1/n\sum_{i=1}^n X_{ik}^2)^{1/2}}, \nonumber
     \end{align}
    and %define the collections of events
    \begin{align}     
        A_n^r(m) =& \{ \max_{(J,k):\abs{J}\le m-1, k\notin J} \abs{\hat{\mu}^r_{J,k} - \mu^r_{J,k}} \le C(\log p/n)^{1/2}\},\text{ and}\label{eq:Anm:approx}\\
        B_n^r(m) =& \{ \min_{0\le j\le m-1} \max_{1\le k\le p} \abs{\mu^r_{\hat{J}_j,k}}>\tilde{\xi}C(\log p/n)^{1/2}\},\label{eq:Bnm:approx}
    \end{align}
    where $\tilde{\xi}=2/(1-\xi)$ for some $0<\xi<1$. 

    % Note that the bounds in \eqref{eq:Anm} and \eqref{eq:Bnm} hold with and without the approximation errors. 
    We will show that \eqref{eq:Anm}, \eqref{eq:Bnm} and Assumption \ref{a:approx} imply \eqref{eq:Anm:approx} and \eqref{eq:Bnm:approx} with appropriate choices on the constants.
    \begin{align*}
        \abs{\hat{\mu}_{J,k}^r - \mu^r_{J,k}} =& \left| \hat{\mu}_{J,k}-\mu_{J,k}+\frac{1/n\sum_{i=1}^n r(X_i)X_{ik}}{(1/n\sum_{i=1}^n X_{ij}^2)^{1/2}} - E[r(X)X_{\cdot k}]/\sigma_k \right|\\
        \le &\left| \hat{\mu}_{J,i}-\mu_{J,i}\right| + \left| \frac{1/n\sum_{i=1}^n r(X_i)X_{ik}}{(1/n\sum_{i=1}^n X_{ij}^2)^{1/2}} - E[r(X)X_{\cdot k}]/\sigma_k \right|\\
        \le &\left| \hat{\mu}_{J,i}-\mu_{J,i}\right| + R_{p,3},
    \end{align*}
    where $R_{p,3} \equiv \max_{1\le k\le p} \left|{1/n\sum_{i=1}^n r(X_i)X_{ik}}/{(1/n\sum_{i=1}^n X_{ij}^2)^{1/2}} - E[r(X)X_{\cdot k}]/\sigma_k\right|$ and $R_{p,3}=o_p(1)$ 
    by Assumption \ref{a:approx} \ref{r4_bdd}, \ref{rX_bdd}, and Lemma \ref{lemma:lmi} in Appendix A.2.

    Conditional on the events $B_n(m)$,
    \begin{align*}
        \abs{\mu_{\hat{J}_j,k}^r} =& \left| \mu_{\hat{J}_j,k} + E[r(X)X_{\cdot k}]/\sigma_k \right|\\
        \le& \left| \mu_{\hat{J}_j,k}\right| + \left| E[r(X)X_{\cdot k}]/\sigma_k \right|\\
        \le& \bar{\xi} C(\log p / n)^{1 / 2} + C(\log p / n)^{1 / 2},
    \end{align*}
    where $C>0$ and $\bar{\xi}$ is so that $\tilde{\xi}\equiv \bar{\xi}+1=2/(1-\xi)$.

    Using the above definitions, it holds for all $1\le q \le m$ on $A_n^r(m)\cap B_n^r(m)$,
    \begin{align*}
        \left| \mu^r_{\hat{J}_{q-1},j_q} \right| \ge& - \left| \hat{\mu}^r_{\hat{J}_{q-1},\hat{j}_q} - \mu^r_{\hat{J}_{q-1},j_q} \right| + \abs{\hat{\mu}^r_{\hat{J}_{q-1},\hat{j}_q}} \\
        \ge& - \max_{(J,i):\#(J)\le m-1,i\notin J}\abs{\hat{\mu}^r_{\hat{J}_{q-1},\hat{j}_q} - \mu^r_{\hat{J}_{q-1},j_q}} +  \abs{\hat{\mu}^r_{\hat{J}_{q-1},\hat{j}_q}} \\
        \ge& -C(\log p_n/n)^{1/2} + \max_{1\le j \le p_n} \abs{\hat{\mu}^r_{\hat{J}_{q-1},j}} \\
        \ge& -2C(\log p_n/n)^{1/2} + \max_{1\le j \le p_n} \abs{\mu^r_{\hat{J}_{q-1},j}} \\
        >& \xi \max_{1\le j \le p_n} \abs{\mu^r_{\hat{J}_{q-1},j}},
    \end{align*} 
    where the first inequality comes from the triangle inequality, the second from taking the maximum, the third from \eqref{eq:Anm:approx} and since $\abs{\hat{\mu}^r_{\hat{J}_{q-1},\hat{j}_q}}=\max_{1\le j \le p_n} \abs{\hat{\mu}^r_{\hat{J}_{q-1},j}}$, the fourth from the triangle inequality and \eqref{eq:Anm:approx}, and the last from $2C(\log p_n/n)^{1/2} < (2/\tilde{\xi}) \max_{1\le j \le p_n} \abs{\mu^r_{\hat{J}_{q-1},j}} $ on $B_n^r(m)$ and $1-\xi=2/\tilde{\xi}$. 
    
    Hence, with Assumption \ref{a:approx} and $\tilde{\xi}= \bar{\xi}+1$, $A_n(m)$ implies $A_n^r(m)$ and thus $\lim_{n\to\infty}P(A_n(m))=1$ implies $\lim_{n\to\infty}P(A_n^r(m))=1$.
    In \textit{Step 3}, we derive the bounds on $A_n^r(m) \cap B_n^r(m)$ and $A_n^r(m) \cap (B_n^r(m))^c$, respectively, and use the fact that $\lim_{n\to\infty}P(A_n(m))=1$ to show that it is always the case that either $A_n^r(m) \cap B_n^r(m)$ or $A_n^r(m) \cap (B_n^r(m))^c$ holds.

    \textit{Step 2.} (OGA part 2: Lemma A.1. from \cite{ing2020})
    % Following \cite{ing2020}, 
    We now establish error bounds for the population OGA under some high-level conditions (for polynomial decay). Recursively define $J_{\xi,m} = J_{\xi,m-1}\cup \{j_{\xi,m}\}$, with $J_{\xi,0} = \emptyset$ and $j_{\xi,m}$ any element $\ell \in \{1,\dots,p\}$ satisfying 
    \begin{align}
        \abs{E[V_{m-1}X_{\cdot \ell}]}\ge \xi \max_{1\le j\le p}\abs{E[V_{m-1}X_{\cdot j}]}. \label{def:xi}
    \end{align}
    Denote $V_m= D^r(X)-D_{J_{\xi,m}}(X)$, then
    \begin{align}
        \begin{aligned}
            E[V_m^2] \le& E\brac{(D^r(X)-D_{J_{\xi,m}}(X))\sum_{j=1}^p \beta_j X_{\cdot j}} \\
            \le& \max_{1\le j\le p}\abs{\mu^r_{J_{\xi,m},j}}\sum_{j=1,j\notin J_{\xi,m}}^p\abs{\beta_j}.\label{ing:A.3}
        \end{aligned}
    \end{align}
    
    Recall it can be written that $V_m = X'\beta(J_{\xi,m}^c)$. Then Assumption \ref{a:2} \ref{additionalA} implies 
    \begin{align}
        E[V_m^2]\ge \lambda_1 \sum_{j=1,j\notin J_{\xi,m}}^p\beta_j^2.\label{ing:A.4}
    \end{align}
    Combining \eqref{ing:A.3} and \eqref{ing:A.4}, we obtain 
    \begin{align}
        \begin{aligned}
            E[V_m^2] \le& C \max_{1\le j\le p}\abs{\mu^r_{J_{\xi,m},j}}\paren{\sum_{j=1,j\notin J_{\xi,m}}^p \beta_j^2}^{(\alpha-1)/(2\gamma-1)}\\
            \le& C \lambda_1^{-(\alpha-1)/(2\alpha-1)}\max_{1\le j\le p} \abs{\mu^r_{J_{\xi,m},j}}\{E[V_m^2]\}^{(\alpha-1)/(2\alpha-1)}.\label{ing:A.5}
        \end{aligned}
    \end{align}
    
    Note that $\beta(J)=\Gamma^{-1}(J)E[X(J)'D]=\argmin_{c\in\R^{|J|}}E[(D-X'c)^2]$. We now have
    \begin{align}
        \begin{aligned}
        E[V_{m+1}^2] =& E\left[(D^r(X)-\sum_{j\in J_{\xi,m}}\beta_j(J_{\xi,m+1})X_{\cdot j} - \beta_{j_{m+1}}(J_{\xi,m+1})X_{\cdot j_{\xi,m+1}} )^{2}\right] \\
        \le& E\left[(D^r(X)-\sum_{j\in J_{\xi,m}}\beta_j(J_{\xi,m+1})X_{\cdot j} - \mu^r_{J_{\xi, m}, j_{\xi, m+1}} X_{\cdot j_{\xi, m+1}})^{2}\right]\\
        =& E\left[(V_m - \mu^r_{J_{\xi, m}, j_{\xi, m+1}} X_{\cdot j_{\xi, m+1}})\right]^{2} \\
        \le& E[V_m^2] -\xi^2 \max_{1 \leq j \leq p} (\mu_{J_{\xi, m}, j}^r)^2 \\
        \le& E[V_m^2] -\xi^{2} \lambda_{1}^{2(\gamma-1) /(2 \gamma-1)} C_{\gamma}^{-2}\left[\mathrm{E}\left(V_{m}^{2}\right)\right]^{2 \gamma /(2 \gamma-1)} \\
        =&\mathrm{E}\left(V_{m}^{2}\right)\left\{1-\xi^{2} \lambda_{1}^{2(\gamma-1) /(2 \gamma-1)} C_{\gamma}^{-2}\left[\mathrm{E}\left(V_{m}^{2}\right)\right]^{1 /(2 \gamma-1)}\right\},\label{ing:A.7}
% In case I get confused in the future again, this A7 is not about the direct relationship between beta and mu: it's just that beta is the minimizer and mu is just another constant, which makes the RHS larger.
        \end{aligned}
    \end{align}
    where the second inequality comes from \eqref{def:xi} and the third inequality comes from \eqref{ing:A.5}. Using \eqref{ing:A.7} and Lemma 1 of \cite{GaoIngYang2013}, we obtain the following bound for $G_1>0$:
    \begin{align}
        E[V_m^2] \le G_1 m^{-2\alpha +1}. \label{ing:A.1.:poly}
    \end{align}
    
    \textit{Step 3.} (OGA part 3: Combining Steps 1 and 2)
    Define the shorthand notations $W_1^N = (Y_i,D_i,X_i')_{i=1}^N$ and $E_{W_1^N} = E[\cdot|W_1^N]$.
    Combining \textit{Step 1} and \textit{Step 2}, we obtain the bound 
    \begin{align}
        E_{W_1^N}[(D^r(X)-D_{\hat{J}_m}(X))^2]\le G_1 m^{-2\alpha+1} \quad \text{on } A_n^r(m)\cap B_n^r(m).\label{ing:2.24}
    \end{align}
    Using Assumption \ref{a:3:approx} \ref{poly} and \eqref{ing:A.4}, we have for any $0\le l \le m-1$,
    \begin{align}
        E_{W_1^N}[(D^r(X)-D_{\hat{J}_l}(X))^2] \le \paren{C_\alpha \max_{1\le j\le p}\abs{\mu^r_{\hat{J}_l,j}}}^{2-1/\alpha}\lambda_1^{-1+1/\alpha}.\label{ing:2.25}
    \end{align}
    By \eqref{ing:2.25}, we have
    \begin{align}
        \begin{aligned}
            E_{W_1^N} \left[(D^r(X)-D_{\hat{J}_{m}}(X))^2\right] \le& \min_{0 \leq l \leq m-1} E_{W_1^N}\left[(D^r(X)-D_{\hat{J}_l}(X))^{2}\right] \\
            \le& C_{\alpha}^{2-1/\alpha} \lambda_{1}^{-1+1/\alpha}\left(\min _{0 \leq l \leq m-1} \max _{1 \leq j \leq p}\left|\mu^r_{\hat{J}_{l}, j}\right|\right)^{2-1/\alpha} \\
            \le& C_{\alpha}^{2-1/\alpha} \lambda_{1}^{-1+1/\alpha}\left(\tilde{\xi} {C}\right)^{2-1/\alpha}\left(\log p/n\right)^{1-1/2\alpha},\label{ing:2.26}
        \end{aligned}
    \end{align}
    where the last inequality holds conditioning on $(B_{n}^r(m))^{c}$.

    Combining \eqref{ing:2.24} and \eqref{ing:2.26}, for all $1\le m \le K_n$ and $C>0$, we have
    \begin{align}
        E_{W_1^N} \left[(D^r(X)-D_{\hat{J}_{m}}(X))^2\right] I_{A^r_{n}\left(K_{n}\right)} \le C \max \left\{m^{-2 \alpha+1},\left\{ \log p/n\right\}^{1-1/2 \alpha}\right\}\label{ing:2.27}.
    \end{align}
    Under Assumption \ref{a:approx}, $\lim_{n\to \infty} P(A_n(m))=1$ as shown in Section S1 of supplementary material of \cite{ing2020}, we  then have $\lim_{n\to \infty} P(A_n^r(m))=1$ following the conclusion of \textit{Step 1}. With \eqref{ing:2.27} we achieve
    \begin{align}
        \max_{1\le m\le K_n} \frac{E_{W_1^N} [(D^r(X)-D_{\hat{J}_{m}}(X))^2]}{\max\left\{m^{-2 \alpha+1},\left\{ \log p/n\right\}^{1-1/2 \alpha}\right\}} \lesssim_P C.
    \end{align}

    Note that we are interested in the conditional mean squared prediction error, $E_{W_1^N} [(D^r(X)-\hat{D}_m(X))^2] = E_{W_1^N} [(D^r(X)-D_{\hat{J}_{m}}(X))^2] + E_{W_1^N}[(D_{\hat{J}_{m}}(X) - \hat{D}_m(X))^2]$.
    The convergence rate for the latter term is
    \begin{align}
        \max_{1\le m\le K_n} \frac{E_{W_1^N}[(D_{\hat{J}_{m}}(X) - \hat{D}_m(X))^2]}{m\log p/n} \lesssim_P C,\label{ing:2.32}
    \end{align}
    where the proof follows exactly the same arguments as in Section S1 of supplementary material in \cite{ing2020} under our current setting. Combining \eqref{ing:2.27} and \eqref{ing:2.32}, we obtain
    \begin{align}
        \max_{1\le m\le K_n} \frac{E_{W_1^N} [(D^r(X)-\hat{D}_m(X))^2]}{m^{-2 \alpha+1} + m\log p/n} \lesssim_P C \label{ing:2.22}.
    \end{align}

    % {\color{blue} (lemma A.1. for expo case)
    % We then derive the error bounds for the exponential decay case. Under Assumption \ref{a:3:approx} \ref{expo}, \eqref{ing:A.3}, and \eqref{ing:A.4}, we have
    % \begin{align*}
    %     \paren{E[V_m^2]}^{1/2}\le \lambda_1^{-1/2} C_1 \max_{1\le j\le p}\abs{\mu^r_{J_{\xi,m},j}},
    % \end{align*}
    % where $C_1$ comes from Assumption \ref{a:3:approx} \ref{expo}.

    % Analogous to how we derived \eqref{ing:A.1.:poly}, 
    % \begin{align*}
    %     E[V_{m+1}^2] \le& E[V_m^2] - \xi^2 \lambda_1 C_1^{-2} E[V_m^2]\\
    %     =& E[V_m^2]( 1 - \xi^2 \lambda_1C_1^{-2},)
    % \end{align*}
    % The following bound follows from $C_1>1$, $0<\lambda_1 \le 1$, and $0<\xi \le 1$: for $G_2,G_3>0$, 
    % \begin{align}
    %     E[V_m^2] \le G_2\exp(-G_3\cdot m). \label{ing:A.2.:expo}
    % \end{align}}
    
    \textit{Step 4.} (OGA+HDAIC) Using the results in the previous steps, we now replace $m$ with $\hat{k}_n$ obtained from HDAIC and establish the convergence rate under such setting.
    Define
    \begin{align*}
        V(J) =& D(X) - X(J)'\beta(J),\\
        V^r(J) =& V(J) + r(X) = D(X) - X(J)'\beta(J) + r(X), \\
        V_i(J) =& D_i - V_i - X_i(J)'\beta(J), \text{ and}\\
        V^r_i(J) =& V_i(J) + r(X_i) = D_i - V_i - \sum_{j\in J}\beta_jX_{ij} + r(X_i).
    \end{align*}
    We will establish the following four inequalities for any $1\le m \le K_n$ %{\color{blue} [TO DO] wordings can be improved here}
    \begin{align}
        \abs{\frac{1}{n}\sum_{i=1}^n (V^r_i(\hat{J}_m))^2 - E_{W_1^N}[(V^r(\hat{J}_m))^2]}\le& C R_{1,p}\{E_{W_1^N}[V^2(\hat{J}_m)]\}^{(\alpha-1)/(2\alpha-1)},\label{ing:3.29}\\
        \abs{\frac{1}{n}\sum_{i=1}^n V_iV_i^r(\hat{J}_m)} \le& C R_{2,p} \{E_{W_1^N}[V^2(\hat{J}_m)]\}^{(\alpha-1)/(2\alpha-1)},\label{ing:3.30}\\
        \max_{1\le m\le K_n} \frac{\norm{\frac{1}{n}\sum_{i=1}^n X_i(\hat{J}_m)V_i^r(\hat{J}_m)}^2_{\hat{\Gamma}^{-1}(\hat{J}_m)}}{ m \{E_{W_1^N}[V^2(\hat{J}_m)]\}^{(2\alpha-2)/(2\alpha-1)}} \le& \left\|\hat{\Gamma}^{-1}(K_n)\right\| C R_{1,p}^2,\label{ing:3.31}
        \\        \max_{1\le m\le K_n} \norm{\frac{1}{n}\sum_{i=1}^n X_i(\hat{J}_m)V_i}^2_{\hat{\Gamma}^{-1}(\hat{J}_m)} \le& \left\|\hat{\Gamma}^{-1}(K_n)\right\| R_{2,p}^2,\label{ing:3.32}
    \end{align}
    where $R_{r,1}\equiv \abs{1/n\sum_{i=1}^n r^2(X_i) - E[r^2(X)]}=o_p(1)$ by Assumption \ref{a:approx} \ref{r4_bdd} and the LLN, $R_{2,p}\equiv \max_{1\le j\le p} \abs{1/n\sum_{i=1}^n X_{ij}V_i}\lesssim_P (\log p/n)^{1/2}$ from \eqref{A1}, and recall that $\|v\|_{A}^{2}=v^{\top} A v$ for a vector $v$ and nonnegative definite matrix $A$.
    
    Among the above, we prove \eqref{ing:3.29}--\eqref{ing:3.31} since they include the variables with the approximation errors as \eqref{ing:3.32} does not depend on the newly introduced approximation error in the current result.
    \begin{align*}
        \begin{aligned}
            &\abs{\frac{1}{n}\sum_{i=1}^n (V^r_i(\hat{J}_m))^2 - E_{W_1^N}[(V^r(\hat{J}_m))^2]} \\
            =& \abs{\frac{1}{n}\sum_{i=1}^n V_i^2(\hat{J}_m) + \frac{2}{n}\sum_{i=1}^n V_i(\hat{J}_m)r(X_i) + \frac{1}{n}\sum_{i=1}^n r^2(X_i) - E_{W_1^N}[V^2(\hat{J}_m)] -E_{W_1^N}[r^2(X)]} \\
            \le& \abs{\frac{1}{n}\sum_{i=1}^n V_i^2(\hat{J}_m) - E_{W_1^N}[V^2(\hat{J}_m)]} + 2\abs{\frac{1}{n}\sum_{i=1}^n V_i(\hat{J}_m)r(X_i)} + \abs{\frac{1}{n}\sum_{i=1}^n r^2(X_i) - E_{W_1^N}[r^2(X)]}\\
            \le& \underbrace{\abs{\frac{1}{n}\sum_{i=1}^n V_i^2(\hat{J}_m) - E_{W_1^N}[V^2(\hat{J}_m)]}}_{(a)} + 2\underbrace{\sqrt{\frac{1}{n}\sum_{i=1}^n V_i^2(\hat{J}_m)}\sqrt{\frac{1}{n}\sum_{i=1}^n r^2(X_i)}}_{(b)} + \underbrace{R_{r,1}}_{(c)},
        \end{aligned}
    \end{align*}
    where we want to show that $(b)\lesssim_P (a)$. Note that from Assumption \ref{a:3:approx} \ref{poly} and \eqref{ing:A.4} as shown in Section S2 of supplementary material of \cite{ing2020}, we have
    \begin{align*}
        \abs{\frac{1}{n}\sum_{i=1}^n V_i^2(\hat{J}_m)} \lesssim_P& E_{W_1^N}[V^2(\hat{J}_m)] + C R_{1,p}\{E_{W_1^N}[V^2(\hat{J}_m)]\}^{(\alpha-1)/(2\alpha-1)},\\
        \lesssim_P& E_{W_1^N}[V^2(\hat{J}_m)],
    \end{align*}
    where the second inequality comes from Assumption \ref{a:1:approx} \ref{Vsq_bdd_below}, and hence
    \begin{align*}
        % \sqrt{\frac{1}{n}\sum_{i=1}^n V_i^2(\hat{J}_m)} \lesssim_P& E_{W_1^N}[V^2(\hat{J}_m)]^{1/2},\\
        \sqrt{\frac{1}{n}\sum_{i=1}^n V_i^2(\hat{J}_m)}\sqrt{\frac{1}{n}\sum_{i=1}^n r^2(X_i)} \lesssim_P& E_{W_1^N}[V^2(\hat{J}_m)]^{1/2} (\log p/n)^{1/2} \lesssim_P (a),
    \end{align*}
    where the first bound comes from Assumption \ref{a:approx} \ref{rsq_bdd} and the last comes from Assumption \ref{a:1:approx} \ref{Vsq_bdd_below}. Since $(b)\lesssim_P (a)$ and $(c) = o_p(1)$ by Assumption \ref{a:approx} \ref{r4_bdd}, we obtain \eqref{ing:3.29}.

    % {\color{blue} [TO DO] verify (3.29) with two cases: Denote $\mathcal{A} = E_{W_1^N}[V^2(\hat{J}_m)]$ and $\mathcal{B} = C R_{1,p}\{E_{W_1^N}[V^2(\hat{J}_m)]\}^{(\alpha-1)/(2\alpha-1)}$.
    % \begin{enumerate}
    %     \item (Case 1): $E_{W_1^N}[V^2(\hat{J}_m)] \to 0$
    %     \begin{align}
    %         \abs{\frac{1}{n}\sum_{i=1}^n V_i^2(\hat{J}_m) - \mathcal{A}} \lesssim_P \mathcal{B}
    %     \end{align}
        
    %     \item (Case 2): $E_{W_1^N}[V^2(\hat{J}_m)] > 1$
    %     \begin{align}
    %         \abs{\frac{1}{n}\sum_{i=1}^n V_i^2(\hat{J}_m) - \mathcal{A}} \lesssim_P \mathcal{B}
    %     \end{align}
   
    % \end{enumerate}}

    Now, note that
    \begin{align*}
        \begin{aligned}
            \abs{\frac{1}{n}\sum_{i=1}^n V_iV_i^r(\hat{J}_m)} =& \abs{\frac{1}{n}\sum_{i=1}^n V_iV_i(\hat{J}_m) + \frac{1}{n}\sum_{i=1}^n V_ir(X_i)}\\
            \le& \abs{\frac{1}{n}\sum_{i=1}^n V_iV_i(\hat{J}_m)} + \abs{\frac{1}{n}\sum_{i=1}^n V_ir(X_i)}\\
            \le& \underbrace{\abs{\frac{1}{n}\sum_{i=1}^n V_i V_i(\hat{J}_m)}}_{(d)} + \underbrace{\sqrt{\frac{1}{n}\sum_{i=1}^n V_i^2} \sqrt{\frac{1}{n}\sum_{i=1}^n r^2(X_i)}}_{(e)}.
        \end{aligned}
    \end{align*}
    By an argument similar to deriving \eqref{ing:3.29} and since term $(e)$ above is bounded by a constant from Assumption \ref{a:1:approx} \ref{qth_bdd} and \ref{a:approx} \ref{r4_bdd}, we obtain \eqref{ing:3.30}.

    Next, observe that
    \begin{align*}
        \begin{aligned}
            &\norm{\frac{1}{n}\sum_{i=1}^n X_i(\hat{J}_m)V_i^r(\hat{J}_m)}_{\hat{\Gamma}^{-1}(m)} \\
            \le& \underbrace{\left\|\hat{\Gamma}^{-1}(K_n)\right\|^{1/2} \left\|\frac{1}{n} \sum_{i=1}^n X_i(\hat{J}_m) V_i(\hat{J}_m)\right\|}_{(f)} + \underbrace{\left\|\hat{\Gamma}^{-1}(K_n)\right\|^{1/2}\left\|\frac{1}{n} \sum_{i=1}^n X_i(\hat{J}_m) r(X_i)\right\|}_{(g)}.
            % \\
            % \le& \left\|\hat{\Gamma}^{-1}(K_n)\right\|^{1/2} C m^{1/2} R_{1,p} \{E_n[V^2(\hat{J}_m)]\}^{(\alpha-1)/(2\alpha-1)},
        \end{aligned}
    \end{align*}
    By a similar manipulations as in \eqref{ing:3.29} and since 
    $$\left\|1/n \sum_{i=1}^n X_i(\hat{J}_m) r(X_i)\right\| \le \norm{1/n \sum_{i=1}^n X_i(\hat{J}_m) r(X_i)}_1 \lesssim_P (\log p/n)^{1/2},$$
    where the last inequality holds from Assumption \ref{a:approx} \ref{r4_bdd}, \ref{rX_bdd}, and Lemma \ref{lemma:lmi} in Appendix A.2., we achieve \eqref{ing:3.31}.

    Recall $M_n^*=\min\{(n/\log p)^{1/2\alpha},\bar{\delta} (n/\log p)^{1/2}\}$ and let $\tilde{k_n}=\min_{1\le k\le K_n}\{E_{W_1^N}(V^2(\hat{J}_k))\le G M_n^{*-2\alpha+1}\}$ where $G > C$ is an appropriate constant that is large enough, where $C$ is defined in \eqref{ing:2.27}.

    Using \eqref{ing:3.29}--\eqref{ing:3.32}, it follows exactly the same the proof shown in \cite{ing2020} under our current setting that
    \begin{align}
        \lim_{n\to\infty}P(\hat{k}_n \le \tilde{k}_n-1) =& 0, \label{ing:3.34}\\
        \lim_{n\to\infty}P(\hat{k}_n \ge C M_n^*) =& 0, \label{ing:3.45}
    \end{align}
    hold and hence we have the following results:
    \begin{align}
        E_{W_1^N} [(D^r(X) - \hat{D}_{\hat{k}_n}(X))^2]I_{\left\{\tilde{k}_{n} \leq \hat{k}_{n}<C {M_{n}^{*}}\right\}}=O_{p}\left(M_{n}^{*-2 \gamma+1}\right),\label{ing:3.46}
        % E_{W_1^N} [(D^r(X) - \hat{D}_{\hat{k}_n}(X))^2] I_{\left\{\tilde{k}_{n} \leq \hat{k}_{n} \leq K_{n}\right\}}=O_{p}\left((\log p / n)^{1 / 2}\right), \quad \gamma=1.\label{ing:3.47}
    \end{align}
    and the desired result follows:
    \begin{align}
        E_{W_1^N} [(D^r(X) - \hat{D}_{\hat{k}_n}(X))^2] \lesssim_P (\log p/N)^{1-1/2\alpha}.\label{ing:3.4}
    \end{align} 

    Following the same argument as in the proof of Theorem \ref{theorem:regression}, the conditions for Assumption 4.1. in \cite{ccddhnr} is satisfied and thus by applying Theorem 4.1. in \cite{ccddhnr} we obtain the desired asymptotic normality results.
\end{proof}
%%%%%%%%%%%%%%%%%%%%%%%%%%%%%%%%%%%%%%%%%%%%%%%%%%%%%%%%%%%%%%%%%%%%%

%%%%%%%%%%%%%%%%%%%%%%%%%%%%%%%%%%%%%%%%%%%%%%%%%%%%%%%%%%%%%%%%%%%%%
\subsection{Proof of Theorem \ref{theorem:fullsample}}\label{sec:theorem:fullsample}
%%%%%%%%%%%%%%%%%%%%%%%%%%%%%%%%%%%%%%%%%%%%%%%%%%%%%%%%%%%%%%%%%%%%%

\begin{proof}
    For notational simplicity, we write $\mathcal{Y}=\left[Y_1, \ldots, Y_N\right]^{\prime}, \mathcal{X}=\left[X_1, \ldots, X_N\right]^{\prime}, \mathcal{D}=\left[D_1, \ldots, D_N\right]^{\prime},$ $\mathcal{V}=\left[V_1, \ldots, V_N\right]^{\prime},\mathcal{U}=\left[U_1, \ldots, U_N\right]^{\prime}$, $\mathcal{R}_{Y_\theta}=\left[r_{Y_\theta}(X_1),\dots,r_{Y_\theta}(X_N)\right]^{\prime}$, $\mathcal{R}_{D}=\left[r_{D}(X_1),\dots,r_{D}(X_N)\right]^{\prime}$, and $f = \left[f(X_1),\dots,f(X_N)\right]^{\prime}$ and $g = \left[g(X_1),\dots,g(X_N)\right]^{\prime}$ with
    \begin{align*}
        f(X) = X'\Lambda_0 + r_{Y_\theta}(X),\qquad
        g(X) = X'\beta_0 + r_D(X).
    \end{align*}
    Define for a non-empty set of coordinate indices $J\subseteq \mathfrak{P}$ that $X_{[N] J} = \{X_{i j},i\in \{1,\dots,N\}, j\in J\}$, $P_J = X_{[N] J}(X_{[N] J}'X_{[N] J})^{-}X_{[N] J}'$, and $M_J = I_N - P_J$, where $I_N$ is an $N$-dimensional identity matrix. Define the indices of chosen coordinates from Algorithm \ref{algorithm:dml_oga_hdaic:fullsample} using $(\mathcal{X},\mathcal{D})$ as $\widetilde{J}_{\hat{m}}^D$, $(\mathcal{X},\mathcal{Y})$ as $\widetilde{J}_{\hat{m}}^Y$, and let $\widetilde{J}=\widetilde{J}_{\hat{m}}^D \cup \widetilde{J}_{\hat{m}}^Y$.
    %This section follows the setting from the previous section, so we focus on the polynomial decay case.

    First note that
    \begin{align*}
        \widetilde{\theta} =& \paren{\frac{1}{N} \mathcal{D}' M_{\widetilde{J}} \mathcal{D}}^{-1} \paren{\frac{1}{N} \mathcal{D}' M_{\widetilde{J}} Y}, \text{ and}\\
        \sqrt{N}\paren{\widetilde{\theta}-\theta_0} =& \underbrace{\paren{\frac{1}{N} \mathcal{D}' M_{\widetilde{J}} \mathcal{D}}^{-1}}_{=\textbf{A}^{-1}} \underbrace{\paren{\frac{1}{\sqrt{N}} \mathcal{D}' M_{\widetilde{J}} \paren{f + \mathcal{U}}}}_{=\textbf{B}}.
    \end{align*}
Thus, if we show 
    \begin{align}
        \textbf{A} = \frac{1}{N}\mathcal{V}'\mathcal{V} + o_p(1)\quad \text{and}\quad \textbf{B} = \frac{1}{\sqrt{N}}\mathcal{V}'\mathcal{U} + o_p(1),\label{fullsample:AandB}
    \end{align}
    then an application of CLT yields the desired conclusion. In \textit{Step 1} we derive component-wise bounds that will be used in the following steps. We show \eqref{fullsample:AandB} in \textit{Step 2} and in \textit{Step 3} we conclude.

    \textit{Step 1.} (Component-wise bounds)
    Note we have $R_{2,p}=\max_{1\le j\le p}\abs{1/N\sumi X_{ij}V_i}\lesssim_P (\log p/N)^{1/2}$ from \eqref{A1}.
    First we derive bounds for $\norm{\frac{1}{N}\mathcal{X}'\mathcal{U}}_{\infty}$:
    \begin{align}
        \norm{\frac{1}{N}\mathcal{X}'\mathcal{V}}_{\infty} \le R_{2,p} \lesssim_P (\log p/N)^{1/2}, \label{fullsample:BD1}
    \end{align}
    and similarly $\norm{\mathcal{X}'\mathcal{U}/N}_{\infty} \lesssim_P (\log p/N)^{1/2}$.

    Note $\widetilde{\beta}(J) = (X_{[N] J}'X_{[N] J})^{-1}X_{[N] J}'\mathcal{D}$. From the convergence rates from \cite{ing2020}, we have
    \begin{align}
        \begin{aligned}
            \norm{\widetilde{\beta}(\widetilde{J})-\beta_0} &\lesssim_P \norm{\widetilde{\beta}(\hat{J}_{\hat{m}}^D)-\beta_0}\\
            % &\le \lambda_{\min} (\Gamma) \norm{\widetilde{\beta}(\hat{J}_{\hat{m}}^D)-\beta_0}\\
            % &\le \norm{\frac{1}{\sqrt{N}}\mathcal{X}\paren{\widetilde{\beta}(\hat{J}_{\hat{m}}^D)-\beta_0}} \\
            &\le  \paren{\lambda_1^{-1} E_{W_1^N} [(D^r(X)-\hat{D}_m(X))^2]}^{1/2}\\
            &\lesssim_P (\log p/N)^{(2\alpha-1)/4\alpha}, \label{fullsample:BD:beta}
        \end{aligned}
    \end{align}
    where the first inequality comes from $\hat{J}_{\hat{m}}^D \subseteq \widetilde{J}$, the second from Assumption \ref{a:2} (a) and Equation (3.16) of \cite{ing2020}, and the last from the results in Theorem 3.1. of \cite{ing2020}.

    In what followings, we establish a bound for $\|\widetilde{\beta}(\widetilde{J}) - \beta_0\|_1$. Recall $\beta_0(J) = ({\beta_0}_j, j=1;\dots,p), \text{ where }{\beta_0}_j=0$ for $j\notin J$. Notice that
    \begin{align*}
        \begin{aligned}
            \norm{\widetilde{\beta}(\widetilde{J}) - \beta_0}_1 =& \norm{\widetilde{\beta}(\widetilde{J}) - \beta_0(\widetilde{J})}_1 + \norm{ \beta_0(\widetilde{J}^c)}_1
             = \norm{\widetilde{\beta}(\widetilde{J}) - \beta_0(\widetilde{J})}_1 + \norm{\beta_0 - \beta_0(\widetilde{J})}_1.
            % \\
            % \le& \sqrt{\abs{\widetilde{J}}}\norm{\widetilde{\beta}(\widetilde{J}) - \beta_0(\widetilde{J})} + \norm{\beta_0 - \beta_0(\widetilde{J})}_1\\
            % \le& \sqrt{\abs{\widetilde{J}}}\norm{\widetilde{\beta}(\widetilde{J}) - \beta_0(\widetilde{J})} + C\sum_{j\notin \widetilde{J}}\abs{{\beta_0}_j}\\
            % \le& \sqrt{\abs{\widetilde{J}}}\norm{\widetilde{\beta}(\widetilde{J}) - \beta_0(\widetilde{J})} + C C_{\alpha}\sum_{j\notin \widetilde{J}}\paren{{\beta_0}^2_j}^{(\alpha-1)/(2\alpha-1)}
        \end{aligned}
    \end{align*}
    The first term on RHS is bounded by
    \begin{align*}
        \norm{\widetilde{\beta}(\widetilde{J}) - \beta_0(\widetilde{J})}_1 
        % \lesssim_P& \norm{\widetilde{\beta}(\hat{J}_{\hat{m}}^D)-\beta_0}_1\\
        &\le \sqrt{\abs{\widetilde{J}}}\norm{\widetilde{\beta}(\widetilde{J}) - \beta_0(\widetilde{J})}\\
        &\le \sqrt{\hat{m}^D+\hat{m}^Y} \norm{\widetilde{\beta}(\hat{J}_{\hat{m}}^D) - \beta_0(\hat{J}_{\hat{m}}^D)}\\
        &\lesssim_P (M_N^*)^{1/2} (\log p/N)^{(2\alpha-1)/4\alpha}\\
        &\lesssim_P (\log p/N)^{-1/4\alpha}(\log p/N)^{(2\alpha-1)/4\alpha}\\
        &\lesssim_P (\log p/N)^{(\alpha-1)/2\alpha},
    \end{align*}
    where the second inequality comes from $\widetilde{J}=\widehat{J}_{\hat{m}}^D \cup \widehat{J}_{\hat{m}}^Y$, and the third from \eqref{ing:3.45}. The fourth comes from the definition of $M_N^*$ defined in \eqref{Mnstar}, and the last follows. On the other hand, the second term on RHS can be controlled by
    \begin{align*}
        \norm{\beta_0 - \beta_0(\widetilde{J})}_1 
        &\le C\sum_{j\notin \widetilde{J}}\abs{{\beta_0}_j}\\
        &\le C C_{\alpha} \paren{\sum_{j\notin \widetilde{J}}{\beta_0}^2_j}^{(\alpha-1)/(2\alpha-1)}\\
        &\le C C_{\alpha} \lambda_1^{(-\alpha+1)/(2\alpha-1)} \paren{E[V(\widetilde{J})^2]}^{(\alpha-1)/(2\alpha-1)}\\
        &\le C C_{\alpha} \lambda_1^{(-\alpha+1)/(2\alpha-1)} G_1 |\widetilde{J}|^{-2\alpha+1}\\
        &\le C C_{\alpha} \lambda_1^{(-\alpha+1)/(2\alpha-1)} G_1 (2M_n^*)^{-2\alpha+1}\\
        &\le C C_{\alpha} \lambda_1^{(-\alpha+1)/(2\alpha-1)} G_1 (2)^{-2\alpha+1} (\log p/N)^{(2\alpha - 1)/2\alpha}\\
        &\lesssim_P (\log p/N)^{(2\alpha - 1)/2\alpha}
    \end{align*}
    The first inequality comes from Assumption \ref{a:2} \ref{IngA5}, where it holds for all $J\subseteq \mathfrak{P}$ such that $\abs{J}\le C(N/\log p)^{1/2}$, as shown in \cite{ing2020} Equation (2.16) and the following equation. The second inequality comes from Assumption \ref{a:3:fullsample}. The third comes from \eqref{ing:A.4}, which holds under Assumption \ref{a:2} \ref{additionalA}.
    % , since it implies . 
    The fourth comes from 
    % Lemma A.1. of \cite{ing2020}, as also shown in 
    \eqref{ing:A.1.:poly} in the previous section's proof. The fifth comes from $|\widetilde{J}|\le |\hat{J}_{\hat{m}}^D| + |\hat{J}_{\hat{m}}^Y|$, where $|\hat{J}_{\hat{m}}|\le M_N^*$. The sixth comes from the definition of $M_N^*$ given in \eqref{Mnstar}, and the last follows.
    By combining the bounds, we conclude that
    \begin{align}
        \begin{aligned}\label{fullsample:BD2:L1norm}
            \norm{\widetilde{\beta}(\widetilde{J}) - \beta_0}_1 \lesssim_P (\log p/N)^{(\alpha-1)/4\alpha} + (\log p/N)^{(2\alpha - 1)/2\alpha}\lesssim_P (\log p/N)^{(\alpha-1)/4\alpha}.
        \end{aligned}
    \end{align}
    
    It also holds that
    \begin{align}
        \begin{aligned}
            \norm{\frac{1}{\sqrt{N}}M_{\widetilde{J}}g} &\le \norm{\frac{1}{\sqrt{N}}M_{\hat{J}_{\hat{m}}^D}g}\\
            &\le \norm{\frac{1}{\sqrt{N}}\paren{\mathcal{X}\widetilde{\beta}(\hat{J}_{\hat{m}}^D)-g}}\\
            &\le \norm{\frac{1}{\sqrt{N}}\mathcal{X}\paren{\widetilde{\beta}(\hat{J}_{\hat{m}}^D)-\beta_0}} + \norm{\frac{1}{\sqrt{N}}\mathcal{R}_D}\\
            &\lesssim_P (\log p/N)^{(2\alpha-1)/4\alpha} + \sqrt{\frac{1}{N}\sumi r^2_D(X_i)} \\
            &\lesssim_P (\log p/N)^{(2\alpha-1)/4\alpha} + \sqrt{\log p/N}\\
            &\lesssim_P (\log p/N)^{(2\alpha-1)/4\alpha}, \label{fullsample:BD3:g}
        \end{aligned}
    \end{align}
    where the fourth comes from the convergence rates from \eqref{ing:3.4}, and the fifth from Assumption \ref{a:approx} \ref{rsq_bdd}, and the remainder follows.

    Similarly, 
    note the convergence rate for $\|\widetilde{\gamma}(\hat{J}_{\hat{m}}^Y)-\gamma_0\|$ is of the same order as in \eqref{fullsample:BD:beta}, hence we have
    \begin{align*}
        \begin{aligned}
            \norm{\frac{1}{\sqrt{N}}M_{\widetilde{J}}\paren{\theta_0 g + f}} &\le \norm{\frac{1}{\sqrt{N}}\paren{\mathcal{X}\widetilde{\gamma}(\widetilde{J})-\paren{\theta_0 g + f}}}\\
            &\le \norm{\frac{1}{\sqrt{N}}\paren{\mathcal{X}\widetilde{\gamma}(\hat{J}_{\hat{m}}^Y)-\paren{\theta_0 g + f}}}\\
            &\le \norm{\frac{1}{\sqrt{N}}\mathcal{X}\paren{\widetilde{\gamma}(\hat{J}_{\hat{m}}^Y)-\gamma_0}} + \norm{\frac{1}{\sqrt{N}}\mathcal{R}_Y}\\
            &\lesssim_P (\log p/N)^{(2\alpha-1)/4\alpha},
        \end{aligned}
    \end{align*}
    where the second inequality comes from $\hat{J}_{\hat{m}}^Y\subseteq \widetilde{J}$, the third from triangle inequality, and the last follows Assumption \ref{a:approx} \ref{rsq_bdd} and the convergence rate of $\|\widetilde{\gamma}(\hat{J}_{\hat{m}}^Y)-\gamma_0\|$. 
    Using triangle inequality,
    \begin{align*}
        \Bigl| \norm{M_{\widetilde{J}}\theta_0g } - \norm{M_{\widetilde{J}}f}\Bigr| \le \norm{M_{\widetilde{J}}\paren{\theta_0 g + f}},
    \end{align*}
    where $\|M_{\widetilde{J}}\theta_0g/\sqrt{N}\| = \|\theta_0\| \|\mathcal{X} (\widetilde{\beta}(\widetilde{J})-\beta_0) /\sqrt{N}\| \lesssim_P (\log p/N)^{(2\alpha-1)/4\alpha}$ by the assumption on bounded $\norm{\theta_0}$ in Assumption \ref{a:3:fullsample}. Therefore, $\norm{M_{\widetilde{J}}f}$ is bounded by the same bound and
    \begin{align}
        \begin{aligned}
            \norm{\widetilde{\Lambda}(\widetilde{J}) -\Lambda_0} \lesssim_P \norm{\frac{1}{\sqrt{N}}\mathcal{X}\paren{\widetilde{\Lambda}(\widetilde{J}) -\Lambda_0}} \lesssim_P \norm{\frac{1}{\sqrt{N}}M_{\widetilde{J}}f} \lesssim_P (\log p/N)^{(2\alpha-1)/4\alpha} \label{fullsample:BD:Lambda}
        \end{aligned}
    \end{align}
    by a similar argument as in \eqref{fullsample:BD3:g}.

    Let $\widetilde{\beta}_V({J})= (\mathcal{X}(J)'\mathcal{X}(J))^{-1}\mathcal{X}(J)'\mathcal{V}$.
    \begin{align}
        \begin{aligned}
            \norm{\widetilde{\beta}_V (\widetilde{J})}_1
            &\le \sqrt{|\widetilde{J}|} \norm{\widetilde{\beta}_V (\widetilde{J})}_2\\
            &\le \sqrt{|\widetilde{J}|} \norm{\widehat{\Gamma}^{-1}(\widetilde{J})}_2 \norm{\frac{1}{N} \mathcal{X}(\widetilde{J})'\mathcal{V}}_2 \\
            &\le |\widetilde{J}| \norm{\widehat{\Gamma}^{-1}(\widetilde{J})}_2 \norm{\frac{1}{N} \mathcal{X}(\widetilde{J})'\mathcal{V}}_{\infty} \\
            % &\lesssim_P \max_{1\le \abs{J}\le \bar{C}(N/\log p)^{1/2}} \abs{\Gamma^{-1}(J)E[X_{iJ} V_i]}\\
            % \le& \norm{\hat{\Gamma}(\widetilde{J})^{-1}}\norm{\frac{1}{N} \mathcal{X}'\mathcal{V}}\\
            % \le& \overline{B} R_{2,p} \lesssim_P (\log p/N)^{1/2},
            &\lesssim_P (\log p/N)^{(\alpha-1)/2\alpha}.
            \label{fullsample:BD4}
        \end{aligned}
    \end{align}
    % where the second inequality comes from the Markov inequality, and the last comes from Assumption \ref{a:2:fullsample} \ref{XXXV_bdd}. 
    The last inequality comes from $\lim_{N\to\infty} P(\|\hat{\Gamma}^{-1}(\hat{J}_{K_N})\|\le \overline{B})=1$ as shown in Section S1 of the Supplementary Material of \cite{ing2020}, which holds under current setting and $\bar{B}$ is some large constant defined in Theorem 2.1. of \cite{ing2020}.

    The last component is
    \begin{align}
        \begin{aligned}
            \abs{\frac{1}{\sqrt{N}} \mathcal{R}_D'\mathcal{U}} 
            % = \abs{\frac{1}{\sqrt{N}}\sumi r_D(X_i)U_i} 
            \lesssim_P \sqrt{E\Bigl[\frac{1}{N} \sumi r_D^2(X_i)\Bigr]}
            % \lesssim_P \sqrt{E [r_D^4(X)]} 
            \lesssim_P (\log p/N)^{1/2} \label{fullsample:BD5},
        \end{aligned}
    \end{align}
    where the first inequality comes from Chebyshev and the last from \ref{a:approx} \ref{rsq_bdd}. 
    The same logic applies to $\abs{\mathcal{R}_{Y_{\theta}}'\mathcal{V}/\sqrt{N}}$.

    \textit{Step 2.} (Bounding \textbf{A} and \textbf{B})
    Decompose the two objects in \eqref{fullsample:AandB} into
    \begin{align*}
        \textbf{A} =& \frac{1}{N} \mathcal{V}^{\prime}\mathcal{V} + \underbrace{\frac{1}{N} f'M_{\widetilde{J}}f}_{(a)} + \underbrace{\frac{2}{N}f'M_{\widetilde{J}}\mathcal{V}}_{(b)} - \underbrace{\frac{1}{N}\mathcal{V}'P_{\widetilde{J}}\mathcal{V}}_{(c)},\\
        \textbf{B} =& \frac{1}{\sqrt{N}}\mathcal{V}^{\prime} \mathcal{U}+\underbrace{\frac{1}{\sqrt{N}} g^{\prime} M_{\widetilde{J}} f }_{(d)}+\underbrace{ \frac{1}{\sqrt{N}} g^{\prime} M_{\widetilde{J}} \mathcal{U} }_{(e)}+\underbrace{\frac{1}{\sqrt{N}} \mathcal{V}^{\prime} M_{\widetilde{J}} f}_{(f)}-\underbrace{ \frac{1}{\sqrt{N}} \mathcal{V}^{\prime} P_{\widetilde{J}} \mathcal{U}}_{(g)},
    \end{align*}
    where the components (a)-(g) can be further controlled by
    \begin{align}\label{fullsample:A}
        \begin{aligned}
            \abs{(a)} &\le \norm{\frac{1}{\sqrt{N}}M_{\widetilde{J}}f}^2 \lesssim_P (\log p/N)^{(2\alpha-1)/2\alpha},\\
            \abs{(b)} &\le 2\abs{\frac{1}{N}\mathcal{R}_{Y_{\theta}}'\mathcal{V}} + 2\abs{\paren{\widetilde{\Lambda}(\widetilde{J})-\Lambda_0}^{\prime}\frac{1}{N}\mathcal{X}^{\prime} \mathcal{V}} \\
            &\le 2\abs{\frac{1}{N}\mathcal{R}_{Y_{\theta}}'\mathcal{V}} + 2\norm{\widetilde{\Lambda}(\widetilde{J})-\Lambda_0}_1 \norm{\frac{1}{N}\mathcal{X}'\mathcal{V}}_{\infty}\\
            &\lesssim_P \sqrt{N}^{-1} (\log p/N)^{1/2} + (\log p/N)^{(\alpha-1)/4\alpha}(\log p/N)^{1/2},\\
            \abs{(c)} &\le \abs{\widetilde{\beta}_V(\widetilde{J})' \frac{1}{N}\mathcal{X}'\mathcal{V}} \le \norm{\widetilde{\beta}_V(\widetilde{J})}_1 \norm{\frac{1}{N}\mathcal{X}'\mathcal{V}}_{\infty} \\
            &\lesssim_P (\log p/N)^{(\alpha-1)/2\alpha} (\log p/N)^{1/2},
        \end{aligned}
    \end{align}  
     and
    \begin{align}\label{fullsample:B}
        \begin{aligned}
            \abs{(d)} &\le \sqrt{N}\norm{\frac{1}{\sqrt{N}} M_{\widetilde{J}} f }\norm{\frac{1}{\sqrt{N}} M_{\widetilde{J}} g } 
            \lesssim_P \sqrt{N}(\log p/N)^{(2\alpha-1)/2\alpha},\\
            \abs{(e)} &\le \abs{\frac{1}{\sqrt{N}}\mathcal{R}_{D}'\mathcal{U}} + \abs{\paren{\widetilde{\beta}(\widetilde{J})-\beta_0}^{\prime}\frac{1}{\sqrt{N}}\mathcal{X}^{\prime} \mathcal{U}} \\
            &\le \abs{\frac{1}{\sqrt{N}}\mathcal{R}_{D}'\mathcal{U}} + \norm{\widetilde{\beta}(\widetilde{J})-\beta_0}_1 \norm{\frac{1}{\sqrt{N}}\mathcal{X}^{\prime} \mathcal{U}}_{\infty}\\
            &\lesssim_P (\log p/N)^{1/2} + \sqrt{N} (\log p/N)^{(\alpha-1)/4\alpha}(\log p/N)^{1/2},\\
            \abs{(f)} &\le \abs{\frac{1}{\sqrt{N}}\mathcal{R}_{Y_{\theta}}'\mathcal{V}} + \abs{\paren{\widetilde{\Lambda}(\widetilde{J})-\Lambda_0}^{\prime}\frac{1}{\sqrt{N}}\mathcal{X}^{\prime} \mathcal{V}} \\
            &\le \abs{\frac{1}{\sqrt{N}}\mathcal{R}_{Y_{\theta}}'\mathcal{V}} + \norm{\widetilde{\Lambda}(\widetilde{J})-\Lambda_0}_1 \norm{\frac{1}{\sqrt{N}}\mathcal{X}^{\prime} \mathcal{V}}_{\infty} \\
            &\lesssim_P (\log p/N)^{1/2} + \sqrt{N} (\log p/N)^{(\alpha-1)/4\alpha}(\log p/N)^{1/2},\\
            \abs{(g)} &\le \abs{\widetilde{\beta}_V(\widetilde{J})' \frac{1}{\sqrt{N}}\mathcal{X}'\mathcal{U}} \le \norm{\widetilde{\beta}_V(\widetilde{J})}_1 \norm{\frac{1}{\sqrt{N}}\mathcal{X}'\mathcal{U}}_{\infty} \\
            &\lesssim_P (\log p/N)^{(\alpha-1)/2\alpha} \sqrt{N}(\log p/N)^{1/2}.
        \end{aligned}
    \end{align}     
     Now, we show that each part in \eqref{fullsample:A} and \eqref{fullsample:B} is $o_p(1)$. Note that $\abs{(a)}$, $\abs{(b)}$, and $\abs{(c)}$ are $o(1)$ if $\abs{(d)}$, $\abs{(e)}$, and $\abs{(g)}$ are all $o(1)$. 
    Since
    \begin{align*}
        \abs{(d)} \lesssim_P& \sqrt{N} (\log p/N)=o(1),\\
        \abs{(e)} \lesssim_P& \sqrt{N} (\log p/N)^{(\alpha-1)/4\alpha}(\log p/N)^{1/2}=o(1),\\
        \abs{(g)} \lesssim_P& (\log p/N)^{(\alpha-1)/2\alpha}(\log p)^{1/2} = o(1),
    \end{align*}
    and as $(e)$ and $(f)$ both share the same upper bound, $\log p = o(N^{(\alpha-1)/(3\alpha-1)})$ is a sufficient condition for all the components in \eqref{fullsample:A} and \eqref{fullsample:B} to be $o_p(1)$, given any $\alpha>1$ for Assumption \ref{a:3:fullsample} \ref{poly}.
    Therefore we achieve \eqref{fullsample:AandB}.

    \textit{Step 3}. (CLT)
    From Assumption \ref{a:1:approx} on the error terms, \eqref{fullsample:AandB} implies
    \begin{align*}
        \begin{aligned}
            \sqrt{N}\paren{\widetilde{\theta}-\theta_0} =& \paren{E[V^2]^{-1} + o_p(1)} \paren{\frac{1}{\sqrt{N}}\mathcal{V}'\mathcal{U} + o_p(1)}\\
            \xrightarrow{d}& E[V^2]^{-1} N(0,E[V^2U^2])
        \end{aligned}
    \end{align*}
    following Lindeberg–L\'evy CLT, which concludes the proof.    
\end{proof}
%%%%%%%%%%%%%%%%%%%%%%%%%%%%%%%%%%%%%%%%%%%%%%%%%%%%%%%%%%%%%%%%%%%%%

%%%%%%%%%%%%%%%%%%%%%%%%%%%%%%%%%%%%%%%%%%%%%%%%%%%%%%%%%%%%%%%%%%%%%
\subsection{Proof of Theorem \ref{theorem:iv:regression}}\label{sec:theorem:iv:regression}
%%%%%%%%%%%%%%%%%%%%%%%%%%%%%%%%%%%%%%%%%%%%%%%%%%%%%%%%%%%%%%%%%%%%%
\begin{proof}
% {\color{blue} [JOOYOUNG: PRESENTED A PROOF OF THEOREM \ref{theorem:iv:regression} FOR THE IV MODEL, SIMILARLY TO PF1]}
In proof of Theorem \ref{theorem:regression} we have shown that the convergence rates from Theorem 3.1. of Ing hold under our assumptions. Hence it is enough to show that Assumption 4.2. in CCDDHNR holds. Note again that Assumption \ref{a:1:iv} (b)--(e) corresponds to Assumption 4.2. (a)--(d) in \cite{ccddhnr}.

We shall thus verify condition (e) in \cite{ccddhnr} is implied by the convergence rates \eqref{poly_convgrate} and \eqref{exp_convgrate}. 
Recall that by \eqref{poly_convgrate} and an argument similar to \eqref{A5}--\eqref{A8},
\begin{align} \label{xi:poly_convgrate}
    \norm{\widehat{\xi} - \xi_0}^2 \lesssim \paren{\frac{\log p}{N}}^{1-1/2\alpha}
\end{align}
holds with probability at least $1-\Delta_N$ for some $\Delta_N=o(1)$ and $\xi = \gamma,\zeta,$ and $\beta$.

Now we will show that with $P-$probability no less than $1-\Delta_N$, $\norm{\widehat{\eta} - \eta_0}_{P,q} \le C$, $\norm{\widehat{\eta} - \eta_0}_{P,2} \le \delta_N$, and $\snorm{\widehat{\beta}-\beta_0}_{P,2}\times (\snorm{\widehat{\gamma} - \gamma_0}_{P,2} + \snorm{\widehat{\zeta} - \zeta_0}_{P,2})$ holds that for $\Delta_N,$ $\delta_N$ such that both are sequences of strictly positive constants converging to zero, where $\eta = (\beta, \gamma, \zeta)$. From \eqref{xi:poly_convgrate}, we have
\begin{align*}
    \norm{\widehat{\eta} - \eta_0 }_{P,q} =& \snorm{\widehat{\beta} - \beta_0}_{P,q} \vee \snorm{\hat{\gamma} - \gamma_0}_{P,q} \vee \snorm{\widehat{\zeta} - \zeta_0}_{P,q} 
    = \paren{E\left[\norm{X'(\hat{\beta} -\beta_0)}^q\mid A_N\right]}^{1/q}\\
    \le& \paren{E[\paren{(\hat{\beta}-\beta_0)'E[XX'](\hat{\beta}-\beta_0)}^{q/2}\mid A_N]}^{1/q}
    = \paren{\paren{\lambda_{\max}(E[XX']) \|\hat{\beta}-\beta_0\|^2}^{q/2}}^{1/q}\\
    =& \paren{\lambda_{\max}(E[XX']) \|\hat{\beta}-\beta_0\|^2}^{1/2},
\end{align*}
where it is bounded by Assumption \ref{a:2:iv}\ref{additionalA}, (\ref{poly_convgrate}), and (\ref{exp_convgrate}).

Thus, $\norm{\hat{\eta} - \eta_0}_{P,q}\le C$ holds with probability at least $1-\Delta_N$ for some $\Delta_N = o(1)$.

Since we use the identical procedure, namely the OGA and HDAIC, in estimating $\beta_0$, $\gamma_0$, and $\eta_0$, the convergence rates apply to all the nuisance parameters. 
Like we did in the proof of Theorem 1, We consider two cases where all the parameters follow the polynomial decay case or the exponential decay case.
\begin{enumerate}[label={Case \arabic*.}]
    \item For the polynomial decay case, let $\delta_N = (\log p)^{1-1/2\alpha}N^{1/2\alpha - 1/2}$. Then $\delta_N = o(1)$ since $\alpha$ is assumed to be strictly larger than $1$ in Assumption \ref{a:3:iv} (a). We have
\begin{align*}
    \snorm{\hat{\eta}-\eta_0}_{P,2} =& \snorm{\widehat{\beta} - \beta_0}_{P,q} \vee \snorm{\hat{\gamma} - \gamma_0}_{P,q} \vee \snorm{\widehat{\zeta} - \zeta_0}_{P,q} \lesssim \left(\frac{\log p}{N}\right)^{1/2-1/4\alpha} \le \delta_N,
\end{align*}
where the last inequality holds since $(\log p/N)^{1/4\alpha}\le \log p$ holds with $\alpha>1$.
Also,
\begin{align*}
    \snorm{\hat{\beta}-\beta_0}_{P,2}\times \paren{\snorm{\hat{\zeta}-\zeta_0}_{P,2}+\snorm{\hat{\gamma}-\gamma_0}_{P,2}} \lesssim \paren{\frac{\log p}{N}}^{1-1/2\alpha} = \delta_N N^{-1/2}
\end{align*}
holds. 
    \item For the exponential decay case, let $\delta_N = N^{-1/2}\log p\log N$. By the assumption $\log p = o(N^{1/4})$, $\delta_N = o(1)$. 
We have
\begin{align*}
    \snorm{\hat{\eta}-\eta_0}_{P,2} \lesssim \sqrt{\frac{\log p \log N}{N}} < \delta_N
\end{align*}
and
\begin{align*}
    \snorm{\hat{\beta}-\beta_0}_{P,2}\times \paren{\snorm{\hat{\zeta}-\zeta_0}_{P,2}+\snorm{\hat{\gamma}-\gamma_0}_{P,2}} \lesssim \frac{\log p \log N}{N} = \delta_N N^{-1/2}.
\end{align*}
\end{enumerate}
A similar argument applies to the cross cases.

We have shown that Assumption 4.2 in \cite{ccddhnr} holds for both sparsity assumptions. 
Therefore, applying Theorem 4.2. in \cite{ccddhnr}, we get the desired results.

\end{proof}
%%%%%%%%%%%%%%%%%%%%%%%%%%%%%%%%%%%%%%%%%%%%%%%%%%%%%%%%%%%%%%%%%%%%%

%%%%%%%%%%%%%%%%%%%%%%%%%%%%%%%%%%%%%%%%%%%%%%%%%%%%%%%%%%%%%%%%%%%%%
\section{Omitted Details}
%%%%%%%%%%%%%%%%%%%%%%%%%%%%%%%%%%%%%%%%%%%%%%%%%%%%%%%%%%%%%%%%%%%%%

This appendix section presents details that are omitted from the main text.

%%%%%%%%%%%%%%%%%%%%%%%%%%%%%%%%%%%%%%%%%%%%%%%%%%%%%%%%%%%%%%%%%%%%%
\subsection{Details of the Method}\label{sec:regression:details}
%%%%%%%%%%%%%%%%%%%%%%%%%%%%%%%%%%%%%%%%%%%%%%%%%%%%%%%%%%%%%%%%%%%%%
This section provide details about two tuning parameters, $C^*$ and $M_n^*$, used in Algorithm \ref{algorithm:dml_oga_hdaic}.
Let $c_1,c_2$ be sufficiently large positive constants that satisfy
\begin{align*}
    P\left(\max_{1\le j \le p}\abs{\frac{1}{N}\sum_{i=1}^N X_{ij}\epsilon_i}\ge c_1 \sqrt{\frac{\log p}{N}} \right)=o(1)
\end{align*}
and
\begin{align*}
    P\left(\max_{1\le j,\ell \le p}\abs{\frac{1}{N}\sum_{i=1}^N X_{ij}X_{i\ell}-E[X_{1j}X_{1\ell}]}\ge c_2 \sqrt{\frac{\log p}{N}} \right)=o(1),
\end{align*}
for each of $\epsilon=V$ and $\epsilon=Y-X'\gamma_0$.
We require $c_1,c_2$ to satisfy the above restrictions with $\epsilon=V$ for the estimation of the part $\beta_0$ of the nuisance parameters and with $\epsilon = Y - X'\gamma_0$ for the estimation of the part $\gamma_0$ of the nuisance parameters. 

Define 
$\Gamma(J) = E[X_{iJ}X_{iJ}']$,
\begin{align}
    M^*_N=\min\Biggl\{\paren{\frac{N}{\log p}}^{1/2\alpha}, \bar{\delta}\left(\frac{N}{\log p} \right)^{1/2}\Biggr\}, \label{Mnstar}
    % ,\:K_N = \bar{\delta}\paren{\frac{N}{\log p}}^{1/2}
\end{align}
 and \begin{align*}
    \bar{\tau} =& \sup\left\{\tau:\tau>0,\, \limsup_{N\to\infty} \frac{\tau c_2}{\min_{\abs{J}\le\tau(N/\log p)^{1/2}} \lambda_{\min}(\Gamma(J))}\le 1 \right\},
\end{align*} where $0<\bar{\delta}<\min\{\bar{\tau},\bar{C}\}$ with an arbitrary strictly positive constant $\bar{C}$ restricted in Assumption \ref{a:2} \ref{IngA5}.
In our simulation and real data analysis, we set $\bar{\delta} = 5$ following \cite{ing2020}.
Let $\bar{B}$ be a positive constant satisfying 
\begin{align*}
    \bar{B} > \frac{1}{\liminf_{N\to\infty} \min_{\abs{J}\le \bar{\delta}(N/\log p)^{1/2}}\lambda_{\min} (\Gamma(J))- c_2\bar{\delta}}.
\end{align*}
Define a sufficiently large positive constant $C^*$ satisfying
\begin{align}
    C^* > \frac{2\bar{B}(c_1^2 + c_2^2)}{\sigma_\epsilon^2}, \label{Cstar}
\end{align} 
for each of $\sigma_{\epsilon}^2 = E[V^2]$ and $\sigma_{\epsilon}^2 = E[(Y-X'\gamma_0)^2]$.
In our simulation and real data analysis, we set $C^*=2$ following \cite{ing2020}.
We require $C^*$ to satisfy this restriction with $\sigma_{\epsilon}^2 = E[V^2]$ for estimation of the part $\beta_0$ of the nuisance parameters and with $\sigma_{\epsilon}^2 = E[(Y-X'\gamma_0)^2]$ for estimation of the part $\gamma_0$ of the nuisance parameters.

%%%%%%%%%%%%%%%%%%%%%%%%%%%%%%%%%%%%%%%%%%%%%%%%%%%%%%%%%%%%%%%%%%%%%
\subsection{Finite Sample Adjustment}\label{eq:finite}
%%%%%%%%%%%%%%%%%%%%%%%%%%%%%%%%%%%%%%%%%%%%%%%%%%%%%%%%%%%%%%%%%%%%%

The DML estimator is random even conditionally on data, because of the random splitting of the sample for cross fitting.
To mitigate the effects of this randomness of the DML, \citet[][Sec. 3.4]{ccddhnr} propose procedures of finite-sample adjustments.
In this section, we present one of these procedures for completeness.

Suppose that we repeat the DML estimation $S$ times to obtain
$
\{\check{\theta}^s\}_{s=1}^S.
$
A robust estimator that incorporates the impact of sample splitting is defined by
$$
\check{\theta}^{Med} = \text{Median}\left\{\check{\theta}^s\right\}_{s=1}^S.
$$
\citet[cf.][Definition 3.3]{ccddhnr}.
Its associated variance estimator is given by
$$
\widehat\Omega^{Med} = \text{Median}\left\{ \widehat\Omega^s + (\check{\theta}^s - \check{\theta}^{Med})(\check{\theta}^s - \check{\theta}^{Med})' \right\}_{s=1}^S.
$$
\citet[cf.][Equation 3.14]{ccddhnr}.
We use these estimators with $S=20$ in reporting the estimation results in Section \ref{eq:application}.

%%%%%%%%%%%%%%%%%%%%%%%%%%%%%%%%%%%%%%%%%%%%%%%%%%%%%%%%%%%%%%%%%%%%%
\subsection{The Hermite Basis}\label{eq:hermite}
%%%%%%%%%%%%%%%%%%%%%%%%%%%%%%%%%%%%%%%%%%%%%%%%%%%%%%%%%%%%%%%%%%%%%

In Section \ref{eq:application}, we use the Hermite basis $(\psi_0,\ldots,\psi_9)$.
The $k$-th basis element $\psi_j$ of the Hermite basis is defined by
$$
\psi_k(x) = e^{-x^2/2} H_k(x),
$$
where $H_k$ is the Hermite polynomial defined by
$$
H_k(x) = (-1)^k e^{x^2/2} \frac{d^k}{dx^k} e^{-x^2/2}
$$
for each $k \in \mathbb{Z}_+$.
The the Hermite polynomial can be also written as
$$
H_k(x) = k! \sum_{\kappa=0}^{\lfloor k/2 \rfloor} \frac{(-1)^\kappa}{\kappa!(k-2\kappa)!} \frac{x^{k-2\kappa}}{2^\kappa}
$$
in a closed form, where $\lfloor \cdot \rfloor$ denotes the floor function.

%%%%%%%%%%%%%%%%%%%%%%%%%%%%%%%%%%%%%%%%%%%%%%%%%%%%%%%%%%%%%%%%%%%%%
\section{Additional Simulations}\label{sec:additional_simulations}
%%%%%%%%%%%%%%%%%%%%%%%%%%%%%%%%%%%%%%%%%%%%%%%%%%%%%%%%%%%%%%%%%%%%%

%%%%%%%%%%%%%%%%%%%%%%%%%%%%%%%%%%%%%%%%%%%%%%%%%%%%%%%%%%%%%%%%%%%%%
\subsection{Alternative Values of the Tuning Parameters}\label{sec:simulation:tuning}
%%%%%%%%%%%%%%%%%%%%%%%%%%%%%%%%%%%%%%%%%%%%%%%%%%%%%%%%%%%%%%%%%%%%%

In Section \ref{sec:simulations} in the main text, we present simulation results using the choice $C^*=2$ of the tuning parameter following \cite{ing2020} -- see Appendix \ref{sec:regression:details} for the implementation details.
In the current appendix section, we show additional simulation results that we obtain by varying the value of $C^*$ up and down by ten percent and twenty percent, and report the sensitivity of the results to this variation.

Table \ref{tab:simulation:tuning} shows the results under $C^* = 1.6$, $1.8$ $2.2$, and $2.4$ along with the baseline value of $C^*=2.0$.
We focus on one DGP, namely the case of $\beta_{0,j} = \gamma_{0,j} = j^{-1.5}$, for simplicity.
Observe that the results are fairly insensitive to variations in the values of $C^*$.
% {\color{blue} [JOOYOUNG: VARY TUNING PARAMETERS FOR LASSO, RF, \& OGA (ON THE $j^{-1.5}$ DESIGN ONLY) DEFAULT TUNING PARAMETER $\pm$10\% OR $\pm$25\% ALONG WITH RESULTS FOR DEFAULT TUNING PARAMETERS]}

\begin{table}
    \centering
    \renewcommand{\arraystretch}{.7} 
	\scalebox{1}{
    \begin{tabular}{lllllllll}\\
        \hline \hline
        & $\beta_{0,j},\gamma_{0,j}$ & $C^*$ & $N$ & $p$ & Bias & SD & RMSE & 95\%\\
        \hline
         &$j^{-1.5}$ & 1.6 & 500 & 500 & -0.009 & 0.045 & 0.047 & 0.923\\
         &           &   & 1000 & 500 & -0.003 & 0.032 & 0.033 & 0.943\\
        \hline
        &           & 1.8 & 500 & 500 & -0.003 & 0.045 & 0.047 & 0.937\\
        &           &   & 1000 & 500 & 0.000 & 0.031 & 0.033 & 0.937\\ \hline
        &           & 2 & 500 & 500 & 0.001 & 0.045 & 0.046 & 0.936\\
        &           &(Baseline)   & 1000 & 500 & 0.002 & 0.031 & 0.033 & 0.938\\
        \hline
        &            & 2.2 & 500 & 500 & 0.004 & 0.044 & 0.046 & 0.930\\
        &            &   & 1000 & 500 & 0.003 & 0.031 & 0.032 & 0.941\\ \hline
        &            & 2.4 & 500 & 500 & 0.006 & 0.044 & 0.047 & 0.927\\
        &            &   & 1000 & 500 & 0.005 & 0.031 & 0.033 & 0.939\\ \hline 
        \hline
    \end{tabular} }
    \caption{\setlength{\baselineskip}{1mm}Monte Carlo simulation results under various values of the tuning parameter $C^*$. Displayed are Monte Carlo simulation statistics including the bias, standard deviation (SD), root mean square error (RMSE), and 95\% coverage frequency.}
    \label{tab:simulation:tuning}
\end{table}

%%%%%%%%%%%%%%%%%%%%%%%%%%%%%%%%%%%%%%%%%%%%%%%%%%%%%%%%%%%%%%%%%%%%%
\subsection{Estimation and Inference without Cross Fitting}\label{sec:simulation:no_cross_fitting}
%%%%%%%%%%%%%%%%%%%%%%%%%%%%%%%%%%%%%%%%%%%%%%%%%%%%%%%%%%%%%%%%%%%%%

In Section \ref{sec:simulations} in the main text, we present simulation results for Algorithm \ref{algorithm:dml_oga_hdaic} which uses sample splitting for cross fitting.
In the current appendix section, we present simulation results for Algorithm \ref{algorithm:dml_oga_hdaic:fullsample} based on the full sample without cross fitting. 
We use the the same DGPs as in Section \ref{sec:simulations} in the main text.

The results presented in Table \ref{tab:simulation:no_cross_fitting} for Algorithm \ref{algorithm:dml_oga_hdaic:fullsample} are almost the same as those presented in Table \ref{tab:simulation} in the main text.
In other words, Algorithm \ref{algorithm:dml_oga_hdaic:fullsample} behaves similarly to Algorithm \ref{algorithm:dml_oga_hdaic} under DGPs, while the latter slightly outperforms especially under less sparse designs.

% {\color{blue} [JOOYOUNG: OGA-HDAIC WITH NO CROSS FITTING -- FOR ALL THE DESIGNS]}

%%%%%%%%%%%%%%%%%%%%%%%%%%%%%%%%%%%%%%%%%%%%%%%%%%%%%%%%%%%%%%%%%%%%%
\begin{table}[h!]
	\centering
	\renewcommand{\arraystretch}{.7} 
	\scalebox{1}{
		\begin{tabular}{ccclcccc}\\
		\hline\hline
		  $\beta_{0,j},\gamma_{0,j}$ & $N$ & $p$ & Method of Preliminary Estimation & Bias & SD & RMSE & 95\%\\
		\hline
			Sparse & 500 & 500 & OGA+HDAIC               & -0.002 & 0.045 & 0.045 & 0.943\\
		% \cline{2-8}
			           &1000 & 500 &                 & 0.000 & 0.032 & 0.032 & 0.943\\
		\hline
			$e^{-j}$   & 500 & 500 & OGA+HDAIC                & 0.001 & 0.045 & 0.045 & 0.937\\
		% \cline{2-8}
			           &1000 & 500 &                 & 0.000 & 0.032 & 0.032 & 0.946\\
		\hline
			$j^{-2}$ & 500 & 500 & OGA+HDAIC                &-0.001 & 0.045 & 0.046 & 0.939\\
		% \cline{2-8}
			         &1000 & 500 &                 & 0.001 & 0.032 & 0.032 & 0.947\\
		\hline
		 $j^{-1.75}$& 500 & 500 & OGA+HDAIC                & 0.000 & 0.045 & 0.046 & 0.942\\
		% \cline{2-8}
			         &1000 & 500 &                 & 0.001 & 0.032 & 0.032 & 0.946\\
		\hline
		 $j^{-1.5}$& 500 & 500 & OGA+HDAIC                & 0.002 & 0.045 & 0.046 & 0.935\\
		% \cline{2-8}
			         &1000 & 500 &                 & 0.003 & 0.032 & 0.033 & 0.936\\
		\hline
		 $j^{-1.25}$& 500 & 500 & OGA+HDAIC                & 0.007 & 0.044 & 0.048 & 0.924\\
		% \cline{2-8}
			         &1000 & 500 &                 & 0.005 & 0.031 & 0.034 & 0.923\\
		\hline
			$j^{-1}$ & 500 & 500 & OGA+HDAIC                & 0.022 & 0.044 & 0.056 & 0.872\\
		% \cline{2-8}
			         &1000 & 500 &                 & 0.015 & 0.031 & 0.037 & 0.898\\
		\hline\hline
		\end{tabular}
	}
	\caption{\setlength{\baselineskip}{1mm}Monte Carlo simulation results without cross fitting. Displayed are Monte Carlo simulation statistics including the bias, standard deviation (SD), root mean square error (RMSE), and 95\% coverage frequency.}
	\label{tab:simulation:no_cross_fitting}
\end{table}
%%%%%%%%%%%%%%%%%%%%%%%%%%%%%%%%%%%%%%%%%%%%%%%%%%%%%%%%%%%%%%%%%%%%%
%%%%%%%%%%%%%%%%%%%%%%%%%%%%%%%%%%%%%%%%%%%%%%%%%%%%%%%%%%%%%%%%%%%%%
\subsection{High-Dimensional IV Regression}\label{sec:simulation:iv}
%%%%%%%%%%%%%%%%%%%%%%%%%%%%%%%%%%%%%%%%%%%%%%%%%%%%%%%%%%%%%%%%%%%%%

% {\color{blue} [JOOYOUNG: IV -- FOR ALL THE DESIGNS]}

We consider the simple setting where the data are generated by the system 
\begin{align*}
    Y =& \theta_0 (D-X'\zeta_0) + X'\gamma_0 + U,\\
    D =& \mu Z + X'\zeta_0 + E,\\
    % Z =& X'\beta_0 + V, \indent V\sim N(0,1),\\
    Z \sim& N(0,1),
\end{align*}
where
\begin{align*}
    (U,E)\sim N\left(0,\begin{pmatrix} 1 & 0.5 \\ 0.5 & 1 \end{pmatrix}\right).
\end{align*}
For our method, we use Algorithm \ref{algorithm:dml_oga_hdaic_iv} for the estimation. For DML methods, we used the R package DoubleML's partially linear IV setting. The results are shown in Table \ref{tab:simulation:iv}.

%%%%%%%%%%%%%%%%%%%%%%%%%%%%%%%%%%%%%%%%%%%%%%%%%%%%%%%%%%%%%%%%%%%%%
\begin{table}
	\centering
	\renewcommand{\arraystretch}{0.7} 
	\scalebox{1}{
		\begin{tabular}{ccclcccc}
		\hline\hline
		  $\gamma_{0,j},\zeta_{0,j}$ & $N$ & $p$ & Method of Preliminary Estimation & Bias & SD & RMSE & 95\%\\
		\hline
			Sparse & 500 & 500 & OGA+HDAIC     & -0.001 & 0.046 & 0.047 & 0.949 \\
			           &1000 & 500 &      & 0.002  & 0.032 & 0.033 & 0.951 \\
		\hline
			$e^{-j}$   & 500 & 500 & OGA+HDAIC     & -0.001 & 0.046 & 0.046 & 0.951 \\
			           &1000 & 500 &      & 0.002  & 0.032 & 0.032 & 0.947 \\
        \hline
            $j^{-2}$ & 500 & 500 & OGA+HDAIC     & -0.001 & 0.046 & 0.047 & 0.943 \\
                &1000 & 500 &      & 0.002  & 0.032 & 0.033 & 0.951 \\
		\hline
			$j^{-1.75}$ & 500 & 500 & OGA+HDAIC     & -0.001 & 0.047 & 0.047 & 0.942 \\
			         &1000 & 500 &      & 0.002  & 0.032 & 0.033 & 0.948 \\
		\hline
		 $j^{-1.5}$& 500 & 500 & OGA+HDAIC     & -0.001 & 0.047 & 0.047 & 0.942 \\
			         &1000 & 500 &      & 0.001  & 0.033 & 0.033 & 0.952\\
		\hline
		 $j^{-1.25}$& 500 & 500 & OGA+HDAIC     & -0.002 & 0.048 & 0.049 & 0.943\\
			         &1000 & 500 &      & 0.002  & 0.033 & 0.033 & 0.934\\
		\hline
			$j^{-1}$ & 500 & 500 & OGA+HDAIC     & -0.001 & 0.051 & 0.051 & 0.951\\
			         &1000 & 500 &     & 0.002  & 0.034 & 0.035 & 0.946\\
		\hline\hline
		\end{tabular}
	}
	\caption{\setlength{\baselineskip}{5.5mm}Monte Carlo simulation results for IV model. Displayed are Monte Carlo simulation statistics including the bias, standard deviation (SD), root mean square error (RMSE), and 95\% coverage frequency.}
	\label{tab:simulation:iv}
\end{table}

\section{Additional Empirical Results}\label{sec:additional_empirical}
%%%%%%%%%%%%%%%%%%%%%%%%%%%%%%%%%%%%%%%%%%%%%%%%%%%%%%%%%%%%%%%%%%%%%

In the current appendix section, we provide additional empirical estimates following Section \ref{eq:application} in the main text.
We provide the following two types of alternative estimates.
First, we vary the value of the tuning parameter $C^*$.
Second, we use Algorithm \ref{algorithm:dml_oga_hdaic:fullsample} without cross fitting, instead of Algorithm \ref{algorithm:dml_oga_hdaic}. 
Table \ref{tab:empirical:tuning_no_cross_fitting} gives the baseline results using our method in row (I).
Rows (II) and (III) show the estimates under the choices $C^*=1.8$ and $2.2$, respectively. 
Row (IV) shows the estimates using the whole sample without cross fitting.
Observe that the results are mostly robust, and our empirical findings qualitatively remain the same as those discussed in the main text.

%%%%%%%%%%%%%%%%%%%%%%%%%%%%%%%%%%%%%%%%%%%%%%%%%%%%%%%%%%%%%%%%%%%%%
%\subsection{Alternative Values of the Tuning Parameters}\label{sec:empirical:tuning}
%%%%%%%%%%%%%%%%%%%%%%%%%%%%%%%%%%%%%%%%%%%%%%%%%%%%%%%%%%%%%%%%%%%%%

% {\color{blue} [JOOYOUNG: VARIOUS TUNING PARAMETERS FOR LASSO, RF \& OGA (TUNING PARAMETERS CORRESPONDING TO SIMULATIONS)]}

\begin{table}
	\centering
	\renewcommand{\arraystretch}{0.85} 
		\begin{tabular}{clccccc}\\
		\hline\hline
		&& \multicolumn{2}{c}{Polynomial Basis} && \multicolumn{2}{c}{Hermite Basis} \\
		\cline{3-4}\cline{6-7}
		&& Unskilled & Skilled && Unskilled & Skilled \\
		&& Labor & Labor && Labor & Labor \\
		\hline
(I)	&	Double Machine Learning with & 0.279 & 0.161 && 0.165 & 0.038\\
		&	OGA+HDAIC (Baseline) &(0.017)&(0.012)&&(0.011)&(0.010)\\
        \hline
(II)&	Double Machine Learning with & 0.251 & 0.158 && 0.166 & 0.040\\
        &	OGA+HDAIC, $C^* = 1.8$ &(0.018)&(0.012)&&(0.011)&(0.010)\\
        \hline
(III)   &	Double Machine Learning with & 0.311 & 0.164 && 0.164 & 0.039 \\
        &	OGA+HDAIC, $C^* = 2.2$ &(0.017)&(0.012)&&(0.011)&(0.010)\\
        \hline
(IV)    &	Double Machine Learning with & 0.250 & 0.153 && 0.163 & 0.036\\
        &	OGA+HDAIC without Cross Fitting &(0.015)&(0.011)&&(0.010)&(0.009)\\
        \hline\hline
		\end{tabular}
	\caption{Estimates of labor elasticities in the 3-digit level industry of food products (311) in Chile based on four alternative methods.}
	\label{tab:empirical:tuning_no_cross_fitting}
\end{table}

%%%%%%%%%%%%%%%%%%%%%%%%%%%%%%%%%%%%%%%%%%%%%%%%%%%%%%%%%%%%%%%%%%%%%
%\subsection{Estimation and Inference without Cross Fitting}\label{sec:empirical:no_cross_fitting}
%%%%%%%%%%%%%%%%%%%%%%%%%%%%%%%%%%%%%%%%%%%%%%%%%%%%%%%%%%%%%%%%%%%%%

% {\color{blue} [JOOYOUNG: WITHOUT CROSS FITTING]}

%%%%%%%%%%%%%%%%%%%%%%%%%%%%%%%%%%%%%%%%%%%%%%%%%%%%%%%%%%%%%%%%%%%%%
%\clearpage
\bibliographystyle{ecta}
\bibliography{biblio}

\begin{thebibliography}{45}
\newcommand{\enquote}[1]{``#1''}
\expandafter\ifx\csname natexlab\endcsname\relax\def\natexlab#1{#1}\fi

\bibitem[\protect\citeauthoryear{Ackerberg, Caves, and Frazer}{Ackerberg
  et~al.}{2015}]{ackerberg2015identification}
\textsc{Ackerberg, D.~A., K.~Caves, and G.~Frazer} (2015):
  \enquote{Identification properties of recent production function estimators,}
  \emph{Econometrica}, 83, 2411--2451.

\bibitem[\protect\citeauthoryear{Belloni, Chen, Chernozhukov, and
  Hansen}{Belloni et~al.}{2012}]{belloni2012sparse}
\textsc{Belloni, A., D.~Chen, V.~Chernozhukov, and C.~Hansen} (2012):
  \enquote{Sparse models and methods for optimal instruments with an
  application to eminent domain,} \emph{Econometrica}, 80, 2369--2429.

\bibitem[\protect\citeauthoryear{Belloni, Chernozhukov, Chetverikov, Hansen,
  and Kato}{Belloni et~al.}{2018{\natexlab{a}}}]{belloni2018high}
\textsc{Belloni, A., V.~Chernozhukov, D.~Chetverikov, C.~Hansen, and K.~Kato}
  (2018{\natexlab{a}}): \enquote{High-dimensional econometrics and regularized
  GMM,} \emph{arXiv preprint arXiv:1806.01888}.

\bibitem[\protect\citeauthoryear{Belloni, Chernozhukov, Chetverikov, and
  Wei}{Belloni et~al.}{2018{\natexlab{b}}}]{BCCW2018}
\textsc{Belloni, A., V.~Chernozhukov, D.~Chetverikov, and Y.~Wei}
  (2018{\natexlab{b}}): \enquote{Uniformly valid post-regularization confidence
  regions for many functional parameters in z-estimation framework,}
  \emph{Annals of statistics}, 46, 3643.

\bibitem[\protect\citeauthoryear{Belloni, Chernozhukov, and Hansen}{Belloni
  et~al.}{2013}]{BCH2014RES}
\textsc{Belloni, A., V.~Chernozhukov, and C.~Hansen} (2013):
  \enquote{{Inference on Treatment Effects after Selection among
  High-Dimensional Controls†},} \emph{The Review of Economic Studies}, 81,
  608--650.

\bibitem[\protect\citeauthoryear{Belloni, Chernozhukov, and Kato}{Belloni
  et~al.}{2015}]{belloni2015uniform}
\textsc{Belloni, A., V.~Chernozhukov, and K.~Kato} (2015): \enquote{Uniform
  post-selection inference for least absolute deviation regression and other
  Z-estimation problems,} \emph{Biometrika}, 102, 77--94.

\bibitem[\protect\citeauthoryear{Bickel, Ritov, and Tsybakov}{Bickel
  et~al.}{2009}]{bickel2009simultaneous}
\textsc{Bickel, P.~J., Y.~Ritov, and A.~B. Tsybakov} (2009):
  \enquote{Simultaneous analysis of Lasso and Dantzig selector,} \emph{The
  Annals of statistics}, 37, 1705--1732.

\bibitem[\protect\citeauthoryear{B{\"u}hlmann and van~de Geer}{B{\"u}hlmann and
  van~de Geer}{2011}]{buhlmann2011statistics}
\textsc{B{\"u}hlmann, P. and S.~van~de Geer} (2011): \emph{Statistics for
  high-dimensional data: methods, theory and applications}, Springer Science \&
  Business Media.

\bibitem[\protect\citeauthoryear{B{\"u}hlmann and Yu}{B{\"u}hlmann and
  Yu}{2003}]{buhlmann2003boosting}
\textsc{B{\"u}hlmann, P. and B.~Yu} (2003): \enquote{Boosting with the L 2
  loss: regression and classification,} \emph{Journal of the American
  Statistical Association}, 98, 324--339.

\bibitem[\protect\citeauthoryear{Candes and Tao}{Candes and
  Tao}{2007}]{candes2007dantzig}
\textsc{Candes, E. and T.~Tao} (2007): \enquote{The Dantzig selector:
  Statistical estimation when p is much larger than n,} \emph{Annals of
  statistics}, 35, 2313--2351.

\bibitem[\protect\citeauthoryear{Caner and Kock}{Caner and
  Kock}{2018{\natexlab{a}}}]{caner2018asymptotically}
\textsc{Caner, M. and A.~B. Kock} (2018{\natexlab{a}}): \enquote{Asymptotically
  honest confidence regions for high dimensional parameters by the desparsified
  conservative lasso,} \emph{Journal of Econometrics}, 203, 143--168.

\bibitem[\protect\citeauthoryear{Caner and Kock}{Caner and
  Kock}{2018{\natexlab{b}}}]{caner2018high}
---\hspace{-.1pt}---\hspace{-.1pt}--- (2018{\natexlab{b}}): \enquote{High
  dimensional linear GMM,} \emph{arXiv preprint arXiv:1811.08779}.

\bibitem[\protect\citeauthoryear{Chen}{Chen}{2007}]{chen2007large}
\textsc{Chen, X.} (2007): \enquote{Large sample sieve estimation of
  semi-nonparametric models,} \emph{Handbook of Econometrics}, 6B, 5549--5632.

\bibitem[\protect\citeauthoryear{Chernozhukov, Chetverikov, Demirer, Duflo,
  Hansen, Newey, and Robins}{Chernozhukov et~al.}{2018}]{ccddhnr}
\textsc{Chernozhukov, V., D.~Chetverikov, M.~Demirer, E.~Duflo, C.~Hansen,
  W.~Newey, and J.~Robins} (2018): \enquote{{Double/debiased machine learning
  for treatment and structural parameters},} \emph{The Econometrics Journal},
  21, C1--C68.

\bibitem[\protect\citeauthoryear{Chernozhukov, Chetverikov, and
  Kato}{Chernozhukov et~al.}{2015}]{CCK2015PTRF}
\textsc{Chernozhukov, V., D.~Chetverikov, and K.~Kato} (2015):
  \enquote{Comparison and anti-concentration bounds for maxima of {G}aussian
  random vectors,} \emph{Probability Theory and Related Fields}, 162, 47--70.

\bibitem[\protect\citeauthoryear{Chernozhukov, Escanciano, Ichimura, Newey, and
  Robins}{Chernozhukov et~al.}{2016}]{chernozhukov2016locally}
\textsc{Chernozhukov, V., J.~C. Escanciano, H.~Ichimura, W.~K. Newey, and J.~M.
  Robins} (2016): \enquote{Locally robust semiparametric estimation,}
  \emph{arXiv preprint arXiv:1608.00033}.

\bibitem[\protect\citeauthoryear{Efron, Hastie, Johnstone, Tibshirani
  et~al.}{Efron et~al.}{2004}]{efron2004least}
\textsc{Efron, B., T.~Hastie, I.~Johnstone, R.~Tibshirani, et~al.} (2004):
  \enquote{Least angle regression,} \emph{Annals of statistics}, 32, 407--499.

\bibitem[\protect\citeauthoryear{Fan}{Fan}{1991}]{fan1991optimal}
\textsc{Fan, J.} (1991): \enquote{On the optimal rates of convergence for
  nonparametric deconvolution problems,} \emph{Annals of Statistics},
  1257--1272.

\bibitem[\protect\citeauthoryear{Fan and Truong}{Fan and
  Truong}{1993}]{fan1993nonparametric}
\textsc{Fan, J. and Y.~K. Truong} (1993): \enquote{Nonparametric regression
  with errors in variables,} \emph{The Annals of Statistics}, 1900--1925.

\bibitem[\protect\citeauthoryear{Galbraith and Zinde-Walsh}{Galbraith and
  Zinde-Walsh}{2020}]{galbraith2020simple}
\textsc{Galbraith, J.~W. and V.~Zinde-Walsh} (2020): \enquote{Simple and
  reliable estimators of coefficients of interest in a model with
  high-dimensional confounding effects,} \emph{Journal of Econometrics}, 218,
  609--632.

\bibitem[\protect\citeauthoryear{Gallant and Nychka}{Gallant and
  Nychka}{1987}]{gallant1987semi}
\textsc{Gallant, A.~R. and D.~W. Nychka} (1987): \enquote{Semi-nonparametric
  maximum likelihood estimation,} \emph{Econometrica}, 55, 363--390.

\bibitem[\protect\citeauthoryear{Gao, Ing, and Yang}{Gao
  et~al.}{2013}]{GaoIngYang2013}
\textsc{Gao, F., C.-K. Ing, and Y.~Yang} (2013): \enquote{Metric entropy and
  sparse linear approximation of {$\ensuremath{\ell_q}$}-hulls for
  {$\ensuremath{0 < q \le 1}$},} \emph{Journal of Approximation Theory}, 166.

\bibitem[\protect\citeauthoryear{Giraud}{Giraud}{2015}]{giraud2015introduction}
\textsc{Giraud, C.} (2015): \enquote{Introduction to high-dimensional
  statistics,} \emph{Monographs on Statistics and Applied Probability}, 139,
  139.

\bibitem[\protect\citeauthoryear{Gold, Lederer, and Tao}{Gold
  et~al.}{2020}]{gold2020inference}
\textsc{Gold, D., J.~Lederer, and J.~Tao} (2020): \enquote{Inference for
  high-dimensional instrumental variables regression,} \emph{Journal of
  Econometrics}, 217, 79--111.

\bibitem[\protect\citeauthoryear{Hastie, Tibshirani, and Wainwright}{Hastie
  et~al.}{2019}]{hastie2019statistical}
\textsc{Hastie, T., R.~Tibshirani, and M.~Wainwright} (2019): \emph{Statistical
  learning with sparsity: the lasso and generalizations}, Chapman and Hall/CRC.

\bibitem[\protect\citeauthoryear{Ing}{Ing}{2007}]{ing2007accumulated}
\textsc{Ing, C.-K.} (2007): \enquote{Accumulated prediction errors, information
  criteria and optimal forecasting for autoregressive time series,}
  \emph{Annals of Statistics}, 35, 1238--1277.

\bibitem[\protect\citeauthoryear{Ing}{Ing}{2020}]{ing2020}
---\hspace{-.1pt}---\hspace{-.1pt}--- (2020): \enquote{Model selection for
  high-dimensional linear regression with dependent observations,} \emph{Ann.
  Statist.}, 48, 1959--1980.

\bibitem[\protect\citeauthoryear{Ing and Lai}{Ing and
  Lai}{2011}]{ing2011stepwise}
\textsc{Ing, C.-K. and T.~L. Lai} (2011): \enquote{A stepwise regression method
  and consistent model selection for high-dimensional sparse linear models,}
  \emph{Statistica Sinica}, 1473--1513.

\bibitem[\protect\citeauthoryear{Javanmard and Montanari}{Javanmard and
  Montanari}{2014}]{javanmard2014confidence}
\textsc{Javanmard, A. and A.~Montanari} (2014): \enquote{Confidence intervals
  and hypothesis testing for high-dimensional regression,} \emph{The Journal of
  Machine Learning Research}, 15, 2869--2909.

\bibitem[\protect\citeauthoryear{Kozbur}{Kozbur}{2017}]{kozbur2017testing}
\textsc{Kozbur, D.} (2017): \enquote{Testing-based forward model selection,}
  \emph{American Economic Review}, 107, 266--69.

\bibitem[\protect\citeauthoryear{Kozbur}{Kozbur}{2020}]{kozbur2020analysis}
---\hspace{-.1pt}---\hspace{-.1pt}--- (2020): \enquote{Analysis of
  Testing-Based Forward Model Selection,} \emph{Econometrica}, 88, 2147--2173.

\bibitem[\protect\citeauthoryear{Kueck, Luo, Spindler, and Wang}{Kueck
  et~al.}{2021}]{kueck2021estimation}
\textsc{Kueck, J., Y.~Luo, M.~Spindler, and Z.~Wang} (2021):
  \enquote{Estimation and Inference of Treatment Effects with $L_2$-Boosting in
  High-Dimensional Settings,} .

\bibitem[\protect\citeauthoryear{Levinsohn and Petrin}{Levinsohn and
  Petrin}{2003}]{levinsohn2003estimating}
\textsc{Levinsohn, J. and A.~Petrin} (2003): \enquote{Estimating production
  functions using inputs to control for unobservables,} \emph{Review of
  Economic Studies}, 70, 317--341.

\bibitem[\protect\citeauthoryear{Liu}{Liu}{1991}]{liu1991entry}
\textsc{Liu, L.} (1991): \enquote{Entry-exit and productivity change: An
  empirical analysis of efficiency frontiers,} Ph.D. thesis, University of
  Michigan.

\bibitem[\protect\citeauthoryear{Marschak and Andrews}{Marschak and
  Andrews}{1944}]{marschak1944random}
\textsc{Marschak, J. and W.~H. Andrews} (1944): \enquote{Random simultaneous
  equations and the theory of production,} \emph{Econometrica}, 143--205.

\bibitem[\protect\citeauthoryear{Olley and Pakes}{Olley and
  Pakes}{1996}]{OlleyPakes1996}
\textsc{Olley, G.~S. and A.~Pakes} (1996): \enquote{The Dynamics of
  Productivity in the Telecommunications Equipment Industry,}
  \emph{Econometrica}, 64, 1263--1297.

\bibitem[\protect\citeauthoryear{Petrin, Poi, and Levinsohn}{Petrin
  et~al.}{2004}]{petrin2004production}
\textsc{Petrin, A., B.~P. Poi, and J.~Levinsohn} (2004): \enquote{Production
  function estimation in Stata using inputs to control for unobservables,}
  \emph{Stata Journal}, 4, 113--123.

\bibitem[\protect\citeauthoryear{Robinson}{Robinson}{1988}]{Robinson}
\textsc{Robinson, P.} (1988): \enquote{Root- N-Consistent Semiparametric
  Regression,} \emph{Econometrica}, 56, 931--54.

\bibitem[\protect\citeauthoryear{Shibata}{Shibata}{1980}]{shibata1980asymptotically}
\textsc{Shibata, R.} (1980): \enquote{Asymptotically efficient selection of the
  order of the model for estimating parameters of a linear process,}
  \emph{Annals of Statistics}, 147--164.

\bibitem[\protect\citeauthoryear{Temlyakov}{Temlyakov}{2000}]{temlyakov2000weak}
\textsc{Temlyakov, V.~N.} (2000): \enquote{Weak greedy algorithms,}
  \emph{Advances in Computational Mathematics}, 12, 213--227.

\bibitem[\protect\citeauthoryear{Tibshirani}{Tibshirani}{1996}]{tibshirani1996regression}
\textsc{Tibshirani, R.} (1996): \enquote{Regression shrinkage and selection via
  the lasso,} \emph{Journal of the Royal Statistical Society: Series B
  (Methodological)}, 58, 267--288.

\bibitem[\protect\citeauthoryear{Tropp}{Tropp}{2004}]{tropp2004greed}
\textsc{Tropp, J.~A.} (2004): \enquote{Greed is Good: Algorithmic Results for
  Sparse Approximation,} \emph{IEEE Trans. Inform. Theory}, 10, 2231--2242.

\bibitem[\protect\citeauthoryear{Tropp and Gilbert}{Tropp and
  Gilbert}{2007}]{tropp2007signal}
\textsc{Tropp, J.~A. and A.~C. Gilbert} (2007): \enquote{Signal Recovery from
  Random Measurements,} \emph{Adaptif Kom{\c{s}}uluk Se{\c{c}}imi ve
  A{\u{g}}{\i}rl{\i}k Atama Y{\"o}ntemleri ile Hiperspektral
  G{\"o}r{\"u}nt{\"u}lerin S{\i}n{\i}fland{\i}r{\i}lmas{\i}}.

\bibitem[\protect\citeauthoryear{van~de Geer, B{\"u}hlmann, Ritov, and
  Dezeure}{van~de Geer et~al.}{2014}]{van2014asymptotically}
\textsc{van~de Geer, S., P.~B{\"u}hlmann, Y.~Ritov, and R.~Dezeure} (2014):
  \enquote{On asymptotically optimal confidence regions and tests for
  high-dimensional models,} \emph{Annals of Statistics}, 42, 1166--1202.

\bibitem[\protect\citeauthoryear{Zhang and Zhang}{Zhang and
  Zhang}{2014}]{zhang2014confidence}
\textsc{Zhang, C.-H. and S.~S. Zhang} (2014): \enquote{Confidence intervals for
  low dimensional parameters in high dimensional linear models,} \emph{Journal
  of the Royal Statistical Society: Series B: Statistical Methodology},
  217--242.

\end{thebibliography}
%%%%%%%%%%%%%%%%%%%%%%%%%%%%%%%%%%%%%%%%%%%%%%%%%%%%%%%%%%%%%%%%%%%%%

\end{document}